\documentclass[longauth]{aa}

\usepackage{graphicx}
\usepackage{natbib}
\usepackage{scalerel}
\usepackage[utf8]{inputenc}

\usepackage[table]{xcolor}

\usepackage{txfonts}
\usepackage[pdfencoding=auto,psdextra]{hyperref}
\hypersetup{
    colorlinks=true,
    linkcolor=blue,
    filecolor=magenta,      
    urlcolor=blue,
    citecolor=blue
}
\usepackage[nameinlink,capitalise]{cleveref}
\crefname{section}{Sect.}{Sects.}
\Crefname{section}{Section}{Sections}
\crefname{figure}{Fig.}{Figs.}
\Crefname{figure}{Figure}{Figures}
\crefname{equation}{Eq.}{Eqs.}
\Crefname{equation}{Equation}{Equations}
\crefname{table}{Table}{Tables}
\crefname{appendix}{Appendix}{Appendices}

\crefformat{equation}{#2Eq.~(#1)#3}
\Crefformat{equation}{#2Eq.~(#1)#3}
\crefrangeformat{equation}{Eqs.~(#3#1#4)--(#5#2#6)}
\Crefrangeformat{equation}{Eqs.~(#3#1#4)--(#5#2#6)}

\makeatletter
\renewcommand*\aa@pageof{, page \thepage{} of \pageref*{LastPage}}
\makeatother

\usepackage[utf8]{inputenc}

\usepackage[switch, modulo]{lineno}

\usepackage{euclid}

\begin{document}
%
%
   \title{\Euclid \/: Photometric redshift calibration with \\ self-organising maps\thanks{This paper is published on
     behalf of the Euclid Consortium}}


\institute{Max Planck Institute for Extraterrestrial Physics, Giessenbachstr. 1, 85748 Garching, Germany\label{aff1}
\and
Ruhr University Bochum, Faculty of Physics and Astronomy, Astronomical Institute (AIRUB), German Centre for Cosmological Lensing (GCCL), 44780 Bochum, Germany\label{aff2}
\and
Universit\"at Bonn, Argelander-Institut f\"ur Astronomie, Auf dem H\"ugel 71, 53121 Bonn, Germany\label{aff3}
\and
Aix-Marseille Universit\'e, CNRS, CNES, LAM, Marseille, France\label{aff4}
\and
Institut de F\'{i}sica d'Altes Energies (IFAE), The Barcelona Institute of Science and Technology, Campus UAB, 08193 Bellaterra (Barcelona), Spain\label{aff5}
\and
Serra H\'unter Fellow, Departament de F\'{\i}sica, Universitat Aut\`onoma de Barcelona, E-08193 Bellaterra, Spain\label{aff6}
\and
INAF-Osservatorio di Astrofisica e Scienza dello Spazio di Bologna, Via Piero Gobetti 93/3, 40129 Bologna, Italy\label{aff7}
\and
Infrared Processing and Analysis Center, California Institute of Technology, Pasadena, CA 91125, USA\label{aff8}
\and
Department of Astronomy, University of Geneva, ch. d'Ecogia 16, 1290 Versoix, Switzerland\label{aff9}
\and
Leiden Observatory, Leiden University, Einsteinweg 55, 2333 CC Leiden, The Netherlands\label{aff10}
\and
ESAC/ESA, Camino Bajo del Castillo, s/n., Urb. Villafranca del Castillo, 28692 Villanueva de la Ca\~nada, Madrid, Spain\label{aff11}
\and
School of Mathematics and Physics, University of Surrey, Guildford, Surrey, GU2 7XH, UK\label{aff12}
\and
INAF-Osservatorio Astronomico di Brera, Via Brera 28, 20122 Milano, Italy\label{aff13}
\and
IFPU, Institute for Fundamental Physics of the Universe, via Beirut 2, 34151 Trieste, Italy\label{aff14}
\and
INAF-Osservatorio Astronomico di Trieste, Via G. B. Tiepolo 11, 34143 Trieste, Italy\label{aff15}
\and
INFN, Sezione di Trieste, Via Valerio 2, 34127 Trieste TS, Italy\label{aff16}
\and
SISSA, International School for Advanced Studies, Via Bonomea 265, 34136 Trieste TS, Italy\label{aff17}
\and
Dipartimento di Fisica e Astronomia, Universit\`a di Bologna, Via Gobetti 93/2, 40129 Bologna, Italy\label{aff18}
\and
INFN-Sezione di Bologna, Viale Berti Pichat 6/2, 40127 Bologna, Italy\label{aff19}
\and
INAF-Osservatorio Astronomico di Padova, Via dell'Osservatorio 5, 35122 Padova, Italy\label{aff20}
\and
Universit\"ats-Sternwarte M\"unchen, Fakult\"at f\"ur Physik, Ludwig-Maximilians-Universit\"at M\"unchen, Scheinerstrasse 1, 81679 M\"unchen, Germany\label{aff21}
\and
Dipartimento di Fisica, Universit\`a di Genova, Via Dodecaneso 33, 16146, Genova, Italy\label{aff22}
\and
INFN-Sezione di Genova, Via Dodecaneso 33, 16146, Genova, Italy\label{aff23}
\and
Department of Physics "E. Pancini", University Federico II, Via Cinthia 6, 80126, Napoli, Italy\label{aff24}
\and
INAF-Osservatorio Astronomico di Capodimonte, Via Moiariello 16, 80131 Napoli, Italy\label{aff25}
\and
Dipartimento di Fisica, Universit\`a degli Studi di Torino, Via P. Giuria 1, 10125 Torino, Italy\label{aff26}
\and
INFN-Sezione di Torino, Via P. Giuria 1, 10125 Torino, Italy\label{aff27}
\and
INAF-Osservatorio Astrofisico di Torino, Via Osservatorio 20, 10025 Pino Torinese (TO), Italy\label{aff28}
\and
European Space Agency/ESTEC, Keplerlaan 1, 2201 AZ Noordwijk, The Netherlands\label{aff29}
\and
Institute Lorentz, Leiden University, Niels Bohrweg 2, 2333 CA Leiden, The Netherlands\label{aff30}
\and
INAF-IASF Milano, Via Alfonso Corti 12, 20133 Milano, Italy\label{aff31}
\and
INAF-Osservatorio Astronomico di Roma, Via Frascati 33, 00078 Monteporzio Catone, Italy\label{aff32}
\and
INFN-Sezione di Roma, Piazzale Aldo Moro, 2 - c/o Dipartimento di Fisica, Edificio G. Marconi, 00185 Roma, Italy\label{aff33}
\and
Centro de Investigaciones Energ\'eticas, Medioambientales y Tecnol\'ogicas (CIEMAT), Avenida Complutense 40, 28040 Madrid, Spain\label{aff34}
\and
Port d'Informaci\'{o} Cient\'{i}fica, Campus UAB, C. Albareda s/n, 08193 Bellaterra (Barcelona), Spain\label{aff35}
\and
Institut d'Estudis Espacials de Catalunya (IEEC),  Edifici RDIT, Campus UPC, 08860 Castelldefels, Barcelona, Spain\label{aff36}
\and
Institute of Space Sciences (ICE, CSIC), Campus UAB, Carrer de Can Magrans, s/n, 08193 Barcelona, Spain\label{aff37}
\and
Institute for Theoretical Particle Physics and Cosmology (TTK), RWTH Aachen University, 52056 Aachen, Germany\label{aff38}
\and
INFN section of Naples, Via Cinthia 6, 80126, Napoli, Italy\label{aff39}
\and
Institute for Astronomy, University of Hawaii, 2680 Woodlawn Drive, Honolulu, HI 96822, USA\label{aff40}
\and
Dipartimento di Fisica e Astronomia "Augusto Righi" - Alma Mater Studiorum Universit\`a di Bologna, Viale Berti Pichat 6/2, 40127 Bologna, Italy\label{aff41}
\and
Instituto de Astrof\'{\i}sica de Canarias, V\'{\i}a L\'actea, 38205 La Laguna, Tenerife, Spain\label{aff42}
\and
Institute for Astronomy, University of Edinburgh, Royal Observatory, Blackford Hill, Edinburgh EH9 3HJ, UK\label{aff43}
\and
Jodrell Bank Centre for Astrophysics, Department of Physics and Astronomy, University of Manchester, Oxford Road, Manchester M13 9PL, UK\label{aff44}
\and
European Space Agency/ESRIN, Largo Galileo Galilei 1, 00044 Frascati, Roma, Italy\label{aff45}
\and
Universit\'e Claude Bernard Lyon 1, CNRS/IN2P3, IP2I Lyon, UMR 5822, Villeurbanne, F-69100, France\label{aff46}
\and
Institut de Ci\`{e}ncies del Cosmos (ICCUB), Universitat de Barcelona (IEEC-UB), Mart\'{i} i Franqu\`{e}s 1, 08028 Barcelona, Spain\label{aff47}
\and
Instituci\'o Catalana de Recerca i Estudis Avan\c{c}ats (ICREA), Passeig de Llu\'{\i}s Companys 23, 08010 Barcelona, Spain\label{aff48}
\and
UCB Lyon 1, CNRS/IN2P3, IUF, IP2I Lyon, 4 rue Enrico Fermi, 69622 Villeurbanne, France\label{aff49}
\and
Mullard Space Science Laboratory, University College London, Holmbury St Mary, Dorking, Surrey RH5 6NT, UK\label{aff50}
\and
Departamento de F\'isica, Faculdade de Ci\^encias, Universidade de Lisboa, Edif\'icio C8, Campo Grande, PT1749-016 Lisboa, Portugal\label{aff51}
\and
Instituto de Astrof\'isica e Ci\^encias do Espa\c{c}o, Faculdade de Ci\^encias, Universidade de Lisboa, Campo Grande, 1749-016 Lisboa, Portugal\label{aff52}
\and
INFN-Padova, Via Marzolo 8, 35131 Padova, Italy\label{aff53}
\and
Aix-Marseille Universit\'e, CNRS/IN2P3, CPPM, Marseille, France\label{aff54}
\and
INAF-Istituto di Astrofisica e Planetologia Spaziali, via del Fosso del Cavaliere, 100, 00100 Roma, Italy\label{aff55}
\and
Universit\'e Paris-Saclay, Universit\'e Paris Cit\'e, CEA, CNRS, AIM, 91191, Gif-sur-Yvette, France\label{aff56}
\and
Space Science Data Center, Italian Space Agency, via del Politecnico snc, 00133 Roma, Italy\label{aff57}
\and
INFN-Bologna, Via Irnerio 46, 40126 Bologna, Italy\label{aff58}
\and
University Observatory, LMU Faculty of Physics, Scheinerstrasse 1, 81679 Munich, Germany\label{aff59}
\and
Institute of Theoretical Astrophysics, University of Oslo, P.O. Box 1029 Blindern, 0315 Oslo, Norway\label{aff60}
\and
Jet Propulsion Laboratory, California Institute of Technology, 4800 Oak Grove Drive, Pasadena, CA, 91109, USA\label{aff61}
\and
Felix Hormuth Engineering, Goethestr. 17, 69181 Leimen, Germany\label{aff62}
\and
Technical University of Denmark, Elektrovej 327, 2800 Kgs. Lyngby, Denmark\label{aff63}
\and
Cosmic Dawn Center (DAWN), Denmark\label{aff64}
\and
Institut d'Astrophysique de Paris, UMR 7095, CNRS, and Sorbonne Universit\'e, 98 bis boulevard Arago, 75014 Paris, France\label{aff65}
\and
Max-Planck-Institut f\"ur Astronomie, K\"onigstuhl 17, 69117 Heidelberg, Germany\label{aff66}
\and
NASA Goddard Space Flight Center, Greenbelt, MD 20771, USA\label{aff67}
\and
Department of Physics and Astronomy, University College London, Gower Street, London WC1E 6BT, UK\label{aff68}
\and
Department of Physics and Helsinki Institute of Physics, Gustaf H\"allstr\"omin katu 2, 00014 University of Helsinki, Finland\label{aff69}
\and
Department of Physics, P.O. Box 64, 00014 University of Helsinki, Finland\label{aff70}
\and
Helsinki Institute of Physics, Gustaf H{\"a}llstr{\"o}min katu 2, University of Helsinki, Helsinki, Finland\label{aff71}
\and
Laboratoire d'etude de l'Univers et des phenomenes eXtremes, Observatoire de Paris, Universit\'e PSL, Sorbonne Universit\'e, CNRS, 92190 Meudon, France\label{aff72}
\and
SKA Observatory, Jodrell Bank, Lower Withington, Macclesfield, Cheshire SK11 9FT, UK\label{aff73}
\and
Dipartimento di Fisica "Aldo Pontremoli", Universit\`a degli Studi di Milano, Via Celoria 16, 20133 Milano, Italy\label{aff74}
\and
INFN-Sezione di Milano, Via Celoria 16, 20133 Milano, Italy\label{aff75}
\and
Dipartimento di Fisica e Astronomia "Augusto Righi" - Alma Mater Studiorum Universit\`a di Bologna, via Piero Gobetti 93/2, 40129 Bologna, Italy\label{aff76}
\and
Department of Physics, Institute for Computational Cosmology, Durham University, South Road, Durham, DH1 3LE, UK\label{aff77}
\and
Universit\'e Paris Cit\'e, CNRS, Astroparticule et Cosmologie, 75013 Paris, France\label{aff78}
\and
CNRS-UCB International Research Laboratory, Centre Pierre Bin\'etruy, IRL2007, CPB-IN2P3, Berkeley, USA\label{aff79}
\and
University of Applied Sciences and Arts of Northwestern Switzerland, School of Engineering, 5210 Windisch, Switzerland\label{aff80}
\and
Institut d'Astrophysique de Paris, 98bis Boulevard Arago, 75014, Paris, France\label{aff81}
\and
Institute of Physics, Laboratory of Astrophysics, Ecole Polytechnique F\'ed\'erale de Lausanne (EPFL), Observatoire de Sauverny, 1290 Versoix, Switzerland\label{aff82}
\and
Telespazio UK S.L. for European Space Agency (ESA), Camino bajo del Castillo, s/n, Urbanizacion Villafranca del Castillo, Villanueva de la Ca\~nada, 28692 Madrid, Spain\label{aff83}
\and
DARK, Niels Bohr Institute, University of Copenhagen, Jagtvej 155, 2200 Copenhagen, Denmark\label{aff84}
\and
Centre National d'Etudes Spatiales -- Centre spatial de Toulouse, 18 avenue Edouard Belin, 31401 Toulouse Cedex 9, France\label{aff85}
\and
Institute of Space Science, Str. Atomistilor, nr. 409 M\u{a}gurele, Ilfov, 077125, Romania\label{aff86}
\and
Consejo Superior de Investigaciones Cientificas, Calle Serrano 117, 28006 Madrid, Spain\label{aff87}
\and
Universidad de La Laguna, Departamento de Astrof\'{\i}sica, 38206 La Laguna, Tenerife, Spain\label{aff88}
\and
Dipartimento di Fisica e Astronomia "G. Galilei", Universit\`a di Padova, Via Marzolo 8, 35131 Padova, Italy\label{aff89}
\and
Institut f\"ur Theoretische Physik, University of Heidelberg, Philosophenweg 16, 69120 Heidelberg, Germany\label{aff90}
\and
Institut de Recherche en Astrophysique et Plan\'etologie (IRAP), Universit\'e de Toulouse, CNRS, UPS, CNES, 14 Av. Edouard Belin, 31400 Toulouse, France\label{aff91}
\and
Universit\'e St Joseph; Faculty of Sciences, Beirut, Lebanon\label{aff92}
\and
Departamento de F\'isica, FCFM, Universidad de Chile, Blanco Encalada 2008, Santiago, Chile\label{aff93}
\and
Universit\"at Innsbruck, Institut f\"ur Astro- und Teilchenphysik, Technikerstr. 25/8, 6020 Innsbruck, Austria\label{aff94}
\and
Satlantis, University Science Park, Sede Bld 48940, Leioa-Bilbao, Spain\label{aff95}
\and
Centre for Electronic Imaging, Open University, Walton Hall, Milton Keynes, MK7~6AA, UK\label{aff96}
\and
Instituto de Astrof\'isica e Ci\^encias do Espa\c{c}o, Faculdade de Ci\^encias, Universidade de Lisboa, Tapada da Ajuda, 1349-018 Lisboa, Portugal\label{aff97}
\and
Cosmic Dawn Center (DAWN)\label{aff98}
\and
Niels Bohr Institute, University of Copenhagen, Jagtvej 128, 2200 Copenhagen, Denmark\label{aff99}
\and
Universidad Polit\'ecnica de Cartagena, Departamento de Electr\'onica y Tecnolog\'ia de Computadoras,  Plaza del Hospital 1, 30202 Cartagena, Spain\label{aff100}
\and
Kapteyn Astronomical Institute, University of Groningen, PO Box 800, 9700 AV Groningen, The Netherlands\label{aff101}
\and
INAF, Istituto di Radioastronomia, Via Piero Gobetti 101, 40129 Bologna, Italy\label{aff102}
\and
Department of Physics, Oxford University, Keble Road, Oxford OX1 3RH, UK\label{aff103}
\and
ICL, Junia, Universit\'e Catholique de Lille, LITL, 59000 Lille, France\label{aff104}}        

\newcommand{\orcid}[1]{} 
\author{W.~Roster\orcid{0000-0002-9149-6528}\thanks{\email{wroster@mpe.mpg.de}}\inst{\ref{aff1}}
\and A.~H.~Wright\orcid{0000-0001-7363-7932}\inst{\ref{aff2}}
\and H.~Hildebrandt\orcid{0000-0002-9814-3338}\inst{\ref{aff2}}
\and R.~Reischke\orcid{0000-0001-5404-8753}\inst{\ref{aff3}}
\and O.~Ilbert\orcid{0000-0002-7303-4397}\inst{\ref{aff4}}
\and W.~d'Assignies~D.~\orcid{0000-0002-9719-1717}\inst{\ref{aff5}}
\and M.~Manera\orcid{0000-0003-4962-8934}\inst{\ref{aff6},\ref{aff5}}
\and M.~Bolzonella\orcid{0000-0003-3278-4607}\inst{\ref{aff7}}
\and D.~C.~Masters\orcid{0000-0001-5382-6138}\inst{\ref{aff8}}
\and S.~Paltani\orcid{0000-0002-8108-9179}\inst{\ref{aff9}}
\and W.~G.~Hartley\inst{\ref{aff9}}
\and Y.~Kang\orcid{0009-0000-8588-7250}\inst{\ref{aff9}}
\and H.~Hoekstra\orcid{0000-0002-0641-3231}\inst{\ref{aff10}}
\and B.~Altieri\orcid{0000-0003-3936-0284}\inst{\ref{aff11}}
\and A.~Amara\inst{\ref{aff12}}
\and S.~Andreon\orcid{0000-0002-2041-8784}\inst{\ref{aff13}}
\and N.~Auricchio\orcid{0000-0003-4444-8651}\inst{\ref{aff7}}
\and C.~Baccigalupi\orcid{0000-0002-8211-1630}\inst{\ref{aff14},\ref{aff15},\ref{aff16},\ref{aff17}}
\and M.~Baldi\orcid{0000-0003-4145-1943}\inst{\ref{aff18},\ref{aff7},\ref{aff19}}
\and A.~Balestra\orcid{0000-0002-6967-261X}\inst{\ref{aff20}}
\and S.~Bardelli\orcid{0000-0002-8900-0298}\inst{\ref{aff7}}
\and P.~Battaglia\orcid{0000-0002-7337-5909}\inst{\ref{aff7}}
\and R.~Bender\orcid{0000-0001-7179-0626}\inst{\ref{aff1},\ref{aff21}}
\and A.~Biviano\orcid{0000-0002-0857-0732}\inst{\ref{aff15},\ref{aff14}}
\and E.~Branchini\orcid{0000-0002-0808-6908}\inst{\ref{aff22},\ref{aff23},\ref{aff13}}
\and M.~Brescia\orcid{0000-0001-9506-5680}\inst{\ref{aff24},\ref{aff25}}
\and S.~Camera\orcid{0000-0003-3399-3574}\inst{\ref{aff26},\ref{aff27},\ref{aff28}}
\and G.~Ca\~nas-Herrera\orcid{0000-0003-2796-2149}\inst{\ref{aff29},\ref{aff30},\ref{aff10}}
\and V.~Capobianco\orcid{0000-0002-3309-7692}\inst{\ref{aff28}}
\and C.~Carbone\orcid{0000-0003-0125-3563}\inst{\ref{aff31}}
\and V.~F.~Cardone\inst{\ref{aff32},\ref{aff33}}
\and J.~Carretero\orcid{0000-0002-3130-0204}\inst{\ref{aff34},\ref{aff35}}
\and R.~Casas\orcid{0000-0002-8165-5601}\inst{\ref{aff36},\ref{aff37}}
\and S.~Casas\orcid{0000-0002-4751-5138}\inst{\ref{aff38}}
\and F.~J.~Castander\orcid{0000-0001-7316-4573}\inst{\ref{aff37},\ref{aff36}}
\and M.~Castellano\orcid{0000-0001-9875-8263}\inst{\ref{aff32}}
\and G.~Castignani\orcid{0000-0001-6831-0687}\inst{\ref{aff7}}
\and S.~Cavuoti\orcid{0000-0002-3787-4196}\inst{\ref{aff25},\ref{aff39}}
\and K.~C.~Chambers\orcid{0000-0001-6965-7789}\inst{\ref{aff40}}
\and A.~Cimatti\inst{\ref{aff41}}
\and C.~Colodro-Conde\inst{\ref{aff42}}
\and G.~Congedo\orcid{0000-0003-2508-0046}\inst{\ref{aff43}}
\and C.~J.~Conselice\orcid{0000-0003-1949-7638}\inst{\ref{aff44}}
\and L.~Conversi\orcid{0000-0002-6710-8476}\inst{\ref{aff45},\ref{aff11}}
\and Y.~Copin\orcid{0000-0002-5317-7518}\inst{\ref{aff46}}
\and A.~Costille\inst{\ref{aff4}}
\and F.~Courbin\orcid{0000-0003-0758-6510}\inst{\ref{aff47},\ref{aff48}}
\and H.~M.~Courtois\orcid{0000-0003-0509-1776}\inst{\ref{aff49}}
\and M.~Cropper\orcid{0000-0003-4571-9468}\inst{\ref{aff50}}
\and A.~Da~Silva\orcid{0000-0002-6385-1609}\inst{\ref{aff51},\ref{aff52}}
\and H.~Degaudenzi\orcid{0000-0002-5887-6799}\inst{\ref{aff9}}
\and S.~de~la~Torre\inst{\ref{aff4}}
\and G.~De~Lucia\orcid{0000-0002-6220-9104}\inst{\ref{aff15}}
\and F.~Dubath\orcid{0000-0002-6533-2810}\inst{\ref{aff9}}
\and C.~A.~J.~Duncan\orcid{0009-0003-3573-0791}\inst{\ref{aff43},\ref{aff44}}
\and X.~Dupac\inst{\ref{aff11}}
\and S.~Dusini\orcid{0000-0002-1128-0664}\inst{\ref{aff53}}
\and S.~Escoffier\orcid{0000-0002-2847-7498}\inst{\ref{aff54}}
\and M.~Farina\orcid{0000-0002-3089-7846}\inst{\ref{aff55}}
\and R.~Farinelli\inst{\ref{aff7}}
\and S.~Farrens\orcid{0000-0002-9594-9387}\inst{\ref{aff56}}
\and F.~Faustini\orcid{0000-0001-6274-5145}\inst{\ref{aff32},\ref{aff57}}
\and S.~Ferriol\inst{\ref{aff46}}
\and F.~Finelli\orcid{0000-0002-6694-3269}\inst{\ref{aff7},\ref{aff58}}
\and P.~Fosalba\orcid{0000-0002-1510-5214}\inst{\ref{aff36},\ref{aff37}}
\and N.~Fourmanoit\orcid{0009-0005-6816-6925}\inst{\ref{aff54}}
\and M.~Frailis\orcid{0000-0002-7400-2135}\inst{\ref{aff15}}
\and E.~Franceschi\orcid{0000-0002-0585-6591}\inst{\ref{aff7}}
\and M.~Fumana\orcid{0000-0001-6787-5950}\inst{\ref{aff31}}
\and S.~Galeotta\orcid{0000-0002-3748-5115}\inst{\ref{aff15}}
\and K.~George\orcid{0000-0002-1734-8455}\inst{\ref{aff59}}
\and W.~Gillard\orcid{0000-0003-4744-9748}\inst{\ref{aff54}}
\and B.~Gillis\orcid{0000-0002-4478-1270}\inst{\ref{aff43}}
\and C.~Giocoli\orcid{0000-0002-9590-7961}\inst{\ref{aff7},\ref{aff19}}
\and J.~Gracia-Carpio\inst{\ref{aff1}}
\and A.~Grazian\orcid{0000-0002-5688-0663}\inst{\ref{aff20}}
\and F.~Grupp\inst{\ref{aff1},\ref{aff21}}
\and S.~V.~H.~Haugan\orcid{0000-0001-9648-7260}\inst{\ref{aff60}}
\and W.~Holmes\inst{\ref{aff61}}
\and F.~Hormuth\inst{\ref{aff62}}
\and A.~Hornstrup\orcid{0000-0002-3363-0936}\inst{\ref{aff63},\ref{aff64}}
\and P.~Hudelot\inst{\ref{aff65}}
\and K.~Jahnke\orcid{0000-0003-3804-2137}\inst{\ref{aff66}}
\and M.~Jhabvala\inst{\ref{aff67}}
\and B.~Joachimi\orcid{0000-0001-7494-1303}\inst{\ref{aff68}}
\and E.~Keih\"anen\orcid{0000-0003-1804-7715}\inst{\ref{aff69}}
\and S.~Kermiche\orcid{0000-0002-0302-5735}\inst{\ref{aff54}}
\and B.~Kubik\orcid{0009-0006-5823-4880}\inst{\ref{aff46}}
\and H.~Kurki-Suonio\orcid{0000-0002-4618-3063}\inst{\ref{aff70},\ref{aff71}}
\and A.~M.~C.~Le~Brun\orcid{0000-0002-0936-4594}\inst{\ref{aff72}}
\and D.~Le~Mignant\orcid{0000-0002-5339-5515}\inst{\ref{aff4}}
\and S.~Ligori\orcid{0000-0003-4172-4606}\inst{\ref{aff28}}
\and P.~B.~Lilje\orcid{0000-0003-4324-7794}\inst{\ref{aff60}}
\and V.~Lindholm\orcid{0000-0003-2317-5471}\inst{\ref{aff70},\ref{aff71}}
\and I.~Lloro\orcid{0000-0001-5966-1434}\inst{\ref{aff73}}
\and D.~Maino\inst{\ref{aff74},\ref{aff31},\ref{aff75}}
\and E.~Maiorano\orcid{0000-0003-2593-4355}\inst{\ref{aff7}}
\and O.~Mansutti\orcid{0000-0001-5758-4658}\inst{\ref{aff15}}
\and O.~Marggraf\orcid{0000-0001-7242-3852}\inst{\ref{aff3}}
\and M.~Martinelli\orcid{0000-0002-6943-7732}\inst{\ref{aff32},\ref{aff33}}
\and N.~Martinet\orcid{0000-0003-2786-7790}\inst{\ref{aff4}}
\and F.~Marulli\orcid{0000-0002-8850-0303}\inst{\ref{aff76},\ref{aff7},\ref{aff19}}
\and R.~J.~Massey\orcid{0000-0002-6085-3780}\inst{\ref{aff77}}
\and E.~Medinaceli\orcid{0000-0002-4040-7783}\inst{\ref{aff7}}
\and S.~Mei\orcid{0000-0002-2849-559X}\inst{\ref{aff78},\ref{aff79}}
\and M.~Melchior\inst{\ref{aff80}}
\and Y.~Mellier\inst{\ref{aff81},\ref{aff65}}
\and M.~Meneghetti\orcid{0000-0003-1225-7084}\inst{\ref{aff7},\ref{aff19}}
\and E.~Merlin\orcid{0000-0001-6870-8900}\inst{\ref{aff32}}
\and G.~Meylan\inst{\ref{aff82}}
\and A.~Mora\orcid{0000-0002-1922-8529}\inst{\ref{aff83}}
\and M.~Moresco\orcid{0000-0002-7616-7136}\inst{\ref{aff76},\ref{aff7}}
\and L.~Moscardini\orcid{0000-0002-3473-6716}\inst{\ref{aff76},\ref{aff7},\ref{aff19}}
\and R.~Nakajima\orcid{0009-0009-1213-7040}\inst{\ref{aff3}}
\and C.~Neissner\orcid{0000-0001-8524-4968}\inst{\ref{aff5},\ref{aff35}}
\and S.-M.~Niemi\orcid{0009-0005-0247-0086}\inst{\ref{aff29}}
\and C.~Padilla\orcid{0000-0001-7951-0166}\inst{\ref{aff5}}
\and F.~Pasian\orcid{0000-0002-4869-3227}\inst{\ref{aff15}}
\and K.~Pedersen\inst{\ref{aff84}}
\and V.~Pettorino\inst{\ref{aff29}}
\and S.~Pires\orcid{0000-0002-0249-2104}\inst{\ref{aff56}}
\and G.~Polenta\orcid{0000-0003-4067-9196}\inst{\ref{aff57}}
\and M.~Poncet\inst{\ref{aff85}}
\and L.~A.~Popa\inst{\ref{aff86}}
\and L.~Pozzetti\orcid{0000-0001-7085-0412}\inst{\ref{aff7}}
\and F.~Raison\orcid{0000-0002-7819-6918}\inst{\ref{aff1}}
\and R.~Rebolo\orcid{0000-0003-3767-7085}\inst{\ref{aff42},\ref{aff87},\ref{aff88}}
\and A.~Renzi\orcid{0000-0001-9856-1970}\inst{\ref{aff89},\ref{aff53}}
\and J.~Rhodes\orcid{0000-0002-4485-8549}\inst{\ref{aff61}}
\and G.~Riccio\inst{\ref{aff25}}
\and E.~Romelli\orcid{0000-0003-3069-9222}\inst{\ref{aff15}}
\and M.~Roncarelli\orcid{0000-0001-9587-7822}\inst{\ref{aff7}}
\and C.~Rosset\orcid{0000-0003-0286-2192}\inst{\ref{aff78}}
\and E.~Rossetti\orcid{0000-0003-0238-4047}\inst{\ref{aff18}}
\and R.~Saglia\orcid{0000-0003-0378-7032}\inst{\ref{aff21},\ref{aff1}}
\and Z.~Sakr\orcid{0000-0002-4823-3757}\inst{\ref{aff90},\ref{aff91},\ref{aff92}}
\and D.~Sapone\orcid{0000-0001-7089-4503}\inst{\ref{aff93}}
\and B.~Sartoris\orcid{0000-0003-1337-5269}\inst{\ref{aff21},\ref{aff15}}
\and M.~Schirmer\orcid{0000-0003-2568-9994}\inst{\ref{aff66}}
\and P.~Schneider\orcid{0000-0001-8561-2679}\inst{\ref{aff3}}
\and T.~Schrabback\orcid{0000-0002-6987-7834}\inst{\ref{aff94}}
\and M.~Scodeggio\inst{\ref{aff31}}
\and A.~Secroun\orcid{0000-0003-0505-3710}\inst{\ref{aff54}}
\and E.~Sefusatti\orcid{0000-0003-0473-1567}\inst{\ref{aff15},\ref{aff14},\ref{aff16}}
\and G.~Seidel\orcid{0000-0003-2907-353X}\inst{\ref{aff66}}
\and S.~Serrano\orcid{0000-0002-0211-2861}\inst{\ref{aff36},\ref{aff95},\ref{aff37}}
\and P.~Simon\inst{\ref{aff3}}
\and C.~Sirignano\orcid{0000-0002-0995-7146}\inst{\ref{aff89},\ref{aff53}}
\and G.~Sirri\orcid{0000-0003-2626-2853}\inst{\ref{aff19}}
\and J.~Skottfelt\orcid{0000-0003-1310-8283}\inst{\ref{aff96}}
\and L.~Stanco\orcid{0000-0002-9706-5104}\inst{\ref{aff53}}
\and J.~Steinwagner\orcid{0000-0001-7443-1047}\inst{\ref{aff1}}
\and P.~Tallada-Cresp\'{i}\orcid{0000-0002-1336-8328}\inst{\ref{aff34},\ref{aff35}}
\and A.~N.~Taylor\inst{\ref{aff43}}
\and H.~I.~Teplitz\orcid{0000-0002-7064-5424}\inst{\ref{aff8}}
\and I.~Tereno\orcid{0000-0002-4537-6218}\inst{\ref{aff51},\ref{aff97}}
\and N.~Tessore\orcid{0000-0002-9696-7931}\inst{\ref{aff68}}
\and S.~Toft\orcid{0000-0003-3631-7176}\inst{\ref{aff98},\ref{aff99}}
\and R.~Toledo-Moreo\orcid{0000-0002-2997-4859}\inst{\ref{aff100}}
\and F.~Torradeflot\orcid{0000-0003-1160-1517}\inst{\ref{aff35},\ref{aff34}}
\and I.~Tutusaus\orcid{0000-0002-3199-0399}\inst{\ref{aff91}}
\and L.~Valenziano\orcid{0000-0002-1170-0104}\inst{\ref{aff7},\ref{aff58}}
\and J.~Valiviita\orcid{0000-0001-6225-3693}\inst{\ref{aff70},\ref{aff71}}
\and T.~Vassallo\orcid{0000-0001-6512-6358}\inst{\ref{aff21},\ref{aff15}}
\and G.~Verdoes~Kleijn\orcid{0000-0001-5803-2580}\inst{\ref{aff101}}
\and A.~Veropalumbo\orcid{0000-0003-2387-1194}\inst{\ref{aff13},\ref{aff23},\ref{aff22}}
\and Y.~Wang\orcid{0000-0002-4749-2984}\inst{\ref{aff8}}
\and J.~Weller\orcid{0000-0002-8282-2010}\inst{\ref{aff21},\ref{aff1}}
\and G.~Zamorani\orcid{0000-0002-2318-301X}\inst{\ref{aff7}}
\and F.~M.~Zerbi\inst{\ref{aff13}}
\and E.~Zucca\orcid{0000-0002-5845-8132}\inst{\ref{aff7}}
\and C.~Burigana\orcid{0000-0002-3005-5796}\inst{\ref{aff102},\ref{aff58}}
\and L.~Gabarra\orcid{0000-0002-8486-8856}\inst{\ref{aff103}}
\and C.~Porciani\orcid{0000-0002-7797-2508}\inst{\ref{aff3}}
\and V.~Scottez\orcid{0009-0008-3864-940X}\inst{\ref{aff81},\ref{aff104}}
\and M.~Sereno\orcid{0000-0003-0302-0325}\inst{\ref{aff7},\ref{aff19}}}

%
%
 \abstract{The \Euclid large-scale weak-lensing survey aims to trace the evolution of cosmic structures up to redshift $z$ $\sim$ 3 and beyond. Its success depends critically on obtaining highly accurate mean redshifts for ensembles of galaxies $n(z)$ in all tomographic bins, essential for deriving robust cosmological constraints. However, photometric redshifts (photo-$z$s) are affected by systematic biases, arising from various sources of uncertainty and dominated by selection effects of the spectroscopic sample used for calibration. To address these challenges, we utilised self-organising maps (SOMs) with mock samples resembling the Euclid Wide Survey (EWS) from the Flagship2 simulation, to validate \Euclid's uncertainty requirement of $|\Delta\langle z \rangle| = \langle z_{\text{est}} \rangle - \langle z \rangle \leq 0.002 (1+z)$ per tomographic bin, assuming DR3-level data. Consequently, we identify the most effective galaxy selection for our tomographic bins, while systematically examining the implementation of quality control cuts to reduce sources of uncertainty. In particular, we observe that defining the redshift tomography using the mean spectroscopic redshift (spec-$z$) per SOM cell, results in none of the ten tomographic redshift bins satisfying the requirement. In contrast, the redshift tomography on the photo-$z$s of the EWS-like sample yields superior results, with eight out of ten bins [$0 < z\leq 2.5$] meeting the \Euclid requirement. To enhance the realism of our study, we morph our calibration sample to mimic the C3R2 survey in incremental steps. In this context, a maximum of six out of ten bins meet the requirement, strongly advocating the adoption of a redshift tomography defined by the photo-$z$s of individual galaxies rather than the commonly used mean spec-$z$ of SOM cells. To examine the impact on the expected biases for $\Omega_{\text{m}}$, $\sigma_{8}$, and $\Delta w_{0}$ measured by \Euclid, we perform a Fisher forecast for cosmic shear only, based on our redshift uncertainties. Here, we find that even under an evaluation of the uncertainty where the impact of the redshift bias is substantial, most absolute biases remain below 0.1$\sigma$ in the idealised scenario and below 0.3$\sigma$ in the more realistic case.}
  
%
%
%
%
   \titlerunning{\Euclid\/: Photo-$z$ calibration via SOMs}
   \authorrunning{W. Roster et al.}
   
   \maketitle
%
%
%
%
   
\section{\label{sc:Intro}Introduction}

Our current understanding of the Universe is primarily guided by the widely accepted Lambda cold dark matter ($\Lambda$CDM) model, recognised for its accuracy and reliability in replicating observations \citep{Frieman_2008, Thomas_2016, 2020A&A...641A...1P}. Despite the success thus far, our knowledge of its implicated components, such as dark energy (DE), remains limited. The primary impact of DE lies in inducing late-time acceleration and influencing the growth rate of cosmic structures. In fact, exploring alternatives to $\Lambda$ with an equation-of-state (EOS) parameter $w(z)=-1$ of a dynamical DE model evolving as $w(z)=w_{0}+w_{a}\frac{z}{1+z}$, where $w_0$ denotes the EOS at $z=0$ and $w_a$ describes its evolution with respect to the scale factor $a$, could show incompatibility, and thus invalidate the $\Lambda$CDM model \citep{Caldwell_1998, Ilbert_2021}. Among other approaches, this theory can be probed by weak gravitational lensing (WL) through cosmic shear surveys \citep[e.g.][]{2017MNRAS.465.1454H, Troxel_2018, 2018Mandel,Hikage_2019, Li_2023,wright2025}. These measure the coherent distortion of galaxy shapes in images by large-scale structures, allowing us to trace the apparent underlying mass distribution along the line of sight \citep{Kilbinger_2015}. Past wide sky precision cosmology surveys, such as the ground-based Canada-France-Hawaii Telescope Lensing Survey \citep[CFHTLenS,][]{Erben_2013}, the Kilo-Degree Survey \citep[KiDS,][]{2017MNRAS.465.1454H,wright2025}, the Hyper-Suprime Camera Survey \citep[HSC,][]{Aihara_2017,Hikage_2019}, or the Dark Energy Survey \citep[DES,][]{2018PhRvD..98d3526A, Troxel_2018} have successfully utilised cosmic shear tomography to reconstruct a three-dimensional layout from two-dimensional observations by dividing the observations into a number of redshift bins, depending on the respective probes \citep{Wong_2025}. Following in their footsteps, we hope to gain improved insight with the  {\it Nancy Grace Roman} Space Telescope \citep{spergel_2015}, the Vera C. Rubin Observatory Legacy Survey of Space and Time \citep[LSST,][]{2019ApJ...873..111I}, and the recently launched optical to near-infrared European Space Agency (ESA) \Euclid satellite \citep{laureijs_2011,euclidcollaboration2024mellier}. Central to these efforts is the sufficiently accurate measurement of galaxy redshifts, which is critical for tomographic binning and, in turn, for interpreting cosmological observations \citep{Bordoloi_2012, Stolzner_2021}. Correspondingly, \Euclid will cover over 14\,000 deg\textsuperscript{2} \citep{Scaramella-EP1} during its six-year mission, charting the extragalactic distance-redshift relation for 1.5 billion galaxies out to $z \sim 2$, thereby covering the period in which the Universe passed from DM to DE domination \citep{2013PhR...530...87W}. 

The rapid expansion of wide-field imaging campaigns has intensified efforts to achieve precise redshift calibration across large datasets. Given the impracticality of obtaining spectroscopic redshifts (spec-$z$) for the vast numbers of distant and often faint galaxies essential to WL studies, redshift estimates, ideally in the form of full probability distributions $p(z)$, are instead inferred from photometric observations using intermediate- or broad-band filters \citep{Newman_2015}. Advantageously, imaging surveys provide more galaxies with redshift estimates per unit of telescope time compared to spectroscopic surveys, resulting in significant growth of such photometric redshift (photo-$z$) methods over the past two decades \citep{salvato19}. Several approaches have been employed to infer, among other things, the galaxy redshift distribution $n(z)$ of photometric samples more efficiently, albeit with significantly reduced precision. These include clustering redshifts, which utilise the spatial distribution of overlapping spectroscopic samples \citep[][]{Newman_2008, McQuinn_2013, Morrison_2017,Gatti_2021,dAssignies25}, template fitting methods \citep{Benitez_2000, Brammer_2008, Lephare2011,Ilbert_2021}, and machine learning algorithms such as fully connected neural networks, decision trees, Gaussian processes, and support vector machines \citep{Carrasco_Kind_2013, Sadeh_2016, desprez_2020}.

Since tomographic cosmic shear studies are primarily sensitive to the average distance of sources, they require accurate knowledge of the $n(z)$ distribution and strongly rely on the ability to bin galaxies by redshift coarsely \citep{Hutterer_2006,Newman_2015,Hildebrandt_2021}. To ensure this, a robust redshift calibration is necessary. However, correctly fine-tuning  photo-$z$ is particularly challenging \citep{Hoyle_2018,Merz_2024}, as systematic uncertainties commonly induce undesired bias and scatter. A bias in the estimated $n(z)$, such as introduced by systematically underestimated redshift values, would unequivocally lead to the incorrect model of an overall denser, more highly clustered gravitational landscape than exists in reality \citep{Wright_2020a}. Therefore, an accurate determination of the mean redshift per bin is necessary to derive the cross-correlation of the source distortions between tomographic redshift planes, which in turn provide estimates on the distribution of mass along the line of sight. This poses an arduous task, since the photo-$z$ calibration efforts, similar to clustering redshifts, are heavily dependent on the available reference samples of spec-$z$ \citep[][]{Stanford_2021,dAssignies25}. Minimising bias and its uncertainty is crucial in order to obtain robust WL constraints in future measurements that, once combined with cosmic microwave background (CMB) measurements from the {\it Planck} satellite \citep{2020A&A...641A...1P}, can test discrepancies with other cosmological constraints on the amount and clustering of (predominantly dark) matter across cosmic time \citep{Asghari_2019,Wright_2020b}.

In this paper we focus on analysing the current state-of-the-art efforts in redshift calibration to validate whether the residual systematic bias for the \Euclid mission can be minimised to satisfy its stringent requirement of $\Delta\langle z \rangle \leq 0.002(1+z)$ per tomographic redshift bin, which is comparable to the precision expected for stage-IV surveys \citep{Bordoloi_2012}. Comparable to the approaches taken in \cite{Buchs_2019}, \cite{Zuntz_2021}, \cite{Myles_2021}, or \cite{Gatti_2021}, for example, the indicated objective is subsequently realised by training an unsupervised, competitive machine learning (ML) algorithm known as a self-organising map \citep[SOM,][]{kohonen} with a large\footnote{Compared to the sample size used in previous photometric WL studies, such as KiDS} mock catalogue to map the galaxies' multi-dimensional magnitude-colour space ($u-g$, $g-r$,..., $J-H$) onto a grid of representative cells \citep{Masters_2015}. The SOM performs dimensionality reduction in colour space, preserving the topological relationships of the input features. Since galaxy colours correlate strongly, though not uniquely, with redshift, the resulting SOM tends to place galaxies with similar redshifts in nearby cells. However, because the colour–redshift relation is non-linear and can be degenerate in specific regions of colour space, neighbouring cells may not necessarily have small proximities in redshift. The baseline of this method is that a deep spectroscopic calibration sample can be reweighted such that it is representative of a wide photometric validation sample of unknown $n(z)$ \citep{Lima_2008,van_den_Busch_2022,Campos_2024}. Furthermore, SOMs allow for the identification and systematic removal of sources in the validation sample that are missing representation in the calibration sample, which, if included in the distribution, would bias the estimates. Under suitable conditions, SOMs have previously been shown to reach residual biases in mean redshifts of $\approx$ 0.01 or better \citep{Hildebrandt_2021,Myles_2021,Giannini_2024}. As \Euclid datasets will be complemented by ground-based optical data, this paper tests various tomographic redshift calibration approaches and establishes a proof of concept under an idealised framework for \Euclid and LSST Year 3 (Y3)-like data, not representative of \Euclid DR1 capabilities. By focusing on this controlled set-up, this study serves as an essential test of robustness, demonstrating the viability and effectiveness of our approach, while setting the stage for future applications.
  
This work is structured as follows. In \cref{chap_2}, we introduce the SOM and the implementation of the calibration methodology. \Cref{chap_3} gives an overview of \Euclid and the complementary mocks. \Cref{chap_4} investigates the alteration in performance for various bias calibration reduction techniques attributed to different redshift tomographies. \Cref{chap_5} covers the scientific outlook and the levels of accuracy to expect from \Euclid based on our findings. The main results are summarised in \cref{chap_6}.

\section{Methodology}
\label{chap_2}

Empirical reconstructions of the colour-redshift space can be used in direct calibration, which aims at estimating the unknown redshift distribution $n(z)$ of a photometric galaxy sample. This method involves matching a deep spectroscopic sample $S$ to a photometric sample $P$, only requiring that $S$ covers, even sparsely, the range of shear observations spanned by $P$ \citep{Wright_2020b}. However, in practice \textit{S} often shows different selection functions compared to \textit{P}, caused by spectroscopic targeting strategies and success rates \citep{Beck_2017, Gruen_2017, Hartley_2020}. Work by \cite{Wright_2020b} has shown that this issue can be addressed by additional cleaning of \textit{P}, resulting in a so-called gold sample. This is implemented by training a SOM on a set of values consisting of magnitudes and/or colours drawn from \textit{S}, before parsing \textit{P} into the same cells. Cells that are not occupied by objects in both samples are rejected, effectively removing critical parts of the colour space that are not adequately represented by \textit{S}  \citep[for more details on recalibration, see][]{Hildebrandt_2021}. In principle, additional clustering of cells with similar photometric properties can improve the balance between cell removal due to colour-space exclusion as well as bias introduced by a flawed representation of the gold sample selection \citep{van_den_Busch_2022}. 

Once trained, a higher-dimensional feature space of multiple attributes is projected onto the SOM. In other words, a SOM trained on input data attributes $a$, such as colours, can be painted by an attribute $b$, such as redshift \citep{Davidzon_2019}. This unsupervised technique does not rely on labels, such as redshifts from a spectroscopic sample, during the training process \citep{Carrasco_Kind_2013}. Since the representation of each data point is fixed by a particular cell after training, redshift can be mapped as a function of colour. However, it should be mentioned that this is not true in the case of colour degeneracy.

The finite binning of the SOM manifold inherently discretises the data space, which can distort the true balance between calibration and validation samples. Since the SOM is not populated evenly, this hurdle can be leveraged to identify and filter out poorly represented photometric data by excluding cells lacking spectroscopic information. Much like for clustering algorithms like $k$-nearest neighbour ($k$-NN) or probabilistic principal component analysis (PCA), objects with similar characteristics are grouped together on the SOM. Cells unoccupied by either photometric or spectroscopic sources are removed, as they indicate an imbalance in representation likely caused by selection effects or intrinsically underpopulated regions of colour space. This trade-off involves sacrificing a portion of the photometric sample but aligns with the overarching objective of mitigating redshift bias, as $k$-NN matching can extend to sources well beyond what one might consider the local region of the $n$-dimensional manifold \citep{Wright_2020a}. Therefore, by painting the canvas of the colour space with the slice of well-represented photometric data, which we refer to as gold sample, one can also recognise regions that are difficult to access by direct spectroscopy and specifically target in spectroscopic surveys to widen the coverage of the colour-space to redshift mapping. The most notable implementation of this type of approach can be found in the Complete Calibration of the Color-Redshift Relation (C3R2) project \citep{ Masters_2015,Masters_2017,Masters_2019,Stanford_2021}. Here, the C3R2 team endeavours to utilise SOMs to identify unexplored parts of the $n$-dimensional colour-redshift hyper-volume due to complex selection functions in the ability to measure distinct spectra \citep{Hoyle_2018, Wright_2020a}. They subsequently observe the spectra and thus, as the name suggests, extend the calibration of the colour-redshift relation for use in weak lensing surveys \citep{Amendola_2013, Wright_2020a,Myles_2021,Campos_2024,euclidcollaboration2024mellier}.  

Unlike other calibration efforts, such as geometry-dependent angular cross-correlation \citep[][]{Moessner_1998,dAssignies25}, the two samples used for calibration and validation need not overlap in the projected sky plane. On the contrary, the spec-$z$ required in the calibration sample can be accumulated from a multitude of surveys observing galaxies unrelated to those used to gather the photometry for both the calibration and validation samples. However, as with any ML algorithm, SOMs lack the ability to reliably extrapolate beyond the feature space covered by the calibration data, for example going to ever fainter magnitudes. As such, the calibration sample must cover the parameter range occupied by the validation sample passed to the SOM. As SOMs focus on the intrinsic properties of galaxies instead of positional parameters used in galaxy clustering, these methods are complementary \citep{Newman_2008}.

\section{Data}
\label{chap_3}

The payload on \Euclid feeds two instruments operating in parallel. Firstly, the panoramic very broad-band (\textit{r,i,z}) visible-light 600 megapixel imager VIS with a single band wavelength range, \IE [$\lambda = 5300$--9000\,\AA] \citep{euclidcollaboration2024cropper}. Secondly, the Near-Infrared 3-band (\YE, \JE, and \HE) Photometer (NISP-P), with wavelength ranges [$\lambda = 9500$--12\,123\,\AA], [$\lambda = 11\,676$--15\,670\,\AA], [$\lambda = 15\,215$--20\,214\,\AA] at a pixel size of 0.3 arcseconds, as well as a slitless Spectrograph (NISP-S) comprising two grisms jointly covering $\lambda = 9260$--18\,920\,\AA \, \citep{euclidcollaboration2024Jahnke}. The Euclid wide survey (EWS), reaching (median) depths of \IE = 24.5, represents the core of the dark energy mission from which weak lensing, baryon acoustic oscillation, and redshift space distortion signals will be measured \citep[see][for more details]{euclidcollaboration2024mellier}.

\subsection{Simulated data: Flagship catalogue}
\label{flagship}

\begin{figure}[t!]
\centering
\includegraphics[width=9.5cm]{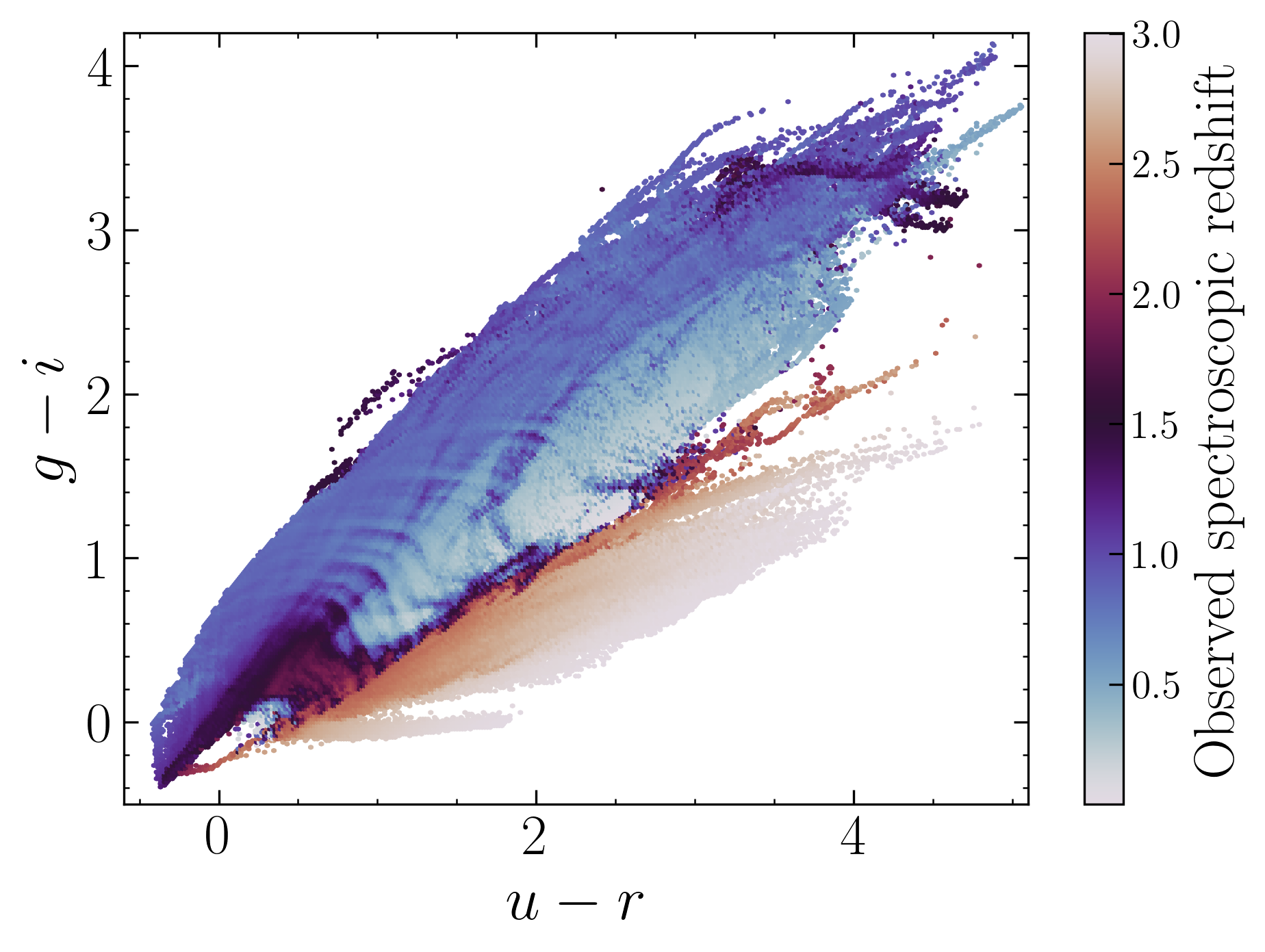}
\caption[]{Colour-colour diagram of Flagship2 galaxies free of flux errors plotted according to the corresponding true redshift. Blue represents the lower-redshift sources and red the higher-redshift sources [$0 \leq z_{\rm obs} \leq 3$], exemplifying the non-linearity and degeneracies of the colour-redshift relation.}
\label{fig:2.03}
\end{figure}

In this section, we describe the development of synthetic observations resembling \Euclid-like datasets, used here to validate the performance of the SOM redshift calibration. The \Euclid Flagship mock galaxy catalogue v.2.1.10 \citep[Flagship2,][]{euclidcollaboration2024euclidvflagshipgalaxy} is the largest cosmologically simulated galaxy catalogue to date. Flagship is based on the $\sim$ 4 $\times$ 10\textsuperscript{12} dark-matter particle $N$-body simulation containing 2.6 billion galaxies over 1/8th of the sky while extending up to redshift $z \leq$ 3 \citep{Potter2017}. It reproduces the temporal evolution of large-scale structures with billions of dark matter halos hosting galaxies and WL observables. From this dark-matter cosmic web  simulation \citep{Potter2017}, a synthetic galaxy catalogue, implementing the halo occupation distribution and halo abundance matching technique \citep{Berlind_2002}, has been generated \citep{euclidcollaboration2024euclidvflagshipgalaxy}. Synthetic galaxies in this simulation mimic, with great detail, the complex properties that observed sources display: ranging from their shapes, luminosities \citep{Blanton2003}, spectral energy distributions \citep[SEDs,][]{Ilbert2006,Polletta2007}, and emission lines in their spectra, to the gravitational lensing distortions that affect the light observed by distant galaxies and noise realisations for observed fluxes. An example of the Flagship2 galaxy colour-colour space is shown in \cref{fig:2.03}. Each galaxy in the catalogue includes entries for observed spectroscopic redshifts ($z_{\text{obs}}$), photometric redshifts ($z_{\text{p}}$), and true redshifts ($z_{\text{t}}$) down to apparent magnitudes of \IE< 25.75 or \HE< 24.25. Here, $z_{\text{obs}}$ accounts for both cosmic expansion and the peculiar velocities of the galaxies, including those of the halo and the galaxy within the halo. The estimation of $z_{\text{p}}$ is achieved using either the Deep Neural Mixture Density Network \citep[{\tt{Deepz}},][]{Eriksen_2020} or the Nearest Neighbours Photometric Redshifts \citep[{\tt{NNPZ}},][]{Tanaka2018} pipeline. {\tt{Deepz}} is trained on 20\,000 simulated noisy galaxies without any colour selection limited to $i_{\rm LSST} < 23$, providing the full $p(z)$ and point estimates, which are defined as the first mode, respectively. Although the property distributions of its simulated training set are considered realistic, it is somewhat optimistic by construction. As a result, the achieved data quality exceeds the expected capabilities of \Euclid. Therefore, to avoid artificially improving our results, we utilise {\tt{NNPZ}} in its standard configuration (using 30 neighbours) throughout this study. {\tt{NNPZ}} is a ML approach that estimates $p(z)$
using a $k$-nearest-neighbour search in flux or colour–magnitude space \citep{Cunha_2009}. For each target galaxy, the method identifies nearby objects from a photometric reference sample that is deeper than the sample under consideration. The $p(z)$ distributions associated with these neighbours, which themselves are obtained from template-fitting photo-$z$ codes applied to the reference sample, are then combined. Each neighbour is assigned a weight based on its distance in colour space, accounting for photometric uncertainties. The final $p(z)$ of the target galaxy is obtained as the weighted redshift histogram of these neighbours. Point estimates are available either as the first mode or the median of the $p(z)$, where the latter was used in this work. However, all analyses in this paper were also performed with {\tt{Deepz}}, yielding similar or slightly improved results. The exclusive use of ML photo-$z$ estimators, rather than a combination of ML and template-fitting methods, is primarily a practical consequence of computational constraints. Running template-fitting codes to derive redshifts or subsequent physical properties for many millions of sources is prohibitively time-consuming, as these methods remain computationally expensive. Moreover, in Flagship2 the galaxy colours are generated directly from SED templates, suggesting such algorithms would perform artificially well in the simulation compared to real observations, where the diversity of galaxy SEDs is significantly larger.

\subsection{Sample requirements}

We start from a set of two isolated and compatible catalogues selected on fundamentally different specifications. The calibration sample used for training the SOM is designed to mimic non-trivial, real-life spectroscopic observations, including all current magnitude restrictions, realistic number counts, and quality indicators. The validation sample consists of a large enough number of data points or area on the sky to reduce sample noise to a negligible influence. The two distributions are not required to share an overlapping region on the sky but merely need to feature a similar parameter space regarding the variables used to train the SOM. In this manner, both samples (with $N$\textsubscript{calib} $\ll$ $N$\textsubscript{valid}) exhibit comparable galaxy attributes, although they were selected using distinct criteria. Flagship2 does not inherently offer two samples of this nature; hence, they must initially be generated.

\subsection{Flagship mocks}

\begin{figure}[t!]
\centering
\includegraphics[width=9cm]{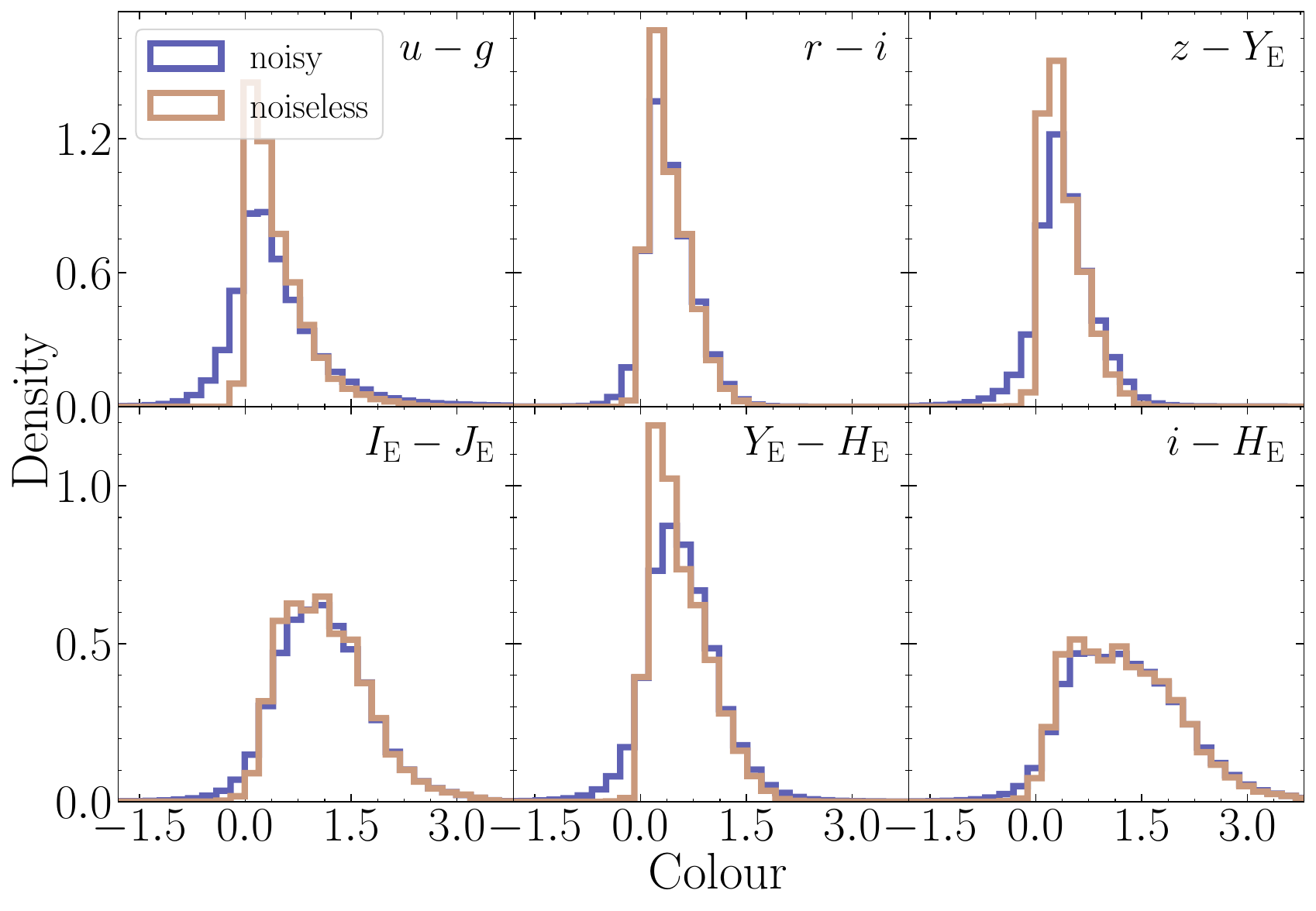}
\includegraphics[width=9cm]{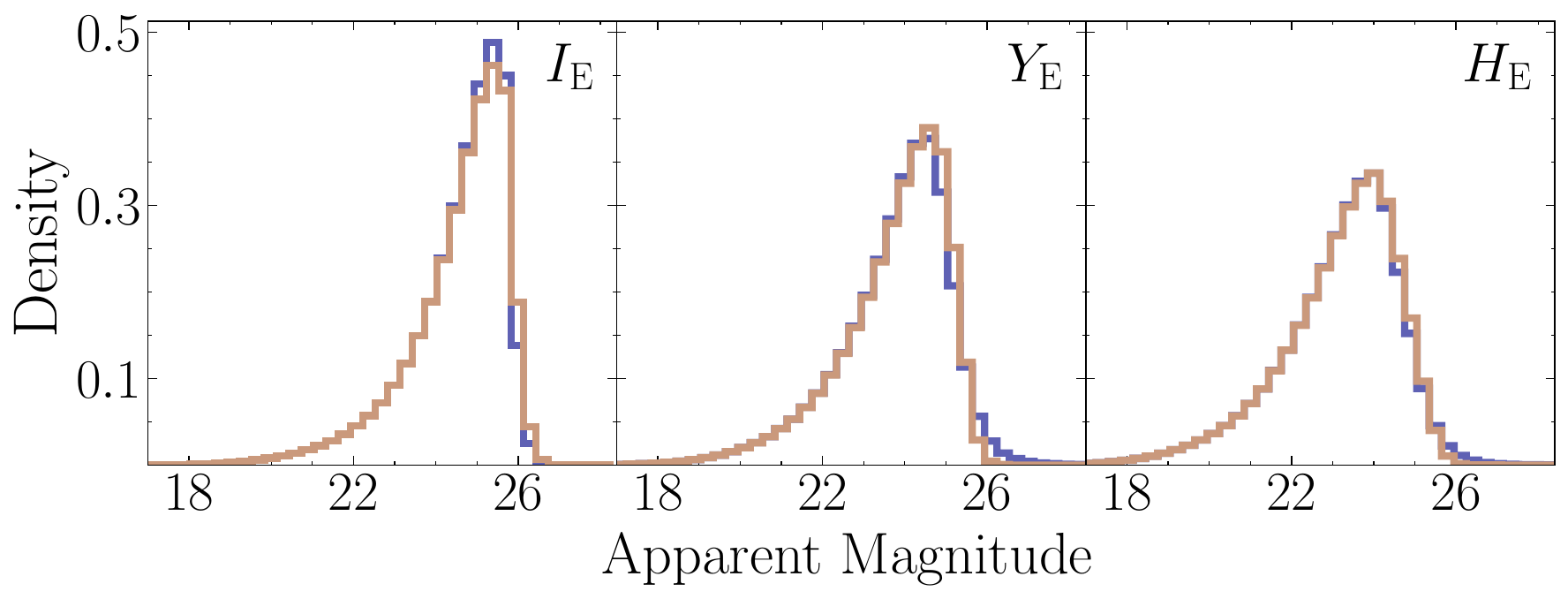}
\caption[Comparison of the noisy and noiseless Flagship2 galaxy catalogue colours (top) and magnitudes (bottom).]{Comparison of some exemplary noisy (blue) and noiseless (brown) Flagship2 EWS-like colours (top) and magnitudes (bottom). The bands comprise LSST $u, g, r, i, z$ and \Euclid \IE, \YE, \JE, \HE at year 3 depth.}
\label{fig:6.1}
\end{figure}

The Flagship2 simulation provides flux measurements of the observed photometry including contributions from both the continuum and emission lines, accounting for internal attenuation and Milky Way extinction, provided in cgs units of erg\,cm$^{-2}$\,s$^{-1}$\,Hz$^{-1}$. To ensure accurate representation of observational constraints, the photometric noise, driven by systematic effects (aperture size, detector filters, size of the PSF, or natural effects such as wavelength-dependent extinction or atmospheric seeing), has been calibrated to match the expected depths of the EWS in the southern hemisphere provided by the complementary ground-based surveys at the DR3 time frame \citep{Scaramella-EP1,euclidcollaboration2024mellier}. Given the flux descriptions above, the observed flux  $F_{\textrm{obs}}$ for each source can be derived as 
\begin{equation}
F_{\textrm{obs}} = F_{\textrm{intrinsic}} + F_{\textrm{error}} \, ,
\end{equation}
with AB magnitudes
\begin{equation}
m_{\rm obs} = -2.5 \, \textrm{log}_{10} (F_{\rm obs}) - 48.6.
\end{equation}
The depth limits are defined at a 10$\sigma$ level for a 2 arcsecond diameter aperture for extended sources. Intrinsic (noiseless) and observed (noisy) colour and magnitude distributions for LSST (\textit{ugriz}), as well as for \Euclid (\IE,\YE,\JE,\HE) at the depths expected to be reached by the third year, are displayed in \cref{fig:6.1}, with limiting magnitudes up to 25.7 AB for extended sources. Note that we do not account for flux magnification, as the effect cancels for colours \citep{Lepori_2022}, or the influence of variable survey depth. Due to computational load, we restrict our base sample to a 1/256 \textsuperscript{th} slice in the Flagship2 mock, corresponding to roughly 19 million sources from CosmoHub \citep{Carretero_2017,Tallada}. After downselecting sources with {\tt{NNPZ}} $z_{\textrm{p}} \leq 3$, we are left with roughly 6 million sources, while being largely unaffected by sample variance. This effect is distinct from shot noise given by the non-homogeneity of galaxies and cosmic variance induced by differing realisations of the observable Universe \citep{Wright_2020b}. 

In addition, we are particularly interested in how large-scale structures in the field affect the observed $n(z)$ and, in turn, the estimation of SOM redshift bias across tomographic bins. For this reason, the sample is further dissected into a grid of $10 \times 10$ equi-sized, completely independent line-of-sight sub-field samples (see \cref{fig:4.1.1}). This allows us to probe the impact of sample variance in the training samples exclusively. Even though every sub-field sample features different galaxies, the influence of photometric noise is not investigated, as all samples use the same noise level. For each sub-field sample, the same training and calibration pipeline, as outlined in the following sections, is executed independently. Subsequently, we derive the mean bias and its uncertainty per tomographic bin, utilising 100 separately analysed samples. 

\begin{table*}[ht!]
\label{noiseless_som}
\centering
\caption[]{Comparison of the characteristics of the noiseless and noisy SOMs.}

\begin{tabular}{lll}
\hline\hline\noalign{\vskip 1.5pt}  
 SOM parameter & Noiseless SOM & Noisy SOM\\
 \hline\noalign{\vskip 1.5pt}
 Training expression & 36 colours \& \IE magnitude  & 36 colours \& \IE magnitude \\ 
 Dimension & 75$\times$150 & 101$\times$101 \\
 Topology & flat and edge limited & toroidal and continuous  \\
 Cell type & hexagon & hexagon \\
 Data threshold & [0,30]\,mag & [0,30]\,mag ; $\rm{S/N(\IE)} > 5$ \\
 Training iterations & 200 & 200 \\ [1ex]

\end{tabular}
\tablefoot{The noiseless SOM \cite{Masters_2015} layout, trained on noiseless Flagship2 photometry. Opposed to the noiseless SOM, the noisy SOM is not trained on a wide field sub-sample, but on the respective calibration sample, generated from the noiseless SOM.}
\label{tab:6.1}
\end{table*}

\section{Calibration efforts}
\label{chap_4}

All SOM analysis in this work relies exclusively on the implementation of the widely used \texttt{kohonen} package \citep{Wehrens10_2018}, as a branched version\footnote{\url{https://github.com/AngusWright/kohonen.git}} within the programming framework \texttt{R}. To emulate \Euclid observations, the Flagship2 sub-field samples are preprocessed using dynamic flux cuts, retaining only galaxies with signal-to-noise ratios (S/N) of S/N > 5 in \IE. This avoids imposing a fixed magnitude limit, such as the EWS threshold as mentioned in \cref{chap_3}, and preserves fainter galaxies present in Flagship2, which is not constrained by this magnitude limit. We end up with a EWS-like catalogue of $\sim$ 1.7 million galaxies. Nonetheless, this is a slightly pessimistic approach, as we assume all \IE detections to make it into the shear sample, which worsens the calibration. Lastly, all galaxies with negative fluxes after error inclusion in any band are removed from the reduced EWS-like sample.

\subsection{SOM architecture}

\cite{Masters_2015} implement a 75$\times$150 SOM of rectangular nature for C3R2, allowing for asymmetrical multi-dimensional parameter space, arguing the manifold gives preferential direction to the principal geometry and therefore improves convergence. In addition to the raw number of cells in each SOM dimension, a different cell shape and surface topology can be chosen. The prior becomes particularly important when dealing with denser areas of the manifold, where it can directly affect the local data distribution. Conversely, SOM topology becomes most important in areas of the manifold which are sparsely populated. Generally, the choice of topology is either flat or toroidal, meaning the edges of the SOM manifold either act as boundaries or reconnect to form a continuous surface \citep{Wright_2020b}. Finally, the fidelity of the SOM in terms of its spectroscopic representation is driven by what information is relevant to parse during training. Leveraging the extensive nine-band photometry available with Flagship2, we harness a rich array of colour information encompassing 36 magnitude combinations to focus on the colour-redshift relation. Additionally, we introduce the \IE magnitude as an extra parameter axis to enhance stability without introducing excessive redundancy. Nonetheless, there are undoubtedly more modifications that one can make to the SOM which are not considered in this work \citep{Carrasco_Kind_2013}. In line with \cite{Wright_2020b}, the SOM shows considerably lower spectroscopic representation when trained solely on magnitudes, as in this scenario the SOM must first learn that the magnitude combinations, such as colours, are indicative features of redshift estimates. Consequently, adding further magnitudes as training parameters has little to no effect. The argumentation of training parameters, however, is to some extent arbitrary, as human reasoning becomes increasingly imperfect for visualisations beyond 3 dimensions. 

\begin{figure}[t!]
\centering
\includegraphics[width=9cm]{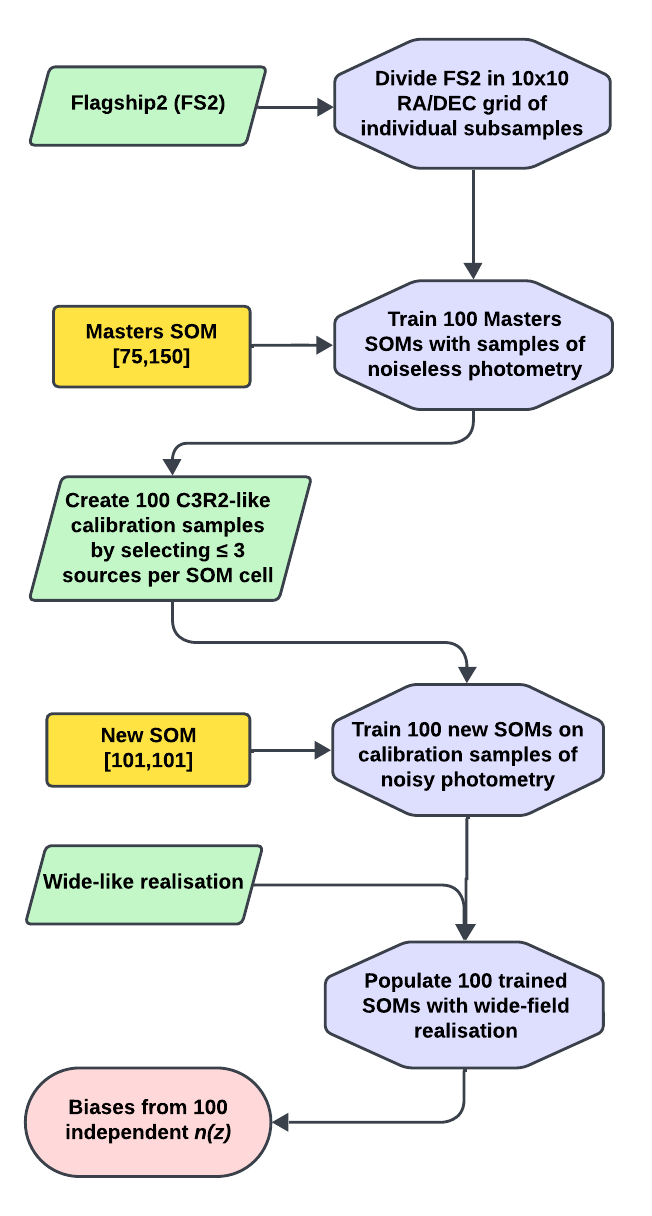}
\caption[]{Flowchart outlining the sequential process of utilising noiseless SOMs to create a set of calibration samples, which are then used to train a secondary set of SOMs using noisy photometric data. Lastly, the trained noisy SOMs are populated by the EWS-like data.}
\label{fig:4.1.1}
\end{figure}

\begin{figure*}[ht!]
\centering
\includegraphics[width=8cm]{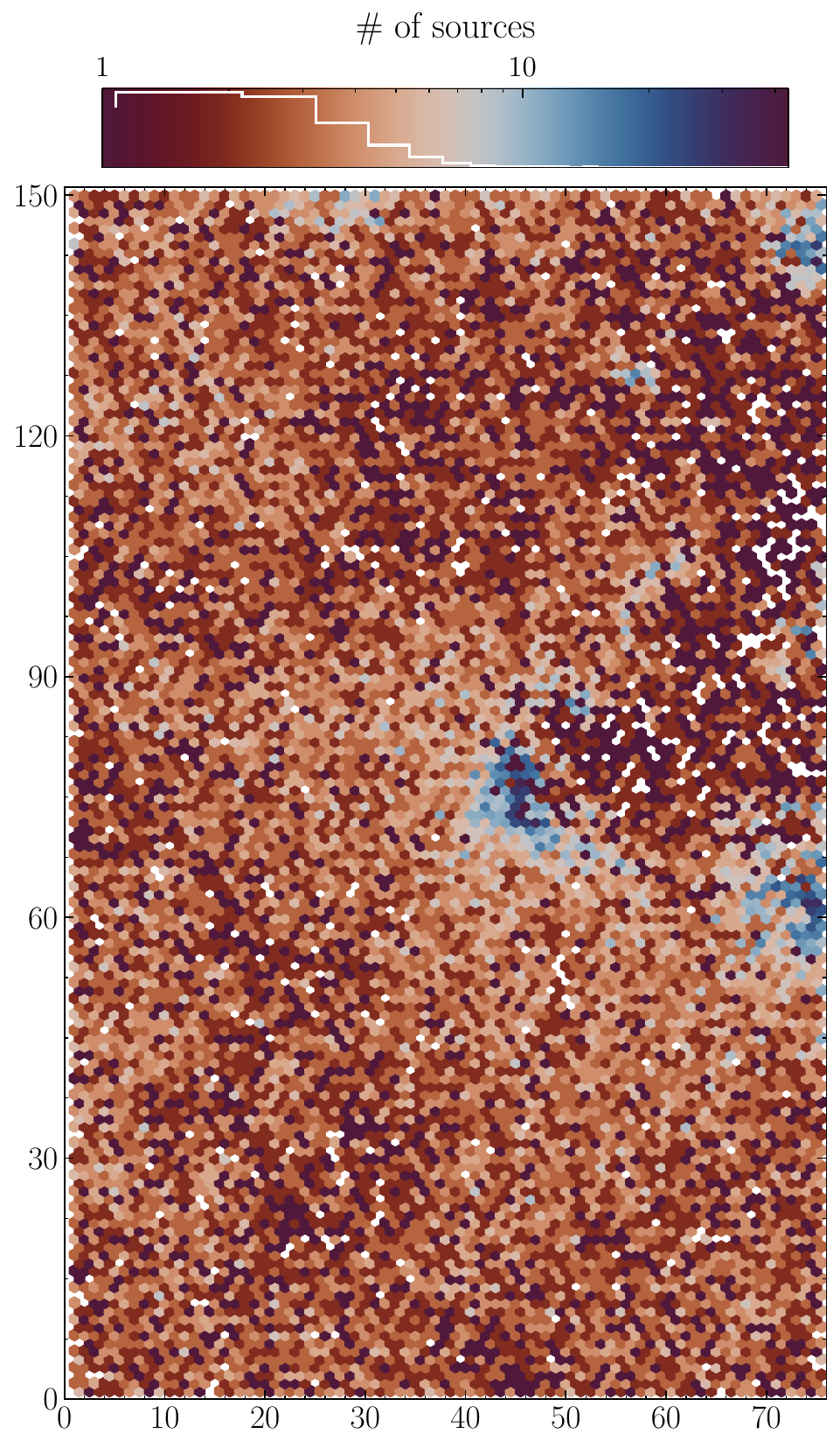}
\hspace{1cm}
\includegraphics[width=8cm]{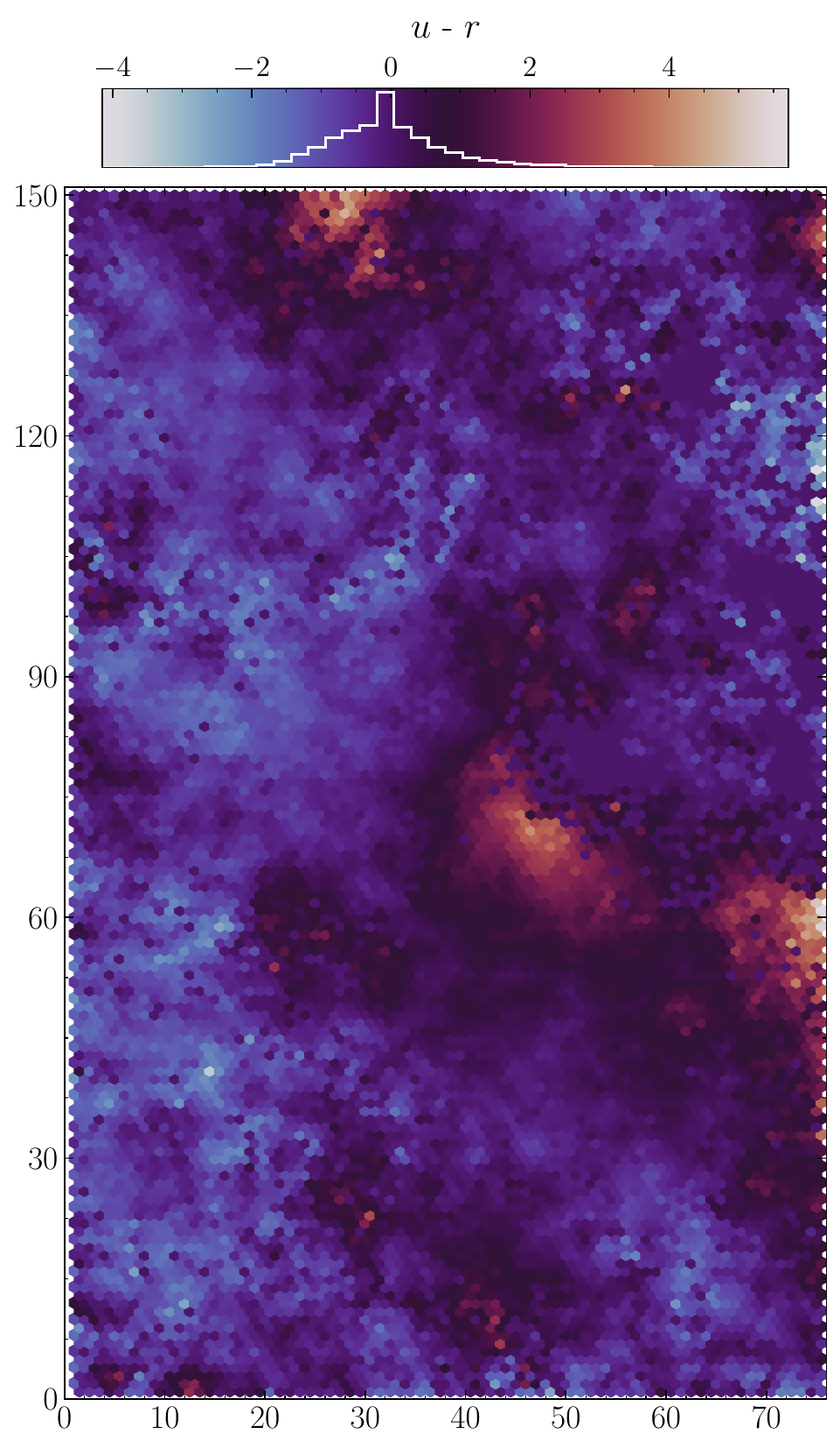}
\caption[]{Masters-like SOM coloured according to different properties of the calibration sample. These include the occupation count (left) and the cells' average $u - r$ colour (right). Furthermore, cells lacking a representation of physical quantities are painted white. The white histograms overlaid within the upper colour bars represent the normalised linear distributions of the respective SOM statistics}.
\label{fig:3.01}
\end{figure*}

\subsection{Calibration sample creation}
\label{CSC}
\Cref{fig:4.1.1} provides an overview of the steps involved in the analysis described in this section. Following the approach taken in C3R2 \citep{Masters_2015}, we utilise Flagship2 with roughly 150\,000 calibration sources per sub-sample to train 100 Masters-like SOMs as characterised by the central column in \cref{tab:6.1}. By sampling the SOMs, we create artificial spec-$z$ calibration samples that closely resemble real-life availability, due to their non-trivial selection functions. Since flux errors are correlated, the introduction of noise to the photometry results in multi-directional shifts across the SOM. However, in the idealised Flagship2 mock catalogue, these higher-dimensional flux error cross-correlations do not exist, making it a more optimistic scenario compared to real \Euclid observations (see Kang et al. in prep.). To prevent this restructuring of the original feature space from happening, and thus securing a uniformly populated and unbiased \citep{Wright_2020b} multi-dimensional parameter volume, the SOMs are instead trained on colours calculated using not entirely unrealistic noiseless photometry, by considering the greater depth of the \Euclid auxiliary fields used for calibration. 
 
Once training commences, the input data is whitened before being presented to the SOM. This involves zero centering, decorrelating, such that the covariance matrix becomes diagonal, and rescaling the data. By standardising the variances in each direction of feature space, whitening typically speeds up convergence and causes models to better capture contributions from low-variance feature directions, as training is not overshadowed by properties which tend to have larger values \citep{Lecun_1998}. The training process is stopped after several hundred iterations, as the mean distance to the closest unit $d_{j}(t)$ drops below $0.005$, as introduced in \cref{TA}. Training could, in principle, be extended to a larger number of iterations; however, continuous training can lead to over-fitting, leading to a drop in model performance. After training, the Masters-like SOM (noiseless SOM, hereafter) cells can be painted by the spatial distribution of numerous variables, such as the inhomogeneous occupation count or training sample-specific attributes, such as colours, as shown in \cref{fig:3.01}. These plots show the noiseless SOM coloured according to its cell properties with red referring to lower values and blue to higher values. Empty cells holding no galaxies, are painted white. To now generate a C3R2-like successful spec-$z$ calibration sample, individual galaxies are randomly drawn from the noiseless SOM. Here we address every cell individually and select $\leq$ 3 sources, respectively. These drawn samples have an average size of $\sim$ 33\,000 sources and will be referred to as calibration samples from here on. By excluding mean magnitude from the selection criteria, we avoid biasing the calibration sample towards brighter sources. Instead, the selection is driven solely by the colour–redshift space of the SOM.

\subsection{SOM spec-\textit{z} based tomography}
\label{SC}

\begin{figure*}[ht!]
\centering
\includegraphics[width=8.5cm]{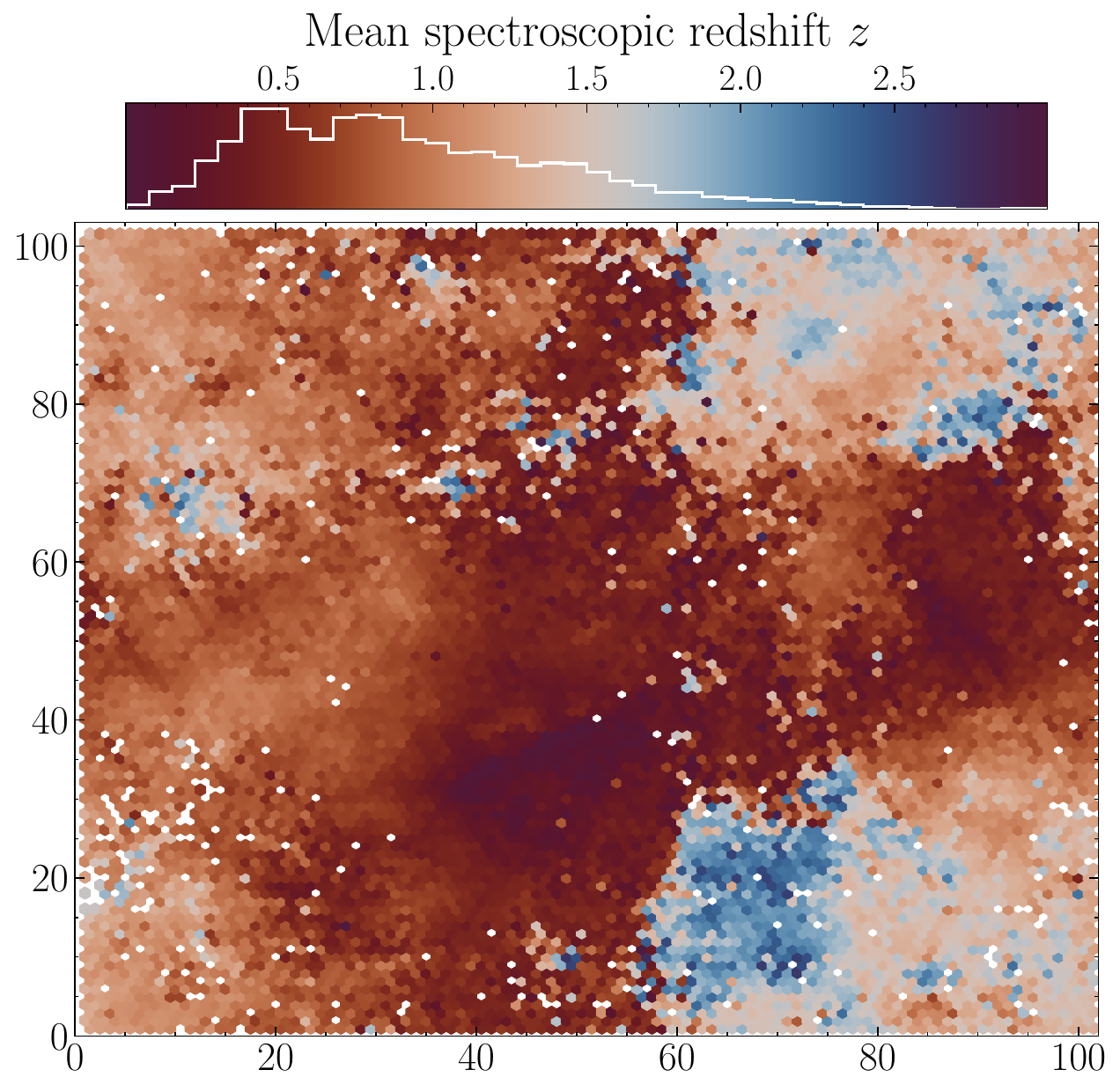}
\includegraphics[width=8.5cm]{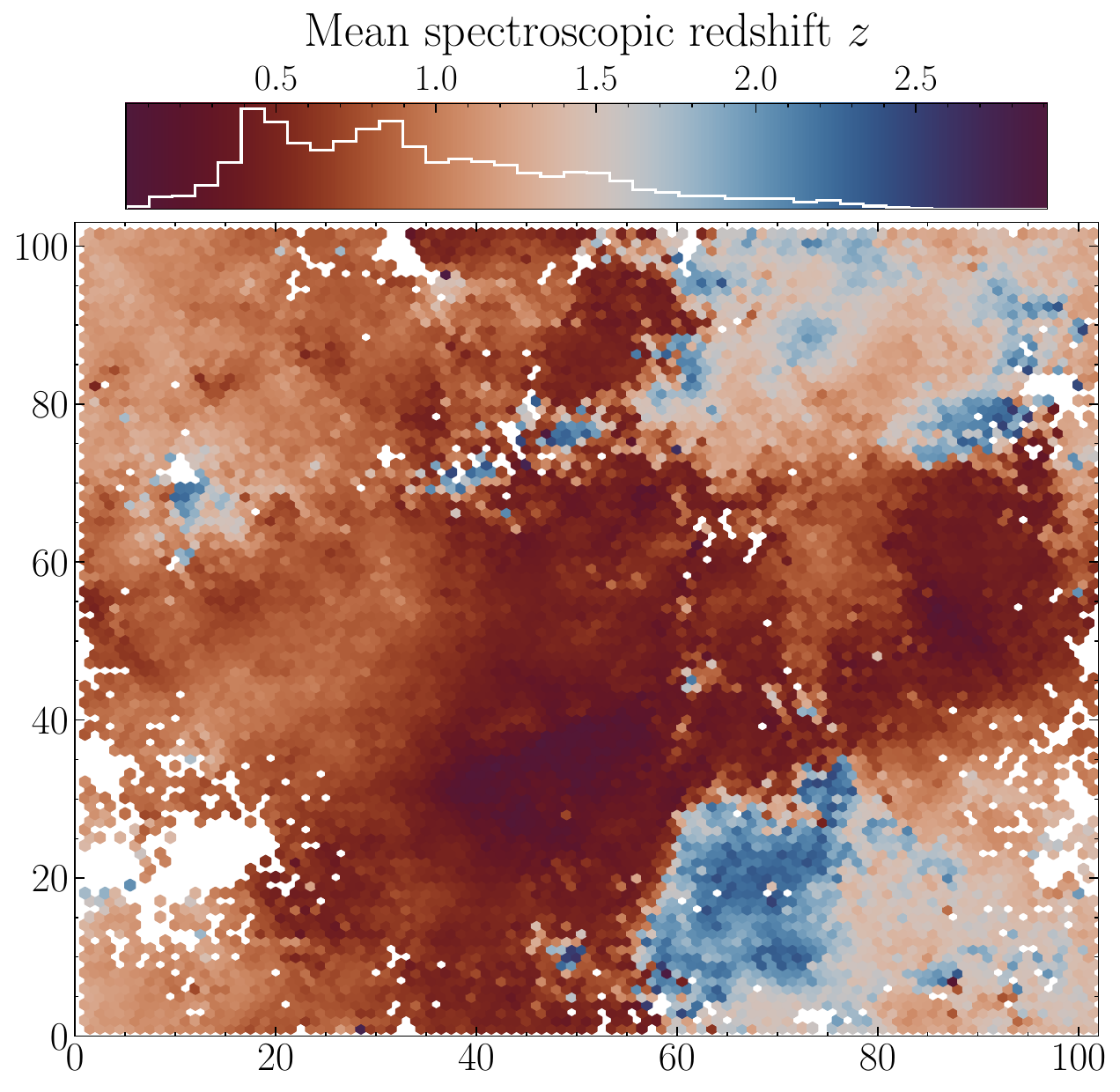}
\caption[]{Noisy SOM painted by the mean spec-$z$ for both the calibration sample used in training (left panel) and the validation sample (right panel). A similar underlying pattern is visible. Areas of spectroscopic incompleteness, coloured in white, cannot be used for calibration purposes as per the gold sample selection.}
\label{fig:6.5}
\end{figure*}

In contrast to training the noiseless SOM on the photometric EWS-like sample, we train a second SOM on the noisy photometry of the calibration sample drawn from the noiseless SOM (see \cref{tab:6.1}). Training a SOM on noiseless data before populating it with noisy data produces misleading results, since this difference in samples introduces covariate shift as introduced in \cref{cv}, rendering this exercise invalid. In contrast to the Masters' SOM, this second, noisy SOM, features a symmetrical 101$\times$101 grid of toroidal topology, meaning there are no hard boundaries limiting the SOM surface plane (see \cref{tab:6.1}). This has the beneficial effect that data parsed to the SOM experiences no prioritised axis. Moreover, the SOM can more effectively encapsulate or wrap around the multi-dimensional data distribution. As shown in \cref{fig:6.5}, the distribution of spec-$z$ found in the validation sample shown on the right, tightly follows the distribution forged during training shown on the left. In addition, areas of spectroscopic incompleteness (i.e. white cells in the left map), which cannot be used for training and areas of unrepresented colour-colour space following the S/N cut in the validation sample (i.e. white cells in the right plot), become detectable. Once the noisy SOM is trained, we propagate the entire EWS-like sample ($\sim$ 1.7 million sources), from now on referred to as validation sample, to the noisy SOM, producing like-for-like groupings. 

\begin{table*}
\centering
\caption[]{Calibration sample distribution on noisy SOMs.}

\vspace{10pt}
\begin{tabular}{ccccccccccc}
\hline\hline\noalign{\vskip 1.5pt}
& \multicolumn{10}{c}{Sample (full) =  33\,602} \\ 
&\multicolumn{10}{c}{$f$\textsubscript{pix} (all, \%) = 96.8} \\ 
\hline\noalign{\vskip 1.5pt}
Bins & bin 1 & bin 2 & bin 3 & bin 4 & bin 5 & bin 6 & bin 7 & bin 8 & bin 9 & bin 10 \\
$z$ $\in$ & (0.05, 0.295] &(-, 0.54] &(-, 0.785] &(-, 1.03] & (-, 1.275] & (-, 1.52] &(-, 1.765] &(-, 2.01] &(-, 2.255] &(-, 2.50] \\ 
$f$\textsubscript{gs} & 98.3 & 97.7 & 93.6 &90.9 &91.1 &93.9 &95.5 &96.4 &97.8 &96.1\\ 
$N$\textsubscript{spec} & 2041  &7404 &6508 &6334 &4177 &3430&1834&962&603&222 \\

\end{tabular}
\tablefoot{The table presents the sample size and fraction of noisy SOM cells occupied by the calibration sample, $f$\textsubscript{pix} (all, \%), along with the fractions of cells representative of the wide sample $f$\textsubscript{gs} and spectroscopic number counts $N$\textsubscript{spec} across 10 equi-sized tomographic bins \citep{Wright_2020b}. The data reveals a notable scarcity of sources for calibration in bins 1, 7, 8, 9, and 10 compared to the other bins.}
\label{tab:6.4}
\end{table*}

We quantify the spectroscopic representation of photometric sources by way of example for a single independent calibration sub-sample in \cref{tab:6.4}. The noisy SOM coverage statistics \citep{Wright_2020b} feature the overall size of the spectroscopic calibration sample (``Sample (full)”; 33\,602 galaxies) and the fraction of cells these sources occupy (``$f$\textsubscript{pix}”; 96.8\%). The latter statistic indicates that $\approx$ 3.2\% of the noisy SOM are unoccupied by the calibration sample. However, the counts in the validation sample exhibit significant variation across the SOM, rendering this value less indicative of the absolute fraction of unrepresented sources in the wide sample. Therefore, \cref{tab:6.4} presents the fraction of cells $f$\textsubscript{gs}, for each of the tomographic bins that contain both calibration and wide sample sources (see \cref{qc}). It should also be mentioned that not all photometric sources hold the same lensing weight in cosmic shear estimates. Given the effective number of weighted photometric sources, merely $\approx$ 1.2\% of the validation sample sources located in cells unoccupied by the calibration sample, lie in relevant parts of the noisy SOM \citep{Wright_2020b}.

\subsubsection{Bias calculation}

The mean spec-$z$ of the calibration sample are compared to the mean spec-$z$ of the validation sample per cell. The measurement within each cell is weighted by the number of galaxies from the validation sample allocated to that specific cell. The mean of these weighted deviations yields the true redshift bias,
\begin{equation}
    \label{eq.8}
    B_{\rm s} = \langle z\textsubscript{\textrm{spec, cal}} - z\textsubscript{\textrm{spec, val}}\rangle \, .
\end{equation}
It is important to clarify that we do not know $B_{\rm s}$, but we desire to constrain it since, in reality, we lack information on the spec-$z$ of the EWS-like validation sample. In a similar fashion, the mean spec-$z$ values from the calibration sample are compared to the mean photo-$z$ of the validation sample per cell. The weighted mean of this quantity gives the photo-$z$ bias  
\begin{equation}
    \label{eq.9}
    B_{\rm p} = \langle z\textsubscript{\textrm{spec, cal}} - z\textsubscript{\textrm{photo, val}}\rangle \, .
\end{equation}

\noindent Detached from the extensively investigated disparity between spectroscopic and photometric redshifts, we utilise \cref{eq.9} to get a handle on the entwined bias denoted by \cref{eq.8}. This is motivated by the near-linear relationship between the biases, given that photo-$z$ estimates typically provide an approximation to spec-$z$s with scatter of order $\sigma_{\Delta z/(1+z)} \sim 0.01-0.03$ and outlier fractions of a few percent for wide-field surveys \citep{Coupon_2009, Tanaka2018,Duncan_2022,Tucci_2025}. For this work, the redshift dimension in the range of 0 $\leq$ $z$ $\leq$ 2.5, is sliced into ten equi-sized tomographic redshift bins. The methodology used to allocate galaxies to their respective bins significantly affects the bias estimation. In an initial approach, the redshift tomography is established by assigning each cell to its corresponding redshift bin, based on the cells' mean spec-$z$ derived from training, $z$\textsubscript{spec, cal}.

\begin{table*}[ht!]
\centering
\caption[]{Biases per bin for redshift tomography defined on the mean spec-$z$ of individual SOM cells. }
\vspace{10pt}
\begin{tabular}{cccccc} 
  \hline\hline\noalign{\vskip 1.5pt}
  Bias ($\Delta\langle z \rangle$) & bin 1  & bin 2  & bin 3  &bin 4  &bin 5  \\ 
 \hline
 QC no & 0.197 $\pm$ 0.029 & 0.122 $\pm$ 0.057 & 0.127 $\pm$ 0.061 & 0.056 $\pm$ 0.038 & 0.032 $\pm$ 0.010 \\
QC yes & 0.087 $\pm$ 0.012 & 0.062 $\pm$ 0.008 & 0.084 $\pm$ 0.056 & 0.050 $\pm$ 0.039 & 0.034 $\pm$ 0.008
\end{tabular}

\vspace{10pt}
\begin{tabular}{cccccc} 
\hline\hline\noalign{\vskip 1.5pt}
 Bias ($\Delta\langle z \rangle$)& bin 6  & bin 7  & bin 8  &bin 9  &bin 10  \\
 \hline
 QC no & 0.034 $\pm$ 0.012 & $-0.013$ $\pm$ 0.014 & $-0.064$ $\pm$ 0.018 & $-0.122$ $\pm$ 0.020 & $-0.253$ $\pm$ 0.034 \\
 QC yes &  0.042 $\pm$ 0.009 & 0.008 $\pm$ 0.010 & $-0.035$ $\pm$ 0.013 & $-0.079$ $\pm$ 0.013 & $-0.143$ $\pm$ 0.015 \\
\end{tabular}
\tablefoot{The values listed are the mean biases over 100 different line of sight sub-samples, including the respective standard deviation. Under the recognition of QC, merely bin 7 comes close to sufficing the dynamic \Euclid bias requirement of $\Delta\langle z \rangle$ < 0.002(1+$z$) when excluding standard deviations and considering the upper redshift bin limits from \cref{tab:6.4}. The expression `QC' refers to the consideration of quality control (see \cref{qual_col}).}
\label{tab:6.7}
\end{table*}

\begin{figure*}
\centering
\includegraphics[width=18cm]{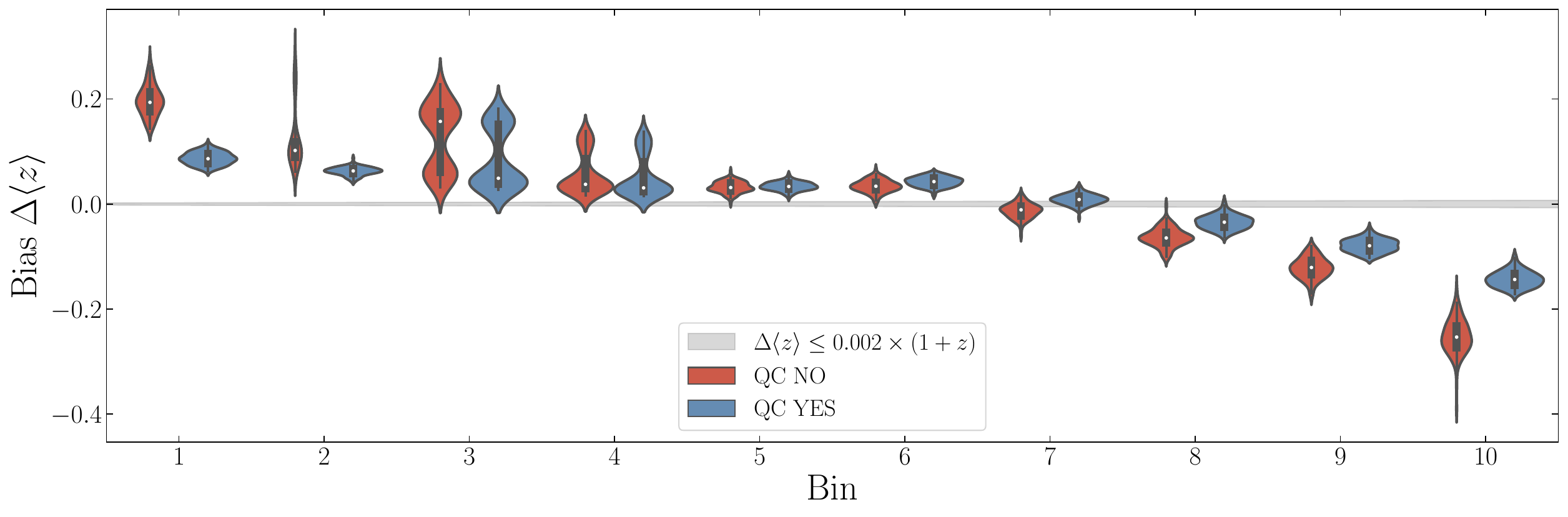} 
\caption[]{Violin plot of biases per bin using tomography defined by SOM spec-$z$ before (red) and after (blue) applying QC. These distributions also include box-and-whisker plots. Lastly, the dynamic ($\Delta\langle z \rangle$) < 0.002(1+$z$) \Euclid requirement are given by the grey shaded area.}
\label{fig:6.10}
\end{figure*}

\subsubsection{Gold sample}
\label{qc}

To enhance the statistical robustness of this approach, we employ gold sample selection. This selection is based on the identification cells without counterparts, whether spectroscopic or photometric, after projecting the EWS-like sample into the noisy SOM. Consequently, we exclusively focus on cells providing a guaranteed representation. This criterion effectively identifies and eliminates empty cells, along with the regions of the colour-colour space where the sample distributions significantly diverge \citep{Wright_2020b}. Given that $N$\textsubscript{calib} $\ll$ $N$\textsubscript{valid}, cells are more likely to lack calibration rather than validation sample sources, making this selection primarily responsible for removing validation sources from each bin. This may not be confused with the larger number of white, empty cells on the right hand side of \cref{fig:6.5}. Although the validation sample is larger than the calibration sample, it spans a smaller feature space, following the one-sided S/N cut and use of noisy as opposed to noiseless photometry, as mentioned at the beginning of \cref{chap_4}.

\begin{table*}[ht!]
\centering
\caption[]{Biases per bin for redshift tomography defined on the mean photo-$z$ of individual sources.}
\vspace{10pt}
\begin{tabular}{cccccc} 
  \hline\hline\noalign{\vskip 1.5pt}
  Bias ($\Delta\langle z \rangle$) & bin 1  & bin 2  & bin 3  &bin 4  &bin 5  \\ 
 \hline
 QC no & $-0.011$ $\pm$ 0.011 & $-0.002$ $\pm$ 0.005 & 0.003 $\pm$ 0.004 & 0.003 $\pm$ 0.004 & $-0.002$ $\pm$ 0.004\\
 
 QC yes & 0.005 $\pm$ 0.007 & $\emptyset$ $\pm$ 0.005 & 0.003 $\pm$ 0.004 & 0.003 $\pm$ 0.004 & $-0.002$ $\pm$ 0.004
\end{tabular}

\vspace{10pt}
\begin{tabular}{cccccc} 
\hline\hline\noalign{\vskip 1.5pt}
 Bias ($\Delta\langle z \rangle$)& bin 6  & bin 7  & bin 8  &bin 9  &bin 10  \\
 \hline
 QC no & 0.002 $\pm$ 0.004 & $-0.00$2 $\pm$ 0.004 & $-0.002$ $\pm$ 0.005 & 0.004 $\pm$ 0.007 & $-0.008$ $\pm$ 0.009 \\
 QC yes & 0.002 $\pm$ 0.004 & $-0.002$ $\pm$ 0.004 & $-0.002$ $\pm$ 0.005 & 0.006 $\pm$ 0.007 & $-0.009$ $\pm$ 0.008 \\
\end{tabular}

\tablefoot{Same as \cref{tab:6.7}, where all but bins 1 and 10 lie within the dynamic \Euclid bias requirement of ($\Delta\langle z \rangle$) < 0.002(1+$z$) when excluding standard deviations. Biases with values ($\Delta\langle z \rangle$) < 0.001 are marked with a null symbol ``$\emptyset$”.}
\label{tab:6.9}
\end{table*}

\begin{figure*}
    \centering
    \includegraphics[width=18cm]{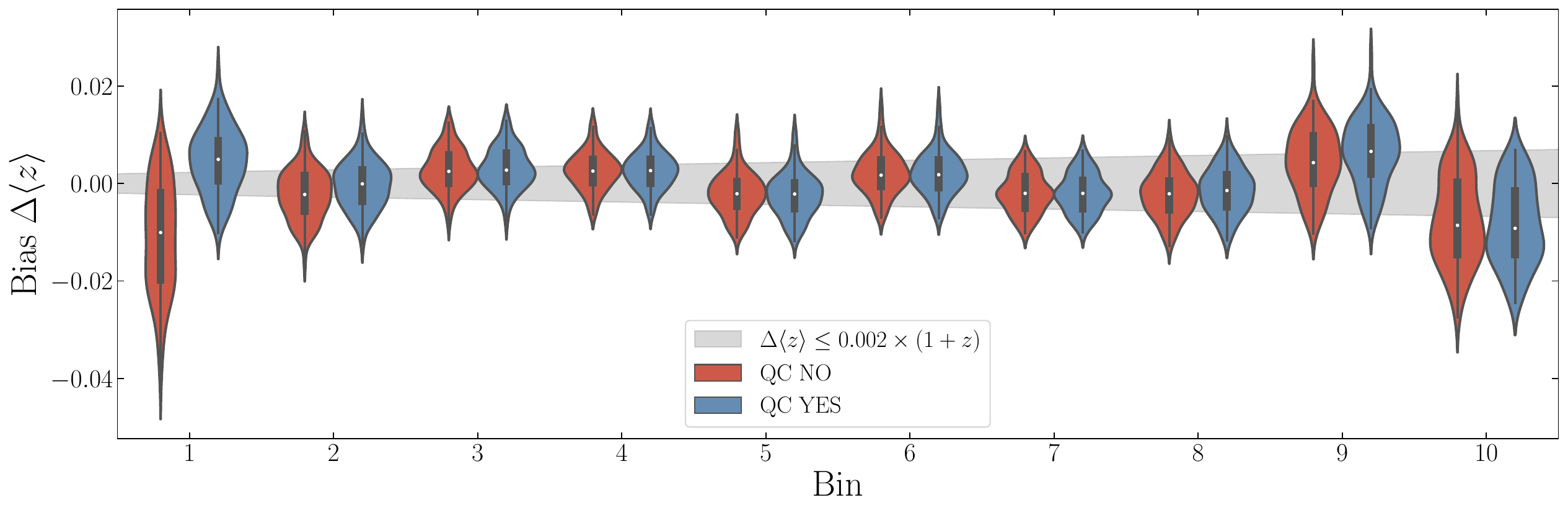}
    \caption[]{Same as \cref{fig:6.10}, but using tomography defined by the photo-$z$ of individual galaxies.}
    \label{fig:6.12}
\end{figure*}

\subsubsection{Quality control}
\label{qual_col}

To further reduce $B_{\rm s}$, we apply an additional, stricter cell selection which we refer to as quality control (QC). It is applied with the goal to remove SOM cells likely to produce biased redshift estimates. Fainter sources generally have larger photometric uncertainties and noisier colours, which can broaden photo-$z$ PDFs and introduce multimodality in the aggregated SOM-cell $p(z)$. However, since SOM cells are defined primarily in colour space, both bright and faint sources can occupy the same cell. Differences between cells passing and failing QC thus reflect a combination of photometric quality, colour-space structure, intrinsic source properties, and sample composition. QC down-selects cells based on their photo-$z$ bias, such that cells are selected if
\begin{equation}
    |B_{\rm p}| < 0.5 \, .
\end{equation}
As elaborated earlier, we can exploit the correlation of the two biases to further constrain $B_{\rm s}$ by restricting the parameter space of $B_{\rm p}$. On average, QC removes roughly 10\% of galaxies from the wide-like sample, with the fraction increasing from $\sim$ 5\% at low redshift to $\sim$ 20\% at high redshift. Given the decreasing fraction of gold sample--selected cells for higher-redshift bins, the QC condition could alternatively be defined dynamically as a function of redshift. Therefore, the implemented QC steps serve merely as a demonstration of the refinement possible using SOMs \citep{Wright_2020b}. The results of the spec-$z$ biases per bin as defined by the mean spec-$z$ of SOM cells, subsequent to gold sample selection and QC consideration, are depicted in \cref{tab:6.7} and \cref{fig:6.10}, respectively. We note that, in line with the issues raised in \cref{cv}, none of the bins in this configuration satisfy the \Euclid requirement outlined in \cref{sc:Intro}. All biases but those in bins 5, 6, 7, and 8 are more than a factor of 10 too large. Furthermore, the biases estimated by this method change sign from positive for the first six bins to negative for the remaining four bins, limiting the impact of the redshift bias on cosmological conclusions somewhat \citep{Wright_2020b}. This trend can be explained by the offset in redshift distributions between the two SOM cell samples, causing the reconstructed n($z$) of the wide sample to be compressed towards the calibration mean, effectively pulling the distribution tails into the middle, due to the lower number of calibration sources in these bins. \cref{tab:6.7} demonstrates that, except for bins 5 and 6, the implementation of QC reduces the redshift bias in the majority of cases. We note that the above-mentioned results do not account for a division of the northern and southern hemispheres. This should be kept in mind, as each hemisphere observes sources with different instruments and therefore bands that may influence the SOM.

\subsection{Photo-\textit{z} based tomography}
\label{PC}

An alternative approach to defining redshift tomography is based on the bin assignment from individual {\tt{NNPZ}} PDF medians as photo-$z$ point estimates of the calibration sample, as briefly discussed in \cref{cv}. SOM cells can contain heterogeneous galaxy populations, which may lead to non-Gaussian or even bimodal $p(z)$ within a cell, since in such cases the mean of the SOM-cell $p(z)$ is not necessarily representative of the dominant population. If, instead, redshift binning is performed using per-galaxy photo-$z$ estimates, then a multimodal SOM cell can naturally be sub-sampled as most individual photo-$z$ PDFs are unimodal, making their medians effectively equivalent to their modes. This is particularly relevant for cells, where the calibration and wide-like redshift distributions differ significantly. While bimodality in individual photo-$z$ PDFs does occur, we find it to be considerably less common than multimodality in the cell-aggregated SOM $p(z)$; therefore, this issue is substantially reduced when using photo-$z$ point estimates.

 In line with the pipeline used for tomography defined on the mean spec-$z$ of the SOM, gold sampling and QC cuts are also applied to the photo-$z$ based tomography. As in \cref{fig:6.10}, the biases estimated with this tomography (see \cref{fig:6.12}) change sign. In this case, bin 1 as well as bins 2, 5, 7, 8 and 10 have a mean negative sign, whereas the remaining bins have a mean positive sign. However, in contrast to \cref{fig:6.10}, there is no degrading bias trend. \Cref{fig:6.12} and \cref{tab:6.9} demonstrate that defining the redshift tomography on photo-$z$ rather than on the mean spec-$z$ of every SOM cell (see \cref{tab:6.7}), provides much improved bias results for noisy photometry. In fact, considering QC, all but two bin means lie within the \Euclid requirement for $\Delta\langle z \rangle$. This enables the use of nearly the entire spectroscopic sample for calibration, as its selection is no longer tied to the mean spec-$z$ of SOM cells (see \cref{fig:43}). To evaluate the effectiveness of galaxy selection strategies in the SOM, we illustrated the two approaches in \cref{fig:6.7}.  Galaxies selected via photo-$z$ binning in the horizontal band exhibit significantly more compact and localised contours, well aligned with the selected redshift slice. In contrast, the SOM cell-based selection leads to a broader distribution that extends well beyond the target bin, as reflected in the larger extent of its contours. This highlights that photo-$z$ binning within the SOM offers greater fidelity in isolating galaxies truly belonging to the desired tomographic redshift range.

\section{Scientific implications for \Euclid}
\label{chap_5}

Given the bias uncertainties expected for redshift measurements performed by \Euclid, one can propagate those into biases of cosmological parameters. The obtained level of accuracy to quantify this task is provided via the margin of uncertainty regarding the cosmological parameters of interest. Consequently, said uncertainties are calculated based on the results obtained in \cref{PC,IR}.

\renewcommand{\arraystretch}{1.1} 
\begin{table*}
    \centering
     \caption{Biases on cosmological parameters due to redshift uncertainties.}
    \begin{tabular}{lcccc}
    \hline\hline\noalign{\vskip 1.5pt}
    case & $\Delta\Omega_\mathrm{m}/\sigma_{\Omega_\mathrm{m}}$ & $\Delta\sigma_8/\sigma_{\sigma_8}$ &$\Delta w_0/\sigma_{w_0}$ &$\Delta w_a/\sigma_{w_a}$ \\ 
    \hline\noalign{\vskip 1.5pt}
realistic and QC yes  &  $0.27\pm0.16$ & $0.11\pm0.06$ & $0.46\pm0.26$ & $0.24\pm0.14$ \\  
realistic and QC no& $0.28\pm0.15$ & $0.11\pm0.06$ & $0.46\pm0.26$ & $0.24\pm0.13$ \\  
best and QC yes & $0.09\pm0.02$ & $0.04\pm0.01$ & $0.16\pm0.04$ & $0.08\pm0.02$ \\  
best and QC no & $0.08\pm0.02$ & $0.03\pm0.01$ & $0.14\pm0.04$ & $0.07\pm0.02$ \\  
    \end{tabular}
    \tablefoot{All values are given in units of $\sigma$ considering the four different cases of redshift calibration (refer to \cref{PC,IR}). The errors are given as standard deviations of the distribution displayed in \cref{fig:biases}. Note that they are given as symmetric only for reference.}
    \label{tab:biases_cosmo}
\end{table*}
\renewcommand{\arraystretch}{1.0}

\subsection{Cosmological parameter biases}
\label{CB}

To estimate the effect on cosmological parameter inference, we sample the shifts of the mean redshift measurements from their covariance matrix. Next, we follow the prescription denoted in \citet{reischke_propagating_2023} and convert those samples into shifts $n^{(m)}(z_n)$ in the functional space of the $n(z)$. Here the redshift support was already discretised, i.e. $\Delta{n}^m(z_n)$ is the shift (uncertainty) of the $m$-th tomographic bin, evaluated at the $n$-th redshift. These shifts can be rearranged into a single vector $\Delta n^\alpha$ and we use Greek indices to denote indices running over the $n(z)$ shifts. By estimating the response of the lensing angular power spectrum, $C_\ell$, to changes in the $n(z)$ via functional derivatives \citep{reischke_propagating_2023}, one can propagate these uncertainties into the shifts of cosmological parameters, $\Delta\theta^{i}$, to be
\begin{equation}
    \Delta\theta^{i} = -(\mathbf{F})^{i}_{\;k}\mathcal{F}^{k}_{\;\beta}\Delta n^\beta\;.
\end{equation}
Here we assume the sum convention, so that repeated indices appearing both as vector and dual vector index are summed over. Latin indices are used as labels for cosmological parameters, while $\mathcal{F}^{k}_{\;\beta}$ is the mixed pseudo-Fisher matrix 
\begin{equation}
    \mathcal{F}^{k}_{\;\beta} \coloneqq - \mathrm{E}\left[\frac{\partial\ln L}{\partial \theta_k}\frac{\delta\ln L}{\delta n^\beta}\mathcal{D}\chi_\beta\right]\;.
\end{equation}
The functional derivative is indicated with ${\delta}$ with $\mathcal{D}\chi_\alpha$ as integration measure that arises from the discretisation \citep[we refer to][for more details]{reischke_propagating_2023}. Note that the index $\beta$ is not summed over in the above equation, as it appears as a dual vector index twice. $\mathbf{F}$ is the Fisher matrix of the cosmological parameters. Lastly,
$L$ is the likelihood, which we assume to be Gaussian for the forecasting done here, by this assumption, all cosmological information is contained in the angular power spectrum, $C_\ell$, where $\ell$ is the multipole. Given a set of tomographic bins, we collect all spectra in a matrix $\mathbf{C}_\ell$, where the Fisher matrix in this case is given by \citep{tegmark_karhunen_1997}
\begin{equation}
    F^{i}_{\; k} = f_\mathrm{sky}\sum_{\ell = \ell_\mathrm{min}}^{\ell_\mathrm{max}} \frac{2\ell + 1}{2}\mathrm{tr}\left[\mathbf{C}^{-1}_\ell\partial^{i}\mathbf{C}_\ell\mathbf{C}^{-1}_\ell\partial_k\mathbf{C}_\ell\right]\;,
\end{equation}
with the sky fraction, $f_\mathrm{sky}$. Deviating from the tomographic bin definition used for calibration purposes in this work, we instead follow the boundaries of the 10 equi-populated bins defined in \cite{2020A&A...642A.191E}.
The final cosmic shear only forecast biases are reported as
\begin{equation}
    b^i = \frac{\Delta\theta^i}{[(\boldsymbol{F}^{-1})_{ii}]^{1/2}}\;,
\end{equation}
in units of the marginalised standard deviation of a forecasted \Euclid setting \citep[labelled as optimistic in][]{2020A&A...642A.191E}. In particular, we focus on the matter density, $\Omega_\mathrm{m}$, the power spectrum amplitude $\sigma_8$, and two equations of state parameters $w_0$ and $w_a$. The following scenarios are assumed: the calibration settings which achieve the lowest biases (best), see \cref{PC} and those which achieve the most realistic biases (real), see \cref{m3}.

\begin{figure}[ht!]
    \centering
    \includegraphics[trim={0.2cm 0 1cm 0},clip,width = 0.5\textwidth]{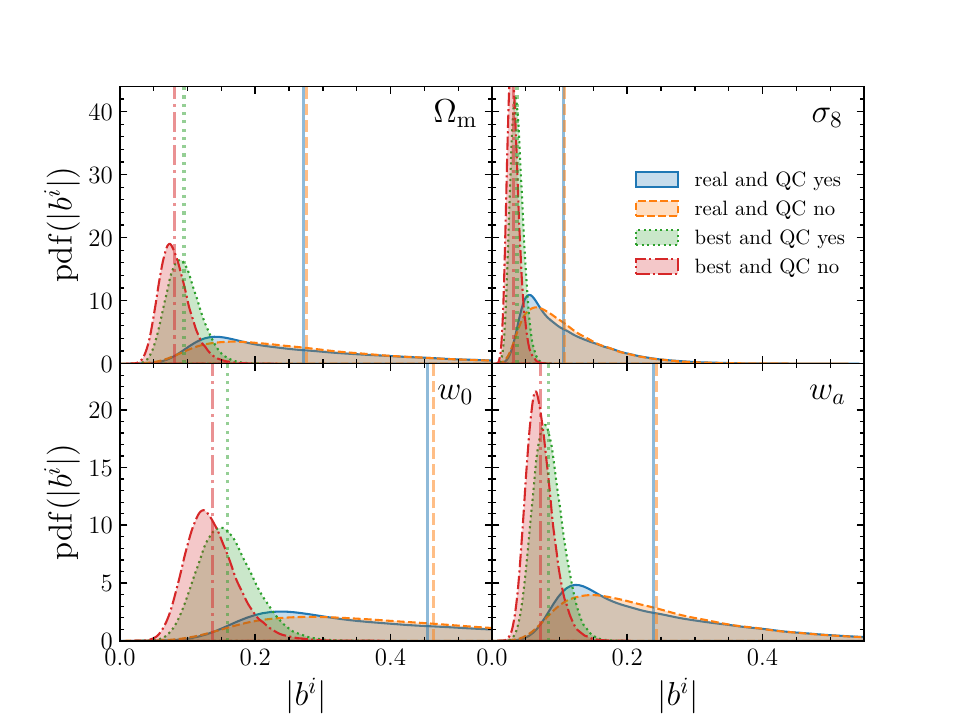}
    \caption{Forecasted absolute biases in units of standard deviations for cosmological parameters. Each panel shows a different cosmological parameter; the different line styles and colours correspond to the different cases presented in \cref{PC} and \cref{IR}. The vertical lines indicate the expected bias corresponding to each case. The kernel density estimate was taken from $10^6$ realisations of the $n(z)$.}
    \label{fig:biases}
\end{figure}

\cref{tab:biases_cosmo} shows mean relative biases across realisations, expressed in units of the statistical uncertainties. Since our forecast is optimistic by design, we expect the real-data scenario to be more favourable due to conservative assumptions in our modelling. In fact, we find that the lower idealised bias outperforms the realistic biases for all cases and each parameter. Furthermore, QC does not necessarily reduce the biases for the cosmological parameters. The lowest bias is induced on $\sigma_8$ which is to be expected, as completely coherent shifts (required for an overall amplitude change leading to a significant $\sigma_8$ bias) during sampling from the $n(z)$ are very unlikely. While both $w_0$ and $w_a$ are affected by uncertainties in the $n(z)$, we find that $w_0$, the present-day value of the DE EOS, is more susceptible to systematic biases caused by redshift errors than $w_a$, which characterises its evolution with redshift. The biases are displayed in \cref{fig:biases} and appear generally asymmetric with a long tail to larger biases, in excess of $1 \sigma$. The vertical lines indicate the expected bias, showing again that $w_0$ is most affected, while all biases are smaller than $1 \sigma$ on average. In general, one should keep in mind that the uncertainties are derived from an optimistic setting without the inclusion of any additional systematic effects. Hence a fair assessment of the performance of the redshift calibration is to treat it as idealised, since a more realistic analysis would produce weaker constraints on cosmological parameters, decreasing the significance of their biases due to the shifts in the $n(z)$.

\section{Discussion and outlook}
\label{chap_6}

In this paper, we validated various approaches regarding the calibration of photometric redshift distributions for \Euclid. The obtained results are indicative as to whether the current aims and goals concerning the implementation of Stage IV precision cosmology are achievable assuming DR3-level data or require updated methodologies. 

Similar to previous work performed for KiDS, such as by \cite{Wright_2020b, wright2025_calibration}, this work focused on implementing a SOM to reduce $n(z)$ biases within 10 tomographic bins [$0 < z \leq 2.5$] to satisfy the \Euclid precision requirement. In extension of this, two types of fundamental tomography layouts were considered, with additional steps undertaken to increase the level of realism. With mocks from the Flagship2 catalogue, we trained a SOM on a calibration sample subject to a non-trivial selection function, before populating it with a large EWS-like photometric sample (see \cref{CSC}). The obtained biases were investigated under the implementation of gold--sample selection and QC control cuts. In the set-up, where the redshift tomography was defined on the mean spec-$z$ per SOM cell, we find that, considering QC, none of the ten biases lie within the \Euclid requirement (see \cref{SC}). In contrast, where the redshift tomography was defined on photo-$z$, we find that eight out of ten biases satisfy the \Euclid requirement (see \cref{PC}). This strongly implies that redshift tomography should generally be defined using photo-$z$s of individual galaxies, rather than the mean spec-$z$ of SOM cells. The level of realism of the calibration sample used in this approach was subsequently increased by adjusting it to mimic the C3R2 survey sample. For this, three different set-ups were implemented, for which we demonstrate the neccessity of spectra in typically under-sampled colour-colour space and the performance increase achievable with photo-$z$ tomography. As a result, in \cref{m3}, six out of ten bins satisfy the \Euclid requirement. In addition, ideas and suggestions to further raise the level of realism were discussed.

Lastly, we utilised a cosmic shear only Fisher forecast to estimate the biases induced into the measurement of cosmological parameters considering the different calibration approaches. By sampling from the covariance of shifts in the mean of the photo-$z$ distribution, we deduced parameter bias distributions for $\Omega_\mathrm{m}$, $\sigma_8$, $w_0$, and $w_a$. We showed, that while the relative bias of $w_0$ displays an increased mean due to the long tail of its distribution, all other biases are observed to be lower than $0.1\sigma$ in the idealised case and $0.3\sigma$ for increased realism (see \cref{CB}). Generally, biases in the photo-$z$ distribution lead to larger biases in parameters controlling the temporal evolution of the background cosmology and perturbations. 

Despite these promising results, it should be in the interest of the precision cosmology community to assemble an extensive spec-$z$ calibration sample. Significant challenges related to survey limitations persist, and preparing such a sample appears crucial to pave the way for successful redshift calibration. As our analysis was conducted under idealised conditions, it serves primarily as a proof of concept to guide the optimal definition of redshift tomography. While we strongly recommend the adoption of our findings, they will need to be adapted and re-evaluated in the context of the data available at the time of the first cosmological analysis with \Euclid DR3.

\begin{acknowledgements}
  WR acknowledges DLR support (Foerderkennzeichen 50002207). 
  \AckEC \AckCosmoHub
\end{acknowledgements}

%
%

\bibliography{my_bib}

@ARTICLE{reischke_propagating_2023,
       author = {{Reischke}, Robert},
        title = "{Propagating photo-z uncertainties: a functional derivative approach}",
      journal = {\mnras},
         year = 2024,
        month = jun,
       volume = {530},
       number = {4},
        pages = {4412--4421},
          doi = {10.1093/mnras/stad3791},
archivePrefix = {arXiv},
       eprint = {2301.04085},
 primaryClass = {astro-ph.CO},
       adsurl = {https://ui.adsabs.harvard.edu/abs/2024MNRAS.530.4412R}
}

@ARTICLE{tegmark_karhunen_1997,
       author = {{Tegmark}, Max and {Taylor}, Andy N. and {Heavens}, Alan F.},
        title = "{Karhunen-Lo{\`e}ve Eigenvalue Problems in Cosmology: How Should We Tackle Large Data Sets?}",
      journal = {\apj},
         year = 1997,
        month = may,
       volume = {480},
       number = {1},
        pages = {22--35},
          doi = {10.1086/303939},
archivePrefix = {arXiv},
       eprint = {astro-ph/9603021},
 primaryClass = {astro-ph},
       adsurl = {https://ui.adsabs.harvard.edu/abs/1997ApJ...480...22T}
}

@ARTICLE{2017MNRAS.465.1454H,
       author = {{Hildebrandt}, H. and {Viola}, M. and {Heymans}, C.and {Joudaki}, S. and {Kuijken}, K. and {Blake}, C. and {Erben}, T. and {Joachimi}, B. and {Klaes}, D. and {Miller}, L. and {Morrison}, C.~B. and {Nakajima}, R.},
        title = "{KiDS-450: cosmological parameter constraints from tomographic weak gravitational lensing}",
      journal = {\mnras},
         year = 2017,
        month = feb,
       volume = {465},
       number = {2},
        pages = {1454--1498},
          doi = {10.1093/mnras/stw2805},
archivePrefix = {arXiv},
       eprint = {1606.05338},
 primaryClass = {astro-ph.CO},
       adsurl = {https://ui.adsabs.harvard.edu/abs/2017MNRAS.465.1454H}
}

@ARTICLE{2018PhRvD..98d3526A,
       author = {{Abbott}, T.~M.~C. and {Abdalla}, F.~B. and {Alarcon}, A. and {Allam}, S. and {Annis}, J. and {Avila}, S. and {Aylor}, K. and {Banerji}, M. and {Banik}, N. and {Baxter}, E.~J. and {Bechtol}, K. and {Becker}},
        title = "{Dark Energy Survey year 1 results: Cosmological constraints from galaxy clustering and weak lensing}",
      journal = {\prd},
         year = 2018,
        month = aug,
       volume = {98},
       number = {4},
          eid = {043526},
        pages = {043526},
          doi = {10.1103/PhysRevD.98.043526},
archivePrefix = {arXiv},
       eprint = {1708.01530},
 primaryClass = {astro-ph.CO},
       adsurl = {https://ui.adsabs.harvard.edu/abs/2018PhRvD..98d3526A}
}

@ARTICLE{2019ApJ...873..111I,
       author = {{Ivezi{\'c}}, {\v{Z}}eljko and {Kahn}, Steven M.},
        title = "{LSST: From Science Drivers to Reference Design and Anticipated Data Products}",
      journal = {\apj},
         year = 2019,
        month = mar,
       volume = {873},
       number = {2},
          eid = {111},
        pages = {111},
          doi = {10.3847/1538-4357/ab042c},
archivePrefix = {arXiv},
       eprint = {0805.2366},
 primaryClass = {astro-ph},
       adsurl = {https://ui.adsabs.harvard.edu/abs/2019ApJ...873..111I}
}

@ARTICLE{2020A&A...642A.191E,
       author = {{Euclid Collaboration: Blanchard}, A. and {Camera}, S. and {Carbone}, C. and {Cardone}, V.~F. and {Casas}, S. and {Clesse}, S. and {Ili{\'c}}, S. and {Kilbinger}, M. and {Kitching}, T. and {Kunz}, M. and {Lacasa}, F. and {Linder}, E. and {Majerotto}, E. and {Markovi{\v{c}}}, K. and {Martinelli}, M. and {Pettorino}, V. and {Pourtsidou}, A. and {Sakr}, Z. and {S{\'a}nchez}, A.~G. and {Sapone}, D. and {Tutusaus}, I. and {Yahia-Cherif}, S. and {Yankelevich}, V. and {Andreon}, S. and {Aussel}, H. and {Balaguera-Antol{\'\i}nez}, A. and {Baldi}, M. and {Bardelli}, S. and {Bender}, R. and {Biviano}, A. and {Bonino}, D. and {Boucaud}, A. and {Bozzo}, E. and {Branchini}, E. and {Brau-Nogue}, S. and {Brescia}, M. and {Brinchmann}, J. and {Burigana}, C. and {Cabanac}, R. and {Capobianco}, V. and {Cappi}, A. and {Carretero}, J. and {Carvalho}, C.~S. and {Casas}, R. and {Castander}, F.~J. and {Castellano}, M. and {Cavuoti}, S. and {Cimatti}, A. and {Cledassou}, R. and {Colodro-Conde}, C. and {Congedo}, G. and {Conselice}, C.~J. and {Conversi}, L. and {Copin}, Y. and {Corcione}, L. and {Coupon}, J. and {Courtois}, H.~M. and {Cropper}, M. and {Da Silva}, A. and {de la Torre}, S. and {Di Ferdinando}, D. and {Dubath}, F. and {Ducret}, F. and {Duncan}, C.~A.~J. and {Dupac}, X. and {Dusini}, S. and {Fabbian}, G. and {Fabricius}, M. and {Farrens}, S. and {Fosalba}, P. and {Fotopoulou}, S. and {Fourmanoit}, N. and {Frailis}, M. and {Franceschi}, E. and {Franzetti}, P. and {Fumana}, M. and {Galeotta}, S. and {Gillard}, W. and {Gillis}, B. and {Giocoli}, C. and {G{\'o}mez-Alvarez}, P. and {Graci{\'a}-Carpio}, J. and {Grupp}, F. and {Guzzo}, L. and {Hoekstra}, H. and {Hormuth}, F. and {Israel}, H. and {Jahnke}, K. and {Keihanen}, E. and {Kermiche}, S. and {Kirkpatrick}, C.~C. and {Kohley}, R. and {Kubik}, B. and {Kurki-Suonio}, H. and {Ligori}, S. and {Lilje}, P.~B. and {Lloro}, I. and {Maino}, D. and {Maiorano}, E. and {Marggraf}, O. and {Martinet}, N. and {Marulli}, F. and {Massey}, R. and {Medinaceli}, E. and {Mei}, S. and {Mellier}, Y. and {Metcalf}, B. and {Metge}, J.~J. and {Meylan}, G. and {Moresco}, M. and {Moscardini}, L. and {Munari}, E. and {Nichol}, R.~C. and {Niemi}, S. and {Nucita}, A.~A. and {Padilla}, C. and {Paltani}, S. and {Pasian}, F. and {Percival}, W.~J. and {Pires}, S. and {Polenta}, G. and {Poncet}, M. and {Pozzetti}, L. and {Racca}, G.~D. and {Raison}, F. and {Renzi}, A. and {Rhodes}, J. and {Romelli}, E. and {Roncarelli}, M. and {Rossetti}, E. and {Saglia}, R. and {Schneider}, P. and {Scottez}, V. and {Secroun}, A. and {Sirri}, G. and {Stanco}, L. and {Starck}, J. -L. and {Sureau}, F. and {Tallada-Cresp{\'\i}}, P. and {Tavagnacco}, D. and {Taylor}, A.~N. and {Tenti}, M. and {Tereno}, I. and {Toledo-Moreo}, R. and {Torradeflot}, F. and {Valenziano}, L. and {Vassallo}, T. and {Verdoes Kleijn}, G.~A. and {Viel}, M. and {Wang}, Y. and {Zacchei}, A. and {Zoubian}, J. and {Zucca}, E.},
        title = "{Euclid preparation. VII. Forecast validation for Euclid cosmological probes}",
      journal = {\aap},
         year = 2020,
        month = oct,
       volume = {642},
          eid = {A191},
        pages = {A191},
          doi = {10.1051/0004-6361/202038071},
archivePrefix = {arXiv},
       eprint = {1910.09273},
 primaryClass = {astro-ph.CO},
       adsurl = {https://ui.adsabs.harvard.edu/abs/2020A\&A...642A.191E}
}

@ARTICLE{2013PhR...530...87W,
       author = {{Weinberg}, David H. and {Mortonson}, Michael J. and {Eisenstein}, Daniel J.},
        title = "{Observational probes of cosmic acceleration}",
      journal = {\physrep},
         year = 2013,
        month = sep,
       volume = {530},
       number = {2},
        pages = {87--255},
          doi = {10.1016/j.physrep.2013.05.001},
archivePrefix = {arXiv},
       eprint = {1201.2434},
 primaryClass = {astro-ph.CO},
       adsurl = {https://ui.adsabs.harvard.edu/abs/2013PhR...530...87W}
}

@ARTICLE{2020A&A...641A...1P,
       author = {{Planck Collaboration: Aghanim}, N. and {Akrami}, Y., et al. },
        title = "{Planck 2018 results. I. Overview and the cosmological legacy of Planck}",
      journal = {\aap},
         year = 2020,
        month = sep,
       volume = {641},
          eid = {A1},
        pages = {A1},
          doi = {10.1051/0004-6361/201833880},
archivePrefix = {arXiv},
       eprint = {1807.06205},
 primaryClass = {astro-ph.CO},
       adsurl = {https://ui.adsabs.harvard.edu/abs/2020A\&A...641A...1P}
}

@ARTICLE{kohonen,
    author = {T. Kohonen},
    title = {Self-organized formation of topologically correct feature maps},
    journal = {Biological Cybernetics},
    volume = {43},
    pages={59--69},
    year = {1982}
}

@ARTICLE{spergel_2015,
       author = {{Spergel}, D. and {Gehrels}, N. and {Baltay}, C. and {Bennett}, D. and {Breckinridge}, J. and {Donahue}, M. and {Dressler}, A. and {Gaudi}, B.~S. and {Greene}, T. and {Guyon}, O. and {Hirata}, C. and {Kalirai}, J. and {Kasdin}, N.~J. and {Macintosh}, B. and {Moos}, W. and {Perlmutter}, S. and {Postman}, M. and {Rauscher}, B. and {Rhodes}, J. and {Wang}, Y. and {Weinberg}, D. and {Benford}, D. and {Hudson}, M. and {Jeong}, W. -S. and {Mellier}, Y. and {Traub}, W. and {Yamada}, T. and {Capak}, P. and {Colbert}, J. and {Masters}, D. and {Penny}, M. and {Savransky}, D. and {Stern}, D. and {Zimmerman}, N. and {Barry}, R. and {Bartusek}, L. and {Carpenter}, K. and {Cheng}, E. and {Content}, D. and {Dekens}, F. and {Demers}, R. and {Grady}, K. and {Jackson}, C. and {Kuan}, G. and {Kruk}, J. and {Melton}, M. and {Nemati}, B. and {Parvin}, B. and {Poberezhskiy}, I. and {Peddie}, C. and {Ruffa}, J. and {Wallace}, J.~K. and {Whipple}, A. and {Wollack}, E. and {Zhao}, F.},
        title = "{Wide-Field InfrarRed Survey Telescope-Astrophysics Focused Telescope Assets WFIRST-AFTA 2015 Report}",
      journal = {arXiv e-prints},
         year = 2015,
        month = mar,
          eid = {arXiv:1503.03757},
        pages = {arXiv:1503.03757},
          doi = {10.48550/arXiv.1503.03757},
archivePrefix = {arXiv},
       eprint = {1503.03757},
 primaryClass = {astro-ph.IM},
       adsurl = {https://ui.adsabs.harvard.edu/abs/2015arXiv150303757S}
}

@article{Erben_2013,
	doi = {10.1093/mnras/stt928},
  
	url = {https://doi.org/10.1093%2Fmnras%2Fstt928},
  
	year = 2013,
	month = {jun},
  
	publisher = {Oxford University Press ({OUP})},
  
	volume = {433},
  
	number = {3},
  
	pages = {2545--2563},
  
	author = {T. Erben and H. Hildebrandt and L. Miller and L. van Waerbeke and C. Heymans and H. Hoekstra and T. D. Kitching and Y. Mellier and J. Benjamin and M. Kilbinger and K. Kuijken},
  
	title = {{CFHTLenS}: the Canada{\textendash}France{\textendash}Hawaii Telescope Lensing Survey {\textendash} imaging data and catalogue products},
  
	journal = {MNRAS}
}

@ARTICLE{Hikage_2019,
       author = {{Hikage}, Chiaki and {Oguri}, Masamune and {Hamana}, Takashi and {More}, Surhud and {Mandelbaum}, Rachel and {Takada}, Masahiro and {K{\"o}hlinger}, Fabian and {Miyatake}, Hironao and {Nishizawa}, Atsushi J. and {Aihara}, Hiroaki and {Armstrong}, Robert and {Bosch}, James and {Coupon}, Jean and {Ducout}, Anne and {Ho}, Paul and {Hsieh}, Bau-Ching and {Komiyama}, Yutaka and {Lanusse}, Fran{\c{c}}ois and {Leauthaud}, Alexie and {Lupton}, Robert H. and {Medezinski}, Elinor and {Mineo}, Sogo and {Miyama}, Shoken and {Miyazaki}, Satoshi and {Murata}, Ryoma and {Murayama}, Hitoshi and {Shirasaki}, Masato and {Sif{\'o}n}, Crist{\'o}bal and {Simet}, Melanie and {Speagle}, Joshua and {Spergel}, David N. and {Strauss}, Michael A. and {Sugiyama}, Naoshi and {Tanaka}, Masayuki and {Utsumi}, Yousuke and {Wang}, Shiang-Yu and {Yamada}, Yoshihiko},
        title = "{Cosmology from cosmic shear power spectra with Subaru Hyper Suprime-Cam first-year data}",
      journal = {\pasj},
         year = 2019,
        month = apr,
       volume = {71},
       number = {2},
          eid = {43},
        pages = {43},
          doi = {10.1093/pasj/psz010},
archivePrefix = {arXiv},
       eprint = {1809.09148},
 primaryClass = {astro-ph.CO},
       adsurl = {https://ui.adsabs.harvard.edu/abs/2019PASJ...71...43H}
}

@ARTICLE{Caldwell_1998,
       author = {{Caldwell}, R.~R. and {Dave}, Rahul and {Steinhardt}, Paul J.},
        title = "{Cosmological Imprint of an Energy Component with General Equation of State}",
      journal = {\prl},
         year = 1998,
        month = feb,
       volume = {80},
       number = {8},
        pages = {1582--1585},
          doi = {10.1103/PhysRevLett.80.1582},
archivePrefix = {arXiv},
       eprint = {astro-ph/9708069},
 primaryClass = {astro-ph},
       adsurl = {https://ui.adsabs.harvard.edu/abs/1998PhRvL..80.1582C}
}

@article{Asghari_2019,
doi = {10.1088/1475-7516/2019/04/042},
url = {https://dx.doi.org/10.1088/1475-7516/2019/04/042},
year = {2019},
month = {apr},
publisher = {},
volume = {04},
pages = {042},
author = {Asghari, Mahnaz and Jiménez, Jose Beltrán and Khosravi, Shahram and Mota, David F.},
title = {On structure formation from a small-scales-interacting dark sector},
journal = {JCAP}
}

@ARTICLE{2018Mandel,
       author = {{Mandelbaum}, Rachel},
        title = "{Weak Lensing for Precision Cosmology}",
      journal = {\araa},
         year = 2018,
        month = sep,
       volume = {56},
        pages = {393--433},
          doi = {10.1146/annurev-astro-081817-051928},
archivePrefix = {arXiv},
       eprint = {1710.03235},
 primaryClass = {astro-ph.CO},
       adsurl = {https://ui.adsabs.harvard.edu/abs/2018ARA&A..56..393M}
}

@ARTICLE{laureijs_2011,
       author = {{Laureijs}, R. and {Amiaux}, J. and {Arduini}, S. and {Augu{\`e}res}, J. -L. and {Brinchmann}, J. and {Cole}, R. and {Cropper}, M. and {Dabin}, C. and {Duvet}, L. and {Ealet}, A. and {Garilli}, B. and {Gondoin}, P. and {Guzzo}, L. and {Hoar}, J. and {Hoekstra}, H. and {Holmes}, R. and {Kitching}, T. and {Maciaszek}, T. and {Mellier}, Y. and {Pasian}, F. and {Percival}, W. and {Rhodes}, J. and {Saavedra Criado}, G. and {Sauvage}, M. and {Scaramella}, R. and {Valenziano}, L. and {Warren}, S. and {Bender}, R. and {Castander}, F. and {Cimatti}, A. and {Le F{\`e}vre}, O. and {Kurki-Suonio}, H. and {Levi}, M. and {Lilje}, P. and {Meylan}, G. and {Nichol}, R. and {Pedersen}, K. and {Popa}, V. and {Rebolo Lopez}, R. and {Rix}, H. -W. and {Rottgering}, H. and {Zeilinger}, W. and {Grupp}, F. and {Hudelot}, P. and {Massey}, R. and {Meneghetti}, M. and {Miller}, L. and {Paltani}, S. and {Paulin-Henriksson}, S. and {Pires}, S. and {Saxton}, C. and {Schrabback}, T. and {Seidel}, G. and {Walsh}, J. and {Aghanim}, N. and {Amendola}, L. and {Bartlett}, J. and {Baccigalupi}, C. and {Beaulieu}, J. -P. and {Benabed}, K. and {Cuby}, J. -G. and {Elbaz}, D. and {Fosalba}, P. and {Gavazzi}, G. and {Helmi}, A. and {Hook}, I. and {Irwin}, M. and {Kneib}, J. -P. and {Kunz}, M. and {Mannucci}, F. and {Moscardini}, L. and {Tao}, C. and {Teyssier}, R. and {Weller}, J. and {Zamorani}, G. and {Zapatero Osorio}, M.~R. and {Boulade}, O. and {Foumond}, J.~J. and {Di Giorgio}, A. and {Guttridge}, P. and {James}, A. and {Kemp}, M. and {Martignac}, J. and {Spencer}, A. and {Walton}, D. and {Bl{\"u}mchen}, T. and {Bonoli}, C. and {Bortoletto}, F. and {Cerna}, C. and {Corcione}, L. and {Fabron}, C. and {Jahnke}, K. and {Ligori}, S. and {Madrid}, F. and {Martin}, L. and {Morgante}, G. and {Pamplona}, T. and {Prieto}, E. and {Riva}, M. and {Toledo}, R. and {Trifoglio}, M. and {Zerbi}, F. and {Abdalla}, F. and {Douspis}, M. and {Grenet}, C. and {Borgani}, S. and {Bouwens}, R. and {Courbin}, F. and {Delouis}, J. -M. and {Dubath}, P. and {Fontana}, A. and {Frailis}, M. and {Grazian}, A. and {Koppenh{\"o}fer}, J. and {Mansutti}, O. and {Melchior}, M. and {Mignoli}, M. and {Mohr}, J. and {Neissner}, C. and {Noddle}, K. and {Poncet}, M. and {Scodeggio}, M. and {Serrano}, S. and {Shane}, N. and {Starck}, J. -L. and {Surace}, C. and {Taylor}, A. and {Verdoes-Kleijn}, G. and {Vuerli}, C. and {Williams}, O.~R. and {Zacchei}, A. and {Altieri}, B. and {Escudero Sanz}, I. and {Kohley}, R. and {Oosterbroek}, T. and {Astier}, P. and {Bacon}, D. and {Bardelli}, S. and {Baugh}, C. and {Bellagamba}, F. and {Benoist}, C. and {Bianchi}, D. and {Biviano}, A. and {Branchini}, E. and {Carbone}, C. and {Cardone}, V. and {Clements}, D. and {Colombi}, S. and {Conselice}, C. and {Cresci}, G. and {Deacon}, N. and {Dunlop}, J. and {Fedeli}, C. and {Fontanot}, F. and {Franzetti}, P. and {Giocoli}, C. and {Garcia-Bellido}, J. and {Gow}, J. and {Heavens}, A. and {Hewett}, P. and {Heymans}, C. and {Holland}, A. and {Huang}, Z. and {Ilbert}, O. and {Joachimi}, B. and {Jennins}, E. and {Kerins}, E. and {Kiessling}, A. and {Kirk}, D. and {Kotak}, R. and {Krause}, O. and {Lahav}, O. and {van Leeuwen}, F. and {Lesgourgues}, J. and {Lombardi}, M. and {Magliocchetti}, M. and {Maguire}, K. and {Majerotto}, E. and {Maoli}, R. and {Marulli}, F. and {Maurogordato}, S. and {McCracken}, H. and {McLure}, R. and {Melchiorri}, A. and {Merson}, A. and {Moresco}, M. and {Nonino}, M. and {Norberg}, P. and {Peacock}, J. and {Pello}, R. and {Penny}, M. and {Pettorino}, V. and {Di Porto}, C. and {Pozzetti}, L. and {Quercellini}, C. and {Radovich}, M. and {Rassat}, A. and {Roche}, N. and {Ronayette}, S. and {Rossetti}, E.},
        title = "{Euclid Definition Study Report}",
      journal = {arXiv e-prints},
         year = 2011,
        month = oct,
          eid = {arXiv:1110.3193},
        pages = {arXiv:1110.3193},
          doi = {10.48550/arXiv.1110.3193},
archivePrefix = {arXiv},
       eprint = {1110.3193},
 primaryClass = {astro-ph.CO},
       adsurl = {https://ui.adsabs.harvard.edu/abs/2011arXiv1110.3193L}
}

@ARTICLE{Troxel_2018,
       author = {{Troxel}, M.~A. and {MacCrann}, N. and {Zuntz}, J. and {Eifler}, T.~F. and {Krause}, E. and {Dodelson}, S. and {Gruen}, D. and {Blazek}, J. and {Friedrich}, O. and {Samuroff}, S. and {Prat}, J. and {Secco}, L.~F. and {Davis}, C. and {Fert{\'e}}, A. and {DeRose}, J. and {Alarcon}, A. and {Amara}, A. and {Baxter}, E. and {Becker}, M.~R. and {Bernstein}, G.~M. and {Bridle}, S.~L. and {Cawthon}, R. and {Chang}, C. and {Choi}, A. and {De Vicente}, J. and {Drlica-Wagner}, A. and {Elvin-Poole}, J. and {Frieman}, J. and {Gatti}, M. and {Hartley}, W.~G. and {Honscheid}, K. and {Hoyle}, B. and {Huff}, E.~M. and {Huterer}, D. and {Jain}, B. and {Jarvis}, M. and {Kacprzak}, T. and {Kirk}, D. and {Kokron}, N. and {Krawiec}, C. and {Lahav}, O. and {Liddle}, A.~R. and {Peacock}, J. and {Rau}, M.~M. and {Refregier}, A. and {Rollins}, R.~P. and {Rozo}, E. and {Rykoff}, E.~S. and {S{\'a}nchez}, C. and {Sevilla-Noarbe}, I. and {Sheldon}, E. and {Stebbins}, A. and {Varga}, T.~N. and {Vielzeuf}, P. and {Wang}, M. and {Wechsler}, R.~H. and {Yanny}, B. and {Abbott}, T.~M.~C. and {Abdalla}, F.~B. and {Allam}, S. and {Annis}, J. and {Bechtol}, K. and {Benoit-L{\'e}vy}, A. and {Bertin}, E. and {Brooks}, D. and {Buckley-Geer}, E. and {Burke}, D.~L. and {Carnero Rosell}, A. and {Carrasco Kind}, M. and {Carretero}, J. and {Castander}, F.~J. and {Crocce}, M. and {Cunha}, C.~E. and {D'Andrea}, C.~B. and {da Costa}, L.~N. and {DePoy}, D.~L. and {Desai}, S. and {Diehl}, H.~T. and {Dietrich}, J.~P. and {Doel}, P. and {Fernandez}, E. and {Flaugher}, B. and {Fosalba}, P. and {Garc{\'\i}a-Bellido}, J. and {Gaztanaga}, E. and {Gerdes}, D.~W. and {Giannantonio}, T. and {Goldstein}, D.~A. and {Gruendl}, R.~A. and {Gschwend}, J. and {Gutierrez}, G. and {James}, D.~J. and {Jeltema}, T. and {Johnson}, M.~W.~G. and {Johnson}, M.~D. and {Kent}, S. and {Kuehn}, K. and {Kuhlmann}, S. and {Kuropatkin}, N. and {Li}, T.~S. and {Lima}, M. and {Lin}, H. and {Maia}, M.~A.~G. and {March}, M. and {Marshall}, J.~L. and {Martini}, P. and {Melchior}, P. and {Menanteau}, F. and {Miquel}, R. and {Mohr}, J.~J. and {Neilsen}, E. and {Nichol}, R.~C. and {Nord}, B. and {Petravick}, D. and {Plazas}, A.~A. and {Romer}, A.~K. and {Roodman}, A. and {Sako}, M. and {Sanchez}, E. and {Scarpine}, V. and {Schindler}, R. and {Schubnell}, M. and {Smith}, M. and {Smith}, R.~C. and {Soares-Santos}, M. and {Sobreira}, F. and {Suchyta}, E. and {Swanson}, M.~E.~C. and {Tarle}, G. and {Thomas}, D. and {Tucker}, D.~L. and {Vikram}, V. and {Walker}, A.~R. and {Weller}, J. and {Zhang}, Y. and {DES Collaboration}},
        title = "{Dark Energy Survey Year 1 results: Cosmological constraints from cosmic shear}",
      journal = {\prd},
         year = 2018,
        month = aug,
       volume = {98},
       number = {4},
          eid = {043528},
        pages = {043528},
          doi = {10.1103/PhysRevD.98.043528},
archivePrefix = {arXiv},
       eprint = {1708.01538},
 primaryClass = {astro-ph.CO},
       adsurl = {https://ui.adsabs.harvard.edu/abs/2018PhRvD..98d3528T}
}

@ARTICLE{Aihara_2017,
       author = {{Aihara}, Hiroaki and {Arimoto}, Nobuo and {Armstrong}, Robert and {Arnouts}, St{\'e}phane and {Bahcall}, Neta A. and {Bickerton}, Steven and {Bosch}, James and {Bundy}, Kevin and {Capak}, Peter L. and {Chan}, James H.~H. and {Chiba}, Masashi and {Coupon}, Jean and {Egami}, Eiichi and {Enoki}, Motohiro and {Finet}, Francois and {Fujimori}, Hiroki and {Fujimoto}, Seiji and {Furusawa}, Hisanori and {Furusawa}, Junko and {Goto}, Tomotsugu and {Goulding}, Andy and {Greco}, Johnny P. and {Greene}, Jenny E. and {Gunn}, James E. and {Hamana}, Takashi and {Harikane}, Yuichi and {Hashimoto}, Yasuhiro and {Hattori}, Takashi and {Hayashi}, Masao and {Hayashi}, Yusuke and {He{\l}miniak}, Krzysztof G. and {Higuchi}, Ryo and {Hikage}, Chiaki and {Ho}, Paul T.~P. and {Hsieh}, Bau-Ching and {Huang}, Kuiyun and {Huang}, Song and {Ikeda}, Hiroyuki and {Imanishi}, Masatoshi and {Inoue}, Akio K. and {Iwasawa}, Kazushi and {Iwata}, Ikuru and {Jaelani}, Anton T. and {Jian}, Hung-Yu and {Kamata}, Yukiko and {Karoji}, Hiroshi and {Kashikawa}, Nobunari and {Katayama}, Nobuhiko and {Kawanomoto}, Satoshi and {Kayo}, Issha and {Koda}, Jin and {Koike}, Michitaro and {Kojima}, Takashi and {Komiyama}, Yutaka and {Konno}, Akira and {Koshida}, Shintaro and {Koyama}, Yusei and {Kusakabe}, Haruka and {Leauthaud}, Alexie and {Lee}, Chien-Hsiu and {Lin}, Lihwai and {Lin}, Yen-Ting and {Lupton}, Robert H. and {Mandelbaum}, Rachel and {Matsuoka}, Yoshiki and {Medezinski}, Elinor and {Mineo}, Sogo and {Miyama}, Shoken and {Miyatake}, Hironao and {Miyazaki}, Satoshi and {Momose}, Rieko and {More}, Anupreeta and {More}, Surhud and {Moritani}, Yuki and {Moriya}, Takashi J. and {Morokuma}, Tomoki and {Mukae}, Shiro and {Murata}, Ryoma and {Murayama}, Hitoshi and {Nagao}, Tohru and {Nakata}, Fumiaki and {Niida}, Mana and {Niikura}, Hiroko and {Nishizawa}, Atsushi J. and {Obuchi}, Yoshiyuki and {Oguri}, Masamune and {Oishi}, Yukie and {Okabe}, Nobuhiro and {Okamoto}, Sakurako and {Okura}, Yuki and {Ono}, Yoshiaki and {Onodera}, Masato and {Onoue}, Masafusa and {Osato}, Ken and {Ouchi}, Masami and {Price}, Paul A. and {Pyo}, Tae-Soo and {Sako}, Masao and {Sawicki}, Marcin and {Shibuya}, Takatoshi and {Shimasaku}, Kazuhiro and {Shimono}, Atsushi and {Shirasaki}, Masato and {Silverman}, John D. and {Simet}, Melanie and {Speagle}, Joshua and {Spergel}, David N. and {Strauss}, Michael A. and {Sugahara}, Yuma and {Sugiyama}, Naoshi and {Suto}, Yasushi and {Suyu}, Sherry H. and {Suzuki}, Nao and {Tait}, Philip J. and {Takada}, Masahiro and {Takata}, Tadafumi and {Tamura}, Naoyuki and {Tanaka}, Manobu M. and {Tanaka}, Masaomi and {Tanaka}, Masayuki and {Tanaka}, Yoko and {Terai}, Tsuyoshi and {Terashima}, Yuichi and {Toba}, Yoshiki and {Tominaga}, Nozomu and {Toshikawa}, Jun and {Turner}, Edwin L. and {Uchida}, Tomohisa and {Uchiyama}, Hisakazu and {Umetsu}, Keiichi and {Uraguchi}, Fumihiro and {Urata}, Yuji and {Usuda}, Tomonori and {Utsumi}, Yousuke and {Wang}, Shiang-Yu and {Wang}, Wei-Hao and {Wong}, Kenneth C. and {Yabe}, Kiyoto and {Yamada}, Yoshihiko and {Yamanoi}, Hitomi and {Yasuda}, Naoki and {Yeh}, Sherry and {Yonehara}, Atsunori and {Yuma}, Suraphong},
        title = "{The Hyper Suprime-Cam SSP Survey: Overview and survey design}",
      journal = {\pasj},
         year = 2018,
        month = jan,
       volume = {70},
          eid = {S4},
        pages = {S4},
          doi = {10.1093/pasj/psx066},
archivePrefix = {arXiv},
       eprint = {1704.05858},
 primaryClass = {astro-ph.IM},
       adsurl = {https://ui.adsabs.harvard.edu/abs/2018PASJ...70S...4A}
}

@article{Thomas_2016,
	doi = {10.3847/0004-637x/830/2/155},
  
	url = {https://doi.org/10.3847%2F0004-637x%2F830%2F2%2F155},
  
	year = 2016,
	month = {oct},
  
	publisher = {American Astronomical Society},
  
	volume = {830},
  
	number = {2},
  
	pages = {155},
  
	author = {Daniel B. Thomas and Michael Kopp and Constantinos Skordis},
  
	title = {{CONSTRAINING} {THE} {PROPERTIES} {OF} {DARK} {MATTER} {WITH} {OBSERVATIONS} {OF} {THE} {COSMIC} {MICROWAVE} {BACKGROUND}
},
  
	journal = {ApJ}
}

@ARTICLE{Wright_2020a,
       author = {{Wright}, Angus H. and {Hildebrandt}, Hendrik and {van den Busch}, Jan Luca and {Heymans}, Catherine},
        title = "{Photometric redshift calibration with self-organising maps}",
      journal = {\aap},
         year = 2020,
        month = may,
       volume = {637},
          eid = {A100},
        pages = {A100},
          doi = {10.1051/0004-6361/201936782},
archivePrefix = {arXiv},
       eprint = {1909.09632},
 primaryClass = {astro-ph.CO},
       adsurl = {https://ui.adsabs.harvard.edu/abs/2020A&A...637A.100W}
}

@article{Newman_2008,
	doi = {10.1086/589982},
  
	url = {https://doi.org/10.1086%2F589982},
  
	year = 2008,
	month = {sep},
  
	publisher = {American Astronomical Society},
  
	volume = {684},
  
	number = {1},
  
	pages = {88--101},
  
	author = {Jeffrey A. Newman},
  
	title = {Calibrating Redshift Distributions beyond Spectroscopic Limits with Cross-Correlations},
  
	journal = {ApJ}
}

@article{McQuinn_2013,
	doi = {10.1093/mnras/stt914},
  
	url = {https://doi.org/10.1093%2Fmnras%2Fstt914},
  
	year = 2013,
	month = {jul},
  
	publisher = {Oxford University Press ({OUP})},
  
	volume = {433},
  
	number = {4},
  
	pages = {2857--2883},
  
	author = {M. McQuinn and M. White},
  
	title = {On using angular cross-correlations to determine source redshift distributions},
  
	journal = {MNRAS}
}

@article{Morrison_2017,
	doi = {10.1093/mnras/stx342},
  
	url = {https://doi.org/10.1093%2Fmnras%2Fstx342},
  
	year = 2017,
	month = {feb},
  
	publisher = {Oxford University Press ({OUP})},
  
	volume = {467},
  
	number = {3},
  
	pages = {3576--3589},
  
	author = {C. B. Morrison and H. Hildebrandt and S. J. Schmidt and I. K. Baldry and M. Bilicki and A. Choi and T. Erben and P. Schneider},
  
	title = {the-wizz: clustering redshift estimation for everyone},
  
	journal = {MNRAS}
}

@article{Benitez_2000,
	doi = {10.1086/308947},
  
	url = {https://doi.org/10.1086%2F308947},
  
	year = 2000,
	month = {jun},
  
	publisher = {American Astronomical Society},
  
	volume = {536},
  
	number = {2},
  
	pages = {571--583},
  
	author = {Narciso Benitez},
  
	title = {Bayesian Photometric Redshift Estimation},
  
	journal = {ApJ}
}

@ARTICLE{desprez_2020,
       author = {{Euclid Collaboration: Desprez}, G. and {Paltani}, S. and {Coupon}, J. and {Almosallam}, I. and {Alvarez-Ayllon}, A. and {Amaro}, V. and {Brescia}, M. and {Brodwin}, M. and {Cavuoti}, S. and {De Vicente-Albendea}, J. and {Fotopoulou}, S. and {Hatfield}, P.~W. and {Hartley}, W.~G. and {Ilbert}, O. and {Jarvis}, M.~J. and {Longo}, G. and {Rau}, M.~M. and {Saha}, R. and {Speagle}, J.~S. and {Tramacere}, A. and {Castellano}, M. and {Dubath}, F. and {Galametz}, A. and {Kuemmel}, M. and {Laigle}, C. and {Merlin}, E. and {Mohr}, J.~J. and {Pilo}, S. and {Salvato}, M. and {Andreon}, S. and {Auricchio}, N. and {Baccigalupi}, C. and {Balaguera-Antol{\'\i}nez}, A. and {Baldi}, M. and {Bardelli}, S. and {Bender}, R. and {Biviano}, A. and {Bodendorf}, C. and {Bonino}, D. and {Bozzo}, E. and {Branchini}, E. and {Brinchmann}, J. and {Burigana}, C. and {Cabanac}, R. and {Camera}, S. and {Capobianco}, V. and {Cappi}, A. and {Carbone}, C. and {Carretero}, J. and {Carvalho}, C.~S. and {Casas}, R. and {Casas}, S. and {Castander}, F.~J. and {Castignani}, G. and {Cimatti}, A. and {Cledassou}, R. and {Colodro-Conde}, C. and {Congedo}, G. and {Conselice}, C.~J. and {Conversi}, L. and {Copin}, Y. and {Corcione}, L. and {Courtois}, H.~M. and {Cuby}, J. -G. and {Da Silva}, A. and {de la Torre}, S. and {Degaudenzi}, H. and {Di Ferdinando}, D. and {Douspis}, M. and {Duncan}, C.~A.~J. and {Dupac}, X. and {Ealet}, A. and {Fabbian}, G. and {Fabricius}, M. and {Farrens}, S. and {Ferreira}, P.~G. and {Finelli}, F. and {Fosalba}, P. and {Fourmanoit}, N. and {Frailis}, M. and {Franceschi}, E. and {Fumana}, M. and {Galeotta}, S. and {Garilli}, B. and {Gillard}, W. and {Gillis}, B. and {Giocoli}, C. and {Gozaliasl}, G. and {Graci{\'a}-Carpio}, J. and {Grupp}, F. and {Guzzo}, L. and {Hailey}, M. and {Haugan}, S.~V.~H. and {Holmes}, W. and {Hormuth}, F. and {Humphrey}, A. and {Jahnke}, K. and {Keihanen}, E. and {Kermiche}, S. and {Kilbinger}, M. and {Kirkpatrick}, C.~C. and {Kitching}, T.~D. and {Kohley}, R. and {Kubik}, B. and {Kunz}, M. and {Kurki-Suonio}, H. and {Ligori}, S. and {Lilje}, P.~B. and {Lloro}, I. and {Maino}, D. and {Maiorano}, E. and {Marggraf}, O. and {Markovic}, K. and {Martinet}, N. and {Marulli}, F. and {Massey}, R. and {Maturi}, M. and {Mauri}, N. and {Maurogordato}, S. and {Medinaceli}, E. and {Mei}, S. and {Meneghetti}, M. and {Metcalf}, R. Benton and {Meylan}, G. and {Moresco}, M. and {Moscardini}, L. and {Munari}, E. and {Niemi}, S. and {Padilla}, C. and {Pasian}, F. and {Patrizii}, L. and {Pettorino}, V. and {Pires}, S. and {Polenta}, G. and {Poncet}, M. and {Popa}, L. and {Potter}, D. and {Pozzetti}, L. and {Raison}, F. and {Renzi}, A. and {Rhodes}, J. and {Riccio}, G. and {Rossetti}, E. and {Saglia}, R. and {Sapone}, D. and {Schneider}, P. and {Scottez}, V. and {Secroun}, A. and {Serrano}, S. and {Sirignano}, C. and {Sirri}, G. and {Stanco}, L. and {Stern}, D. and {Sureau}, F. and {Tallada Cresp{\'\i}}, P. and {Tavagnacco}, D. and {Taylor}, A.~N. and {Tenti}, M. and {Tereno}, I. and {Toledo-Moreo}, R. and {Torradeflot}, F. and {Valenziano}, L. and {Valiviita}, J. and {Vassallo}, T. and {Viel}, M. and {Wang}, Y. and {Welikala}, N. and {Whittaker}, L. and {Zacchei}, A. and {Zamorani}, G. and {Zoubian}, J. and {Zucca}, E.},
        title = "{Euclid preparation. X. The Euclid photometric-redshift challenge}",
      journal = {\aap},
         year = 2020,
        month = dec,
       volume = {644},
          eid = {A31},
        pages = {A31},
          doi = {10.1051/0004-6361/202039403},
archivePrefix = {arXiv},
       eprint = {2009.12112},
 primaryClass = {astro-ph.GA},
       adsurl = {https://ui.adsabs.harvard.edu/abs/2020A\&A...644A..31E}
}

@ARTICLE{lepori_2022,
       author = {{Euclid Collaboration: Lepori}, F. and {Tutusaus}, I. and {Viglione}, C. and {Bonvin}, C. and {Camera}, S. and {Castander}, F.~J. and {Durrer}, R. and {Fosalba}, P. and {Jelic-Cizmek}, G. and {Kunz}, M. and {Adamek}, J. and {Casas}, S. and {Martinelli}, M. and {Sakr}, Z. and {Sapone}, D. and {Amara}, A. and {Auricchio}, N. and {Bodendorf}, C. and {Bonino}, D. and {Branchini}, E. and {Brescia}, M. and {Brinchmann}, J. and {Capobianco}, V. and {Carbone}, C. and {Carretero}, J. and {Castellano}, M. and {Cavuoti}, S. and {Cimatti}, A. and {Cledassou}, R. and {Congedo}, G. and {Conselice}, C.~J. and {Conversi}, L. and {Copin}, Y. and {Corcione}, L. and {Courbin}, F. and {Da Silva}, A. and {Degaudenzi}, H. and {Douspis}, M. and {Dubath}, F. and {Dupac}, X. and {Dusini}, S. and {Ealet}, A. and {Farrens}, S. and {Ferriol}, S. and {Franceschi}, E. and {Fumana}, M. and {Garilli}, B. and {Gillard}, W. and {Gillis}, B. and {Giocoli}, C. and {Grazian}, A. and {Grupp}, F. and {Guzzo}, L. and {Haugan}, S.~V.~H. and {Holmes}, W. and {Hormuth}, F. and {Hudelot}, P. and {Jahnke}, K. and {Kermiche}, S. and {Kiessling}, A. and {Kilbinger}, M. and {Kitching}, T. and {K{\"u}mmel}, M. and {Kurki-Suonio}, H. and {Ligori}, S. and {Lilje}, P.~B. and {Lloro}, I. and {Mansutti}, O. and {Marggraf}, O. and {Markovic}, K. and {Marulli}, F. and {Massey}, R. and {Maurogordato}, S. and {Melchior}, M. and {Meneghetti}, M. and {Merlin}, E. and {Meylan}, G. and {Moresco}, M. and {Moscardini}, L. and {Munari}, E. and {Nakajima}, R. and {Niemi}, S.~M. and {Padilla}, C. and {Paltani}, S. and {Pasian}, F. and {Pedersen}, K. and {Percival}, W.~J. and {Pettorino}, V. and {Pires}, S. and {Poncet}, M. and {Popa}, L. and {Pozzetti}, L. and {Raison}, F. and {Rhodes}, J. and {Roncarelli}, M. and {Rossetti}, E. and {Saglia}, R. and {Schneider}, P. and {Secroun}, A. and {Seidel}, G. and {Serrano}, S. and {Sirignano}, C. and {Sirri}, G. and {Stanco}, L. and {Starck}, J. -L. and {Tallada-Cresp{\'\i}}, P. and {Taylor}, A.~N. and {Tereno}, I. and {Toledo-Moreo}, R. and {Torradeflot}, F. and {Valentijn}, E.~A. and {Valenziano}, L. and {Wang}, Y. and {Weller}, J. and {Zamorani}, G. and {Zoubian}, J. and {Andreon}, S. and {Bardelli}, S. and {Fabbian}, G. and {Graci{\'a}-Carpio}, J. and {Maino}, D. and {Medinaceli}, E. and {Mei}, S. and {Renzi}, A. and {Romelli}, E. and {Sureau}, F. and {Vassallo}, T. and {Zacchei}, A. and {Zucca}, E. and {Baccigalupi}, C. and {Balaguera-Antol{\'\i}nez}, A. and {Bernardeau}, F. and {Biviano}, A. and {Blanchard}, A. and {Bolzonella}, M. and {Borgani}, S. and {Bozzo}, E. and {Burigana}, C. and {Cabanac}, R. and {Cappi}, A. and {Carvalho}, C.~S. and {Castignani}, G. and {Colodro-Conde}, C. and {Coupon}, J. and {Courtois}, H.~M. and {Cuby}, J. -G. and {Davini}, S. and {de la Torre}, S. and {Di Ferdinando}, D. and {Farina}, M. and {Ferreira}, P.~G. and {Finelli}, F. and {Galeotta}, S. and {Ganga}, K. and {Garcia-Bellido}, J. and {Gaztanaga}, E. and {Gozaliasl}, G. and {Hook}, I.~M. and {Ili{\'c}}, S. and {Joachimi}, B. and {Kansal}, V. and {Keihanen}, E. and {Kirkpatrick}, C.~C. and {Lindholm}, V. and {Mainetti}, G. and {Maoli}, R. and {Martinet}, N. and {Maturi}, M. and {Metcalf}, R.~B. and {Monaco}, P. and {Morgante}, G. and {Nightingale}, J. and {Nucita}, A. and {Patrizii}, L. and {Popa}, V. and {Potter}, D. and {Riccio}, G. and {S{\'a}nchez}, A.~G. and {Schirmer}, M. and {Schultheis}, M. and {Scottez}, V. and {Sefusatti}, E. and {Tramacere}, A. and {Valiviita}, J. and {Viel}, M. and {Hildebrandt}, H.},
        title = "{Euclid preparation. XIX. Impact of magnification on photometric galaxy clustering}",
      journal = {\aap},
         year = 2022,
        month = jun,
       volume = {662},
          eid = {A93},
        pages = {A93},
          doi = {10.1051/0004-6361/202142419},
archivePrefix = {arXiv},
       eprint = {2110.05435},
 primaryClass = {astro-ph.CO},
       adsurl = {https://ui.adsabs.harvard.edu/abs/2022A&A...662A..93E}
}

@article{Brammer_2008,
	doi = {10.1086/591786},
  
	url = {https://doi.org/10.1086%2F591786},
  
	year = 2008,
	month = {oct},
  
	publisher = {American Astronomical Society},
  
	volume = {686},
  
	number = {2},
  
	pages = {1503--1513},
  
	author = {Gabriel B. Brammer and Pieter G. van Dokkum and Paolo Coppi},
  
	title = {{EAZY}: A Fast, Public Photometric Redshift Code},
  
	journal = {ApJ}
}

@misc{Lephare2011,
  author       = {{Arnouts}, S. and {Ilbert}, O.},
  howpublished = {ASCL, record ascl:1108.009},
  year         = {2011},
  month        = aug,
  note         = {ascl:1108.009},
  url          = {https://ui.adsabs.harvard.edu/abs/2011ascl.soft08009A}
}

@article{Carrasco_Kind_2013,
	doi = {10.1093/mnras/stt574},
  
	url = {https://doi.org/10.1093%2Fmnras%2Fstt574},
  
	year = 2013,
	month = {may},
  
	publisher = {Oxford University Press ({OUP})},
  
	volume = {432},
  
	number = {2},
  
	pages = {1483--1501},
  
	author = {Matias Carrasco Kind and Robert J. Brunner},
  
	title = {{TPZ}: photometric redshift {PDFs} and ancillary information by using prediction trees and random forests},
  
	journal = {MNRAS}
}

@article{Sadeh_2016,
	doi = {10.1088/1538-3873/128/968/104502},
  
	url = {https://doi.org/10.1088%2F1538-3873%2F128%2F968%2F104502},
  
	year = 2016,
	month = {aug},
  
	publisher = {{IOP} Publishing},
  
	volume = {128},
  
	number = {968},
  
	pages = {104502},
  
	author = {I. Sadeh and F. B. Abdalla and O. Lahav},
  
	title = {{ANNz}2: Photometric Redshift and Probability Distribution Function Estimation using Machine Learning},
  
	journal = {PASP}
}

@article{Stanford_2021,
	doi = {10.3847/1538-4365/ac0833},
  
	url = {https://doi.org/10.3847%2F1538-4365%2Fac0833},
  
	year = 2021,
	month = {aug},
  
	publisher = {American Astronomical Society},
  
	volume = {256},
  
	number = {1},
  
	pages = {9},
  
	author = {S. A. Stanford and D. Masters and B. Darvish and D. Stern and J. G. Cohen and P. Capak and N. Hernitschek and I. Davidzon and J. Rhodes and D. B. Sanders and B. Mobasher and F. J. Castander and S. Paltani and N. Aghanim and A. Amara and N. Auricchio and A. Balestra and R. Bender},
  
	title = {Euclid Preparation. {XIV}. The Complete Calibration of the Color{\textendash}Redshift Relation (C3R2) Survey: Data Release 3},
  
	journal = {ApJS}
}

@ARTICLE{Wright_2020b,
       author = {{Wright}, Angus H. and {Hildebrandt}, Hendrik and {van den Busch}, Jan Luca and {Heymans}, Catherine and {Joachimi}, Benjamin and {Kannawadi}, Arun and {Kuijken}, Konrad},
        title = "{KiDS+VIKING-450: Improved cosmological parameter constraints from redshift calibration with self-organising maps}",
      journal = {\aap},
         year = 2020,
        month = aug,
       volume = {640},
          eid = {L14},
        pages = {L14},
          doi = {10.1051/0004-6361/202038389},
archivePrefix = {arXiv},
       eprint = {2005.04207},
 primaryClass = {astro-ph.CO},
       adsurl = {https://ui.adsabs.harvard.edu/abs/2020A&A...640L..14W}
}

@article{Bordoloi_2012,
	doi = {10.1111/j.1365-2966.2012.20427.x},
  
	url = {https://doi.org/10.1111%2Fj.1365-2966.2012.20427.x},
  
	year = 2012,
	month = {feb},
  
	publisher = {Oxford University Press ({OUP})},
  
	volume = {421},
  
	number = {2},
  
	pages = {1671--1677},
  
	author = {R. Bordoloi and S. J. Lilly and A. Amara and P. A. Oesch and S. Bardelli and E. Zucca and D. Vergani and T. Nagao and T. Murayama and Y. Shioya and Y. Taniguchi},
  
	title = {Photo-$\less$i$\greater$z$\less$/i$\greater$performance for precision cosmology - {II}. Empirical verification$\less$sup$\greater$1{\ding{72}
}$\less$/sup$\greater$},
  
	journal = {MNRAS}
}

@article{Masters_2015,
	doi = {10.1088/0004-637x/813/1/53},
  
	url = {https://doi.org/10.1088\%2F0004-637x\%2F813\%2F1\%2F53},
  
	year = 2015,
	month = {oct},
  
	publisher = {American Astronomical Society},
  
	volume = {813},
  
	number = {1},
  
	pages = {53},
  
	author = {Daniel Masters and Peter Capak and Daniel Stern and Olivier Ilbert and Mara Salvato and Samuel Schmidt and Henk Hoekstra and Hendrik Hildebrandt and Jean Coupon},
  
	title = {{MAPPING} {THE} {GALAXY} {COLOR}{\textendash}{REDSHIFT} {RELATION}: {OPTIMAL} {PHOTOMETRIC} {REDSHIFT} {CALIBRATION} {STRATEGIES} {FOR} {COSMOLOGY} {SURVEYS}
},
  
	journal = {ApJ}
}

@article{Geach_2011,
	doi = {10.1111/j.1365-2966.2011.19913.x},
  
	url = {https://doi.org/10.1111\%2Fj.1365-2966.2011.19913.x},
  
	year = 2011,
	month = {nov},
  
	publisher = {Oxford University Press ({OUP})},
  
	volume = {419},
  
	number = {3},
  
	pages = {2633--2645},
  
	author = {James E. Geach},
  
	title = {Unsupervised self-organized mapping: a versatile empirical tool for object selection, classification and redshift estimation in large surveys},
  
	journal = {MNRAS}
}

@ARTICLE{salvato19,
       author = {{Salvato}, Mara and {Ilbert}, Olivier and {Hoyle}, Ben},
        title = "{The many flavours of photometric redshifts}",
      journal = {Nature Astronomy},
         year = 2019,
        month = jun,
       volume = {3},
        pages = {212--222},
          doi = {10.1038/s41550-018-0478-0},
archivePrefix = {arXiv},
       eprint = {1805.12574},
 primaryClass = {astro-ph.GA},
       adsurl = {https://ui.adsabs.harvard.edu/abs/2019NatAs...3..212S}
}

@article{Hoyle_2018,
	doi = {10.1093/mnras/sty957},
  
	url = {https://doi.org/10.1093%2Fmnras%2Fsty957},
  
	year = 2018,
	month = {apr},
  
	publisher = {Oxford University Press ({OUP})},
  
	volume = {478},
  
	number = {1},
  
	pages = {592--610},
  
	author = {B Hoyle and D Gruen and G M Bernstein and M M Rau and J De~Vicente and W G Hartley and E Gaztanaga and J DeRose and M A Troxel},
  
	title = {Dark Energy Survey Year 1 Results: redshift distributions of the weak-lensing source galaxies},
  
	journal = {MNRAS}
}

@ARTICLE{van_den_Busch_2022,
       author = {{van den Busch}, J.~L. and {Wright}, A.~H. and {Hildebrandt}, H. and {Bilicki}, M. and {Asgari}, M. and {Joudaki}, S. and {Blake}, C. and {Heymans}, C. and {Kannawadi}, A. and {Shan}, H.~Y. and {Tr{\"o}ster}, T.},
        title = "{KiDS-1000: Cosmic shear with enhanced redshift calibration}",
      journal = {\aap},
         year = 2022,
        month = aug,
       volume = {664},
          eid = {A170},
        pages = {A170},
          doi = {10.1051/0004-6361/202142083},
archivePrefix = {arXiv},
       eprint = {2204.02396},
 primaryClass = {astro-ph.CO},
       adsurl = {https://ui.adsabs.harvard.edu/abs/2022A&A...664A.170V}
}

@ARTICLE{Frieman_2008,
       author = {{Frieman}, J.~A. and {Turner}, M.~S. and {Huterer}, D.},
        title = "{Dark energy and the accelerating universe.}",
      journal = {\araa},
         year = 2008,
        month = sep,
       volume = {46},
        pages = {385--432},
          doi = {10.1146/annurev.astro.46.060407.145243},
archivePrefix = {arXiv},
       eprint = {0803.0982},
 primaryClass = {astro-ph},
       adsurl = {https://ui.adsabs.harvard.edu/abs/2008ARA&A..46..385F}
}

@article{Kilbinger_2015,
	doi = {10.1088/0034-4885/78/8/086901},
  
	url = {https://doi.org/10.1088%2F0034-4885%2F78%2F8%2F086901},
  
	year = 2015,
	month = {jul},
  
	publisher = {{IOP} Publishing},
  
	volume = {78},
  
	number = {8},
  
	pages = {086901},
  
	author = {Martin Kilbinger},
  
	title = {Cosmology with cosmic shear observations: a review},
  
	journal = {Reports on Progress in Physics}
}

@article{Newman_2015,
   title={Spectroscopic needs for imaging dark energy experiments},
   volume={63},
   ISSN={0927-6505},
   url={http://dx.doi.org/10.1016/j.astropartphys.2014.06.007},
   DOI={10.1016/j.astropartphys.2014.06.007},
   journal={Astroparticle Physics},
   publisher={Elsevier BV},
   author={Newman, Jeffrey A. and Abate, Alexandra and Abdalla, Filipe B. and Allam, Sahar and Allen, Steven W. and Ansari, Réza and Bailey, Stephen and Barkhouse, Wayne A. and Beers, Timothy C. and Blanton, Michael R. and Brodwin},
   year={2015},
   month=mar, pages={81--100} }

@phdthesis{vdb_2021,
  author       = {J.L. van den Busch},
  title        = {A Study in Redshift -
Simulating and calibrating cosmic shear surveys},
  school       = {University of Bochum},
  year         = {2021},
}

@article{Masters_2017,
	doi = {10.3847/1538-4357/aa6f08},
  
	url = {https://doi.org/10.3847%2F1538-4357%2Faa6f08},
  
	year = 2017,
	month = {may},
  
	publisher = {American Astronomical Society},
  
	volume = {841},
  
	number = {2},
  
	pages = {111},
  
	author = {Daniel C. Masters and Daniel K. Stern and Judith G. Cohen and Peter L. Capak and Jason D. Rhodes and Francisco J. Castander and St{\'{e}
}phane Paltani},
  
	title = {The Complete Calibration of the Color{\textendash}Redshift Relation (C3R2) Survey: Survey Overview and Data Release 1},
  
	journal = {ApJ}
}

@article{Lima_2008,
	doi = {10.1111/j.1365-2966.2008.13510.x},
  
	url = {https://doi.org/10.1111%2Fj.1365-2966.2008.13510.x},
  
	year = 2008,
	month = {oct},
  
	publisher = {Oxford University Press ({OUP})},
  
	volume = {390},
  
	number = {1},
  
	pages = {118--130},
  
	author = {Marcos Lima and Carlos E. Cunha and Hiroaki Oyaizu and Joshua Frieman and Huan Lin and Erin S. Sheldon},
  
	title = {Estimating the redshift distribution of photometric galaxy samples},
  
	journal = {MNRAS}
}

@article{Masters_2019,
	doi = {10.3847/1538-4357/ab184d},
  
	url = {https://doi.org/10.3847%2F1538-4357%2Fab184d},
  
	year = 2019,
	month = {may},
  
	publisher = {American Astronomical Society},
  
	volume = {877},
  
	number = {2},
  
	pages = {81},
  
	author = {Daniel C. Masters and Daniel K. Stern and Judith G. Cohen and Peter L. Capak and S. Adam Stanford and Nina Hernitschek and Audrey Galametz and Iary Davidzon and Jason D. Rhodes and Dave Sanders and Bahram Mobasher and Francisco Castander and Kerianne Pruett and Sotiria Fotopoulou},
  
	title = {The Complete Calibration of the Color{\textendash}Redshift Relation (C3R2) Survey: Analysis and Data Release 2},
  
	journal = {ApJ}
}

@ARTICLE{Amendola_2013,
       author = {{Amendola}, Luca and {Appleby}, Stephen and {Bacon}, David and {Baker}, Tessa and {Baldi}, Marco and {Bartolo}, Nicola and {Blanchard}, Alain and {Bonvin}, Camille and {Borgani}, Stefano and {Branchini}, Enzo and {Burrage}, Clare and {Camera}, Stefano and {Carbone}, Carmelita and {Casarini}, Luciano and {Cropper}, Mark and {de Rham}, Claudia and {Di Porto}, Cinzia and {Ealet}, Anne and {Ferreira}, Pedro G. and {Finelli}, Fabio and {Garc{\'\i}a-Bellido}, Juan and {Giannantonio}, Tommaso and {Guzzo}, Luigi and {Heavens}, Alan and {Heisenberg}, Lavinia and {Heymans}, Catherine and {Hoekstra}, Henk and {Hollenstein}, Lukas and {Holmes}, Rory and {Horst}, Ole and {Jahnke}, Knud and {Kitching}, Thomas D. and {Koivisto}, Tomi and {Kunz}, Martin and {La Vacca}, Giuseppe and {March}, Marisa and {Majerotto}, Elisabetta and {Markovic}, Katarina and {Marsh}, David and {Marulli}, Federico and {Massey}, Richard and {Mellier}, Yannick and {Mota}, David F. and {Nunes}, Nelson J. and {Percival}, Will and {Pettorino}, Valeria and {Porciani}, Cristiano and {Quercellini}, Claudia and {Read}, Justin and {Rinaldi}, Massimiliano and {Sapone}, Domenico and {Scaramella}, Roberto and {Skordis}, Constantinos and {Simpson}, Fergus and {Taylor}, Andy and {Thomas}, Shaun and {Trotta}, Roberto and {Verde}, Licia and {Vernizzi}, Filippo and {Vollmer}, Adrian and {Wang}, Yun and {Weller}, Jochen and {Zlosnik}, Tom and {The Euclid Theory Working Group}},
        title = "{Cosmology and Fundamental Physics with the Euclid Satellite}",
      journal = {Living Reviews in Relativity},
         year = 2013,
        month = dec,
       volume = {16},
       number = {1},
          eid = {6},
        pages = {6},
          doi = {10.12942/lrr-2013-6},
archivePrefix = {arXiv},
       eprint = {1206.1225},
 primaryClass = {astro-ph.CO},
       adsurl = {https://ui.adsabs.harvard.edu/abs/2013LRR....16....6A}
}

@ARTICLE{Hildebrandt_2021,
       author = {{Hildebrandt}, H. and {van den Busch}, J.~L. and {Wright}, A.~H. and {Blake}, C. and {Joachimi}, B. and {Kuijken}, K. and {Tr{\"o}ster}, T. and {Asgari}, M. and {Bilicki}, M. and {de Jong}, J.~T.~A. and {Dvornik}, A. and {Erben}, T. and {Getman}, F. and {Giblin}, B. and {Heymans}, C. and {Kannawadi}, A. and {Lin}, C. -A. and {Shan}, H. -Y.},
        title = "{KiDS-1000 catalogue: Redshift distributions and their calibration}",
      journal = {\aap},
         year = 2021,
        month = mar,
       volume = {647},
          eid = {A124},
        pages = {A124},
          doi = {10.1051/0004-6361/202039018},
archivePrefix = {arXiv},
       eprint = {2007.15635},
 primaryClass = {astro-ph.CO},
       adsurl = {https://ui.adsabs.harvard.edu/abs/2021A&A...647A.124H}
}

@article{Wehrens10_2018,
author = {Wehrens, Ron and Kruisselbrink, Johannes},
year = {2018},
month = {10},
pages = {},
title = {Flexible Self-Organizing Maps in kohonen 3.0},
volume = {87},
journal = {JSS},
doi = {10.18637/jss.v087.i07}
}

@ARTICLE{H_johnston2021,
       author = {{Johnston}, Harry and {Wright}, Angus H. and {Joachimi}, Benjamin and {Bilicki}, Maciej and {Elisa Chisari}, Nora and {Dvornik}, Andrej and {Erben}, Thomas and {Giblin}, Benjamin and {Heymans}, Catherine and {Hildebrandt}, Hendrik and {Hoekstra}, Henk and {Joudaki}, Shahab and {Vakili}, Mohammadjavad},
        title = "{Organised randoms: Learning and correcting for systematic galaxy clustering patterns in KiDS using self-organising maps}",
      journal = {\aap},
         year = 2021,
        month = apr,
       volume = {648},
          eid = {A98},
        pages = {A98},
          doi = {10.1051/0004-6361/202040136},
archivePrefix = {arXiv},
       eprint = {2012.08467},
 primaryClass = {astro-ph.CO},
       adsurl = {https://ui.adsabs.harvard.edu/abs/2021A&A...648A..98J}
}

@article{Davidzon_2019,
	doi = {10.1093/mnras/stz2486},
  
	url = {https://doi.org/10.1093%2Fmnras%2Fstz2486},
  
	year = 2019,
	month = {sep},
  
	publisher = {Oxford University Press ({OUP})},
  
	volume = {489},
  
	number = {4},
  
	pages = {4817--4835},
  
	author = {I Davidzon and C Laigle and P L Capak and O Ilbert and D C Masters and S Hemmati and N Apostolakos and J Coupon and S de~la~Torre and J Devriendt and Y Dubois and D Kashino and S Paltani and C Pichon},
  
	title = {horizon-{AGN} virtual observatory {\textendash} 2. Template-free estimates of galaxy properties from colours},
  
	journal = {MNRAS}
}

@inproceedings{luo2022bayesian,
  title={Bayesian Neural Networks with Covariate Shift Correction For Classification in $\gamma$-ray Astrophysics},
  author={Luo, Shengda and Luo, Jing and Chen, Yue and Kim, Sangin and Hui, David and Zhang, Jianguo and Leung, Alex and Bugiolacchi, Roberto},
  booktitle={Pattern Recognition and Computer Vision: 5th Chinese Conference, PRCV 2022, Shenzhen, China, November 4--7, 2022, Proceedings, Part II},
  pages={706--719},
  year={2022},
  organization={Springer}
}

@ARTICLE{autenrieth2021stratified,
       author = {{Autenrieth}, Maximilian and {van Dyk}, David A. and {Trotta}, Roberto and {Stenning}, David C.},
        title = "{Stratified Learning: A General-Purpose Statistical Method for Improved Learning under Covariate Shift}",
      journal = {arXiv e-prints},
         year = 2021,
        month = jun,
          eid = {arXiv:2106.11211},
        pages = {arXiv:2106.11211},
          doi = {10.48550/arXiv.2106.11211},
archivePrefix = {arXiv},
       eprint = {2106.11211},
 primaryClass = {stat.ML},
       adsurl = {https://ui.adsabs.harvard.edu/abs/2021arXiv210611211A}
}

@article{Gruen_2017,
	doi = {10.1093/mnras/stx471},
  
	url = {https://doi.org/10.1093%2Fmnras%2Fstx471},
  
	year = 2017,
	month = {feb},
  
	publisher = {Oxford University Press ({OUP})},
  
	volume = {468},
  
	number = {1},
  
	pages = {769--782},
  
	author = {D. Gruen and F. Brimioulle},
  
	title = {Selection biases in empirical p(z) methods for weak lensing},
  
	journal = {MNRAS}
}

@article{Hartley_2020,
	doi = {10.1093/mnras/staa1812},
  
	url = {https://doi.org/10.1093%2Fmnras%2Fstaa1812},
  
	year = 2020,
	month = {jun},
  
	publisher = {Oxford University Press ({OUP})},
  
	volume = {496},
  
	number = {4},
  
	pages = {4769--4786},
  
	author = {W G Hartley and C Chang and S Samani and A Carnero~Rosell and T M Davis and B Hoyle and D Gruen and J Asorey},
  
	title = {The impact of spectroscopic incompleteness in direct calibration of redshift distributions for weak lensing surveys},
  
	journal = {MNRAS}
}

@article{Beck_2017,
    author = {Beck, R. and Lin, C.-A. and Ishida, E. E. O. and Gieseke, F. and de Souza, R. S. and Costa-Duarte, M. V. and Hattab, M. W. and Krone-Martins, A. and for the COIN Collaboration},
    title = "{On the realistic validation of photometric redshifts}",
    journal = {MNRAS},
    volume = {468},
    number = {4},
    pages = {4323--4339},
    year = {2017},
    month = {03},
    issn = {0035-8711},
    doi = {10.1093/mnras/stx687},
    url = {https://doi.org/10.1093/mnras/stx687},
    eprint = {https://academic.oup.com/mnras/article-pdf/468/4/4323/14077646/stx687.pdf},
}

@article{Moessner_1998,
    author = {Moessner, R. and Jain, Bhuvnesh},
    title = "{Angular cross-correlation of galaxies: a probe of gravitational lensing by large-scale structure}",
    journal = {MNRAS},
    volume = {294},
    number = {1},
    pages = {L18--L24},
    year = {1998},
    month = {02},
    issn = {0035-8711},
    doi = {10.1046/j.1365-8711.1998.01378.x},
    url = {https://doi.org/10.1046/j.1365-8711.1998.01378.x},
    eprint = {https://academic.oup.com/mnras/article-pdf/294/1/L18/2903908/294-1-L18.pdf},
}

@article{Berlind_2002,
	doi = {10.1086/341469},
  
	url = {https://doi.org/10.1086%2F341469},
  
	year = 2002,
	month = {aug},
  
	publisher = {American Astronomical Society},
  
	volume = {575},
  
	number = {2},
  
	pages = {587--616},
  
	author = {Andreas A. Berlind and David H. Weinberg},
  
	title = {The Halo Occupation Distribution: Toward an Empirical Determination of the Relation between Galaxies and Mass},
  
	journal = {ApJ}
}

@ARTICLE{Tallada,
       author = {{Tallada}, P. and {Carretero}, J. and {Casals}, J. and {Acosta-Silva}, C. and {Serrano}, S. and {Caubet}, M. and {Castander}, F.~J. and {C{\'e}sar}, E. and {Crocce}, M. and {Delfino}, M. and {Eriksen}, M. and {Fosalba}, P. and {Gazta{\~n}aga}, E. and {Merino}, G. and {Neissner}, C. and {Tonello}, N.},
        title = "{CosmoHub: Interactive exploration and distribution of astronomical data on Hadoop}",
      journal = {AC},
         year = 2020,
        month = jul,
       volume = {32},
          eid = {100391},
        pages = {100391},
          doi = {10.1016/j.ascom.2020.100391},
archivePrefix = {arXiv},
       eprint = {2003.03217},
 primaryClass = {astro-ph.IM},
       adsurl = {https://ui.adsabs.harvard.edu/abs/2020A&C....3200391T}
}

@ARTICLE{Lecun_1998,
  author={Lecun, Y. and Bottou, L. and Bengio, Y. and Haffner, P.},
  journal={Proceedings of the IEEE}, 
  title={Gradient-based learning applied to document recognition}, 
  year={1998},
  volume={86},
  number={11},
  pages={2278--2324},
  doi={10.1109/5.726791}}

@ARTICLE{euclidcollaboration2024euclidvflagshipgalaxy,
       author = {{Euclid Collaboration: Castander}, F.~J. and {Fosalba}, P. and {Stadel}, J. and {Potter}, D. and {Carretero}, J. and {Tallada-Cresp{\'\i}}, P. and {Pozzetti}, L. and {Bolzonella}, M. and {Mamon}, G.~A. and {Blot}, L. and {Hoffmann}, K. and {Huertas-Company}, M. and {Monaco}, P. and {Gonzalez}, E.~J. and {De Lucia}, G. and {Scarlata}, C. and {Breton}, M. -A. and {Linke}, L. and {Viglione}, C. and {Li}, S. -S. and {Zhai}, Z. and {Baghkhani}, Z. and {Pardede}, K. and {Neissner}, C. and {Teyssier}, R. and {Crocce}, M. and {Tutusaus}, I. and {Miller}, L. and {Congedo}, G. and {Biviano}, A. and {Hirschmann}, M. and {Pezzotta}, A. and {Aussel}, H. and {Hoekstra}, H. and {Kitching}, T. and {Percival}, W.~J. and {Guzzo}, L. and {Mellier}, Y. and {Oesch}, P.~A. and {Bowler}, R.~A.~A. and {Bruton}, S. and {Allevato}, V. and {Gonzalez-Perez}, V. and {Manera}, M. and {Avila}, S. and {Kov{\'a}cs}, A. and {Aghanim}, N. and {Altieri}, B. and {Amara}, A. and {Amendola}, L. and {Andreon}, S. and {Auricchio}, N. and {Baccigalupi}, C. and {Baldi}, M. and {Balestra}, A. and {Bardelli}, S. and {Bender}, R. and {Bernardeau}, F. and {Bodendorf}, C. and {Bonino}, D. and {Branchini}, E. and {Brescia}, M. and {Brinchmann}, J. and {Camera}, S. and {Capobianco}, V. and {Carbone}, C. and {Casas}, S. and {Castellano}, M. and {Castignani}, G. and {Cavuoti}, S. and {Cimatti}, A. and {Colodro-Conde}, C. and {Conselice}, C.~J. and {Conversi}, L. and {Copin}, Y. and {Corcione}, L. and {Courbin}, F. and {Courtois}, H.~M. and {Da Silva}, A. and {Degaudenzi}, H. and {Di Giorgio}, A.~M. and {Dinis}, J. and {Douspis}, M. and {Dubath}, F. and {Duncan}, C.~A.~J. and {Dupac}, X. and {Dusini}, S. and {Ealet}, A. and {Farina}, M. and {Farrens}, S. and {Ferriol}, S. and {Fotopoulou}, S. and {Fourmanoit}, N. and {Frailis}, M. and {Franceschi}, E. and {Franzetti}, P. and {Galeotta}, S. and {Gillard}, W. and {Gillis}, B. and {Giocoli}, C. and {G{\'o}mez-Alvarez}, P. and {Granett}, B.~R. and {Grazian}, A. and {Grupp}, F. and {Haugan}, S.~V.~H. and {Holliman}, M.~S. and {Holmes}, W. and {Hook}, I. and {Hormuth}, F. and {Hornstrup}, A. and {Hudelot}, P. and {Ili{\'c}}, S. and {Jahnke}, K. and {Jhabvala}, M. and {Joachimi}, B. and {Keih{\"a}nen}, E. and {Kermiche}, S. and {Kiessling}, A. and {Kilbinger}, M. and {Kohley}, R. and {Kubik}, B. and {K{\"u}mmel}, M. and {Kunz}, M. and {Kurki-Suonio}, H. and {Lahav}, O. and {Laureijs}, R. and {Le Mignant}, D. and {Liebing}, P. and {Ligori}, S. and {Lilje}, P.~B. and {Lindholm}, V. and {Lloro}, I. and {Maino}, D. and {Maiorano}, E. and {Mansutti}, O. and {Marcin}, S. and {Marggraf}, O. and {Markovic}, K. and {Martinelli}, M. and {Martinet}, N. and {Marulli}, F. and {Massey}, R. and {Masters}, D.~C. and {Maurogordato}, S. and {McCracken}, H.~J. and {Medinaceli}, E. and {Mei}, S. and {Melchior}, M. and {Meneghetti}, M. and {Merlin}, E. and {Meylan}, G. and {Mohr}, J.~J. and {Moresco}, M. and {Moscardini}, L. and {Munari}, E. and {Nakajima}, R. and {Nichol}, R.~C. and {Niemi}, S. -M. and {Padilla}, C. and {Paech}, K. and {Paltani}, S. and {Pasian}, F. and {Peacock}, J.~A. and {Pedersen}, K. and {Pettorino}, V. and {Pires}, S. and {Polenta}, G. and {Poncet}, M. and {Popa}, L.~A. and {Raison}, F. and {Rebolo}, R. and {Renzi}, A. and {Rhodes}, J. and {Riccio}, G. and {Romelli}, E. and {Roncarelli}, M. and {Rosset}, C. and {Rossetti}, E. and {Rusholme}, B. and {Saglia}, R. and {Sakr}, Z. and {S{\'a}nchez}, A.~G. and {Sapone}, D. and {Schewtschenko}, J.~A. and {Schirmer}, M. and {Schneider}, P. and {Schrabback}, T. and {Scodeggio}, M. and {Secroun}, A. and {Sefusatti}, E. and {Seidel}, G. and {Serrano}, S. and {Sirignano}, C. and {Sirri}, G. and {Stanco}, L. and {Starck}, J. -L. and {Steinwagner}, J. and {Taylor}, A.~N. and {Teplitz}, H.~I.},
        title = "{Euclid: V. The Flagship galaxy mock catalogue: A comprehensive simulation for the Euclid mission}",
      journal = {\aap},
         year = 2025,
        month = may,
       volume = {697},
          eid = {A5},
        pages = {A5},
          doi = {10.1051/0004-6361/202450853},
archivePrefix = {arXiv},
       eprint = {2405.13495},
 primaryClass = {astro-ph.CO},
       adsurl = {https://ui.adsabs.harvard.edu/abs/2025A&A...697A...5E}
}

@article{Potter2017,
author = {Potter, Dazzy and Stadel, Joachim and Teyssier, Romain},
year = {2017},
month = {05},
pages = {},
title = {PKDGRAV3: Beyond Trillion Particle Cosmological Simulations for the Next Era of Galaxy Surveys},
volume = {4},
journal = {CAC},
doi = {10.1186/s40668-017-0021-1}
}

@ARTICLE{euclidcollaboration2024mellier,
       author = {{Euclid Collaboration: Mellier}, Y. and {Abdurro'uf} and {Acevedo Barroso}, J.~A. and {Ach{\'u}carro}, A. and {Adamek}, J. and {Adam}, R. and {Addison}, G.~E. and {Aghanim}, N. and {Aguena}, M. and {Ajani}, V. and {Akrami}, Y. and {Al-Bahlawan}, A. and {Alavi}, A. and {Albuquerque}, I.~S. and {Alestas}, G. and {Alguero}, G. and {Allaoui}, A. and {Allen}, S.~W. and {Allevato}, V. and {Alonso-Tetilla}, A.~V. and {Altieri}, B. and {Alvarez-Candal}, A. and {Alvi}, S. and {Amara}, A. and {Amendola}, L. and {Amiaux}, J. and {Andika}, I.~T. and {Andreon}, S. and {Andrews}, A. and {Angora}, G. and {Angulo}, R.~E. and {Annibali}, F. and {Anselmi}, A. and {Anselmi}, S. and {Arcari}, S. and {Archidiacono}, M. and {Aric{\`o}}, G. and {Arnaud}, M. and {Arnouts}, S. and {Asgari}, M. and {Asorey}, J. and {Atayde}, L. and {Atek}, H. and {Atrio-Barandela}, F. and {Aubert}, M. and {Aubourg}, E. and {Auphan}, T. and {Auricchio}, N. and {Aussel}, B. and {Aussel}, H. and {Avelino}, P.~P. and {Avgoustidis}, A. and {Avila}, S. and {Awan}, S. and {Azzollini}, R. and {Baccigalupi}, C. and {Bachelet}, E. and {Bacon}, D. and {Baes}, M. and {Bagley}, M.~B. and {Bahr-Kalus}, B. and {Balaguera-Antolinez}, A. and {Balbinot}, E. and {Balcells}, M. and {Baldi}, M. and {Baldry}, I. and {Balestra}, A. and {Ballardini}, M. and {Ballester}, O. and {Balogh}, M. and {Ba{\~n}ados}, E. and {Barbier}, R. and {Bardelli}, S. and {Baron}, M. and {Barreiro}, T. and {Barrena}, R. and {Barriere}, J. -C. and {Barros}, B.~J. and {Barthelemy}, A. and {Bartolo}, N. and {Basset}, A. and {Battaglia}, P. and {Battisti}, A.~J. and {Baugh}, C.~M. and {Baumont}, L. and {Bazzanini}, L. and {Beaulieu}, J. -P. and {Beckmann}, V. and {Belikov}, A.~N. and {Bel}, J. and {Bellagamba}, F. and {Bella}, M. and {Bellini}, E. and {Benabed}, K. and {Bender}, R. and {Benevento}, G. and {Bennett}, C.~L. and {Benson}, K. and {Bergamini}, P. and {Bermejo-Climent}, J.~R. and {Bernardeau}, F. and {Bertacca}, D. and {Berthe}, M. and {Berthier}, J. and {Bethermin}, M. and {Beutler}, F. and {Bevillon}, C. and {Bhargava}, S. and {Bhatawdekar}, R. and {Bianchi}, D. and {Bisigello}, L. and {Biviano}, A. and {Blake}, R.~P. and {Blanchard}, A. and {Blazek}, J. and {Blot}, L. and {Bosco}, A. and {Bodendorf}, C. and {Boenke}, T. and {B{\"o}hringer}, H. and {Boldrini}, P. and {Bolzonella}, M. and {Bonchi}, A. and {Bonici}, M. and {Bonino}, D. and {Bonino}, L. and {Bonvin}, C. and {Bon}, W. and {Booth}, J.~T. and {Borgani}, S. and {Borlaff}, A.~S. and {Borsato}, E. and {Bose}, B. and {Botticella}, M.~T. and {Boucaud}, A. and {Bouche}, F. and {Boucher}, J.~S. and {Boutigny}, D. and {Bouvard}, T. and {Bouwens}, R. and {Bouy}, H. and {Bowler}, R.~A.~A. and {Bozza}, V. and {Bozzo}, E. and {Branchini}, E. and {Brando}, G. and {Brau-Nogue}, S. and {Brekke}, P. and {Bremer}, M.~N. and {Brescia}, M. and {Breton}, M. -A. and {Brinchmann}, J. and {Brinckmann}, T. and {Brockley-Blatt}, C. and {Brodwin}, M. and {Brouard}, L. and {Brown}, M.~L. and {Bruton}, S. and {Bucko}, J. and {Buddelmeijer}, H. and {Buenadicha}, G. and {Buitrago}, F. and {Burger}, P. and {Burigana}, C. and {Busillo}, V. and {Busonero}, D. and {Cabanac}, R. and {Cabayol-Garcia}, L. and {Cagliari}, M.~S. and {Caillat}, A. and {Caillat}, L. and {Calabrese}, M. and {Calabro}, A. and {Calderone}, G. and {Calura}, F. and {Camacho Quevedo}, B. and {Camera}, S. and {Campos}, L. and {Ca{\~n}as-Herrera}, G. and {Candini}, G.~P. and {Cantiello}, M. and {Capobianco}, V. and {Cappellaro}, E. and {Cappelluti}, N. and {Cappi}, A. and {Caputi}, K.~I. and {Cara}, C. and {Carbone}, C. and {Cardone}, V.~F. and {Carella}, E. and {Carlberg}, R.~G. and {Carle}, M. and {Carminati}, L. and {Caro}, F. and {Carrasco}, J.~M. and {Carretero}, J. and {Carrilho}, P. and {Carron Duque}, J. and {Carry}, B.},
        title = "{Euclid: I. Overview of the Euclid mission}",
      journal = {\aap},
         year = 2025,
        month = may,
       volume = {697},
          eid = {A1},
        pages = {A1},
          doi = {10.1051/0004-6361/202450810},
archivePrefix = {arXiv},
       eprint = {2405.13491},
 primaryClass = {astro-ph.CO},
       adsurl = {https://ui.adsabs.harvard.edu/abs/2025A&A...697A...1E}
}

@ARTICLE{euclidcollaboration2024cropper,
       author = {{Euclid Collaboration: Cropper}, M.~S. and {Al-Bahlawan}, A. and {Amiaux}, J. and {Awan}, S. and {Azzollini}, R. and {Benson}, K. and {Berthe}, M. and {Boucher}, J. and {Bozzo}, E. and {Brockley-Blatt}, C. and {Candini}, G.~P. and {Cara}, C. and {Chaudery}, R.~A. and {Cole}, R.~E. and {Danto}, P. and {Denniston}, J. and {Di Giorgio}, A.~M. and {Dryer}, B. and {Dubois}, J. -P. and {Endicott}, J. and {Farina}, M. and {Galli}, E. and {Genolet}, L. and {Gow}, J.~P.~D. and {Guttridge}, P. and {Hailey}, M. and {Hall}, D. and {Harper}, C. and {Hoekstra}, H. and {Holland}, A.~D. and {Horeau}, B. and {Hu}, D. and {James}, R.~E. and {Khalil}, A. and {King}, R. and {Kitching}, T. and {Kohley}, R. and {Larcheveque}, C. and {Lawrenson}, A. and {Liebing}, P. and {Liu}, S.~J. and {Martignac}, J. and {Massey}, R. and {McCracken}, H.~J. and {Miller}, L. and {Murray}, N. and {Nakajima}, R. and {Niemi}, S. -M. and {Nightingale}, J.~W. and {Paltani}, S. and {Pendem}, A. and {Philippon}, A. and {Plana}, C. and {Pool}, P. and {Pottinger}, S. and {Racca}, G.~D. and {Rhodes}, J. and {Rousseau}, A. and {Ruane}, K. and {Salatti}, M. and {Salvignol}, J. -C. and {Sciortino}, A. and {Short}, A. and {Skottfelt}, J. and {Smit}, S.~J.~A. and {Swindells}, I. and {Szafraniec}, M. and {Thomas}, P.~D. and {Thomas}, W. and {Tommasi}, E. and {Tosti}, S. and {Visticot}, F. and {Walton}, D.~M. and {Willis}, G. and {Winter}, B. and {Aghanim}, N. and {Altieri}, B. and {Amara}, A. and {Andreon}, S. and {Auricchio}, N. and {Aussel}, H. and {Baccigalupi}, C. and {Baldi}, M. and {Balestra}, A. and {Bardelli}, S. and {Basset}, A. and {Bender}, R. and {Bernardeau}, F. and {Bodendorf}, C. and {Boenke}, T. and {Bonino}, D. and {Branchini}, E. and {Brescia}, M. and {Brinchmann}, J. and {Camera}, S. and {Capobianco}, V. and {Carbone}, C. and {Cardone}, V.~F. and {Carretero}, J. and {Casas}, R. and {Casas}, S. and {Castander}, F.~J. and {Castellano}, M. and {Castignani}, G. and {Cavuoti}, S. and {Cimatti}, A. and {Colodro-Conde}, C. and {Congedo}, G. and {Conselice}, C.~J. and {Conversi}, L. and {Copin}, Y. and {Courbin}, F. and {Courtois}, H.~M. and {Crocce}, M. and {Cuby}, J. -G. and {Cuillandre}, J. -C. and {Da Silva}, A. and {Degaudenzi}, H. and {De Lucia}, G. and {Dinis}, J. and {Dolding}, C. and {Douspis}, M. and {Duncan}, C.~A.~J. and {Dupac}, X. and {Dusini}, S. and {Ealet}, A. and {Fabricius}, M. and {Farrens}, S. and {Ferriol}, S. and {Fosalba}, P. and {Fotopoulou}, S. and {Frailis}, M. and {Franceschi}, E. and {Franzetti}, P. and {Frugier}, P. -A. and {Fumana}, M. and {Galeotta}, S. and {Garilli}, B. and {George}, K. and {Gillard}, W. and {Gillis}, B. and {Giocoli}, C. and {G{\'o}mez-Alvarez}, P. and {Granett}, B.~R. and {Grazian}, A. and {Grupp}, F. and {Guzzo}, L. and {Haugan}, S.~V.~H. and {Herent}, O. and {Hoar}, J. and {Holliman}, M.~S. and {Holmes}, W. and {Hook}, I. and {Hormuth}, F. and {Hornstrup}, A. and {Hudelot}, P. and {Ili{\'c}}, S. and {Jahnke}, K. and {Jhabvala}, M. and {Joachimi}, B. and {Keih{\"a}nen}, E. and {Kermiche}, S. and {Kiessling}, A. and {Kilbinger}, M. and {Kubik}, B. and {Kuijken}, K. and {K{\"u}mmel}, M. and {Kunz}, M. and {Kurki-Suonio}, H. and {Lahav}, O. and {Laureijs}, R. and {Ligori}, S. and {Lilje}, P.~B. and {Lindholm}, V. and {Lloro}, I. and {Lorenzo Alvarez}, J. and {Mainetti}, G. and {Maino}, D. and {Maiorano}, E. and {Mansutti}, O. and {Marcin}, S. and {Marggraf}, O. and {Markovic}, K. and {Martinelli}, M. and {Martinet}, N. and {Marulli}, F. and {Masters}, D.~C. and {Maurogordato}, S. and {Medinaceli}, E. and {Mei}, S. and {Melchior}, M. and {Mellier}, Y. and {Meneghetti}, M. and {Merlin}, E. and {Meylan}, G. and {Mohr}, J.~J. and {Moresco}, M. and {Moscardini}, L. and {Neissner}, C.},
        title = "{Euclid: II. The VIS instrument}",
      journal = {\aap},
         year = 2025,
        month = may,
       volume = {697},
          eid = {A2},
        pages = {A2},
          doi = {10.1051/0004-6361/202450996},
archivePrefix = {arXiv},
       eprint = {2405.13492},
 primaryClass = {astro-ph.IM},
       adsurl = {https://ui.adsabs.harvard.edu/abs/2025A&A...697A...2E}
}

@ARTICLE{euclidcollaboration2024Jahnke,
       author = {{Euclid Collaboration: Jahnke}, K. and {Gillard}, W. and {Schirmer}, M. and {Ealet}, A. and {Maciaszek}, T. and {Prieto}, E. and {Barbier}, R. and {Bonoli}, C. and {Corcione}, L. and {Dusini}, S. and {Grupp}, F. and {Hormuth}, F. and {Ligori}, S. and {Martin}, L. and {Morgante}, G. and {Padilla}, C. and {Toledo-Moreo}, R. and {Trifoglio}, M. and {Valenziano}, L. and {Bender}, R. and {Castander}, F.~J. and {Garilli}, B. and {Lilje}, P.~B. and {Rix}, H. -W. and {Andersen}, M.~I. and {Auricchio}, N. and {Balestra}, A. and {Barriere}, J. -C. and {Battaglia}, P. and {Berthe}, M. and {Bodendorf}, C. and {Boenke}, T. and {Bon}, W. and {Bonnefoi}, A. and {Caillat}, A. and {Capobianco}, V. and {Carle}, M. and {Casas}, R. and {Cho}, H. and {Costille}, A. and {Ducret}, F. and {Ferriol}, S. and {Franceschi}, E. and {Gimenez}, J. -L. and {Holmes}, W. and {Hornstrup}, A. and {Jhabvala}, M. and {Kohley}, R. and {Kubik}, B. and {Laureijs}, R. and {Le Mignant}, D. and {Lloro}, I. and {Medinaceli}, E. and {Mellier}, Y. and {Polenta}, G. and {Racca}, G.~D. and {Renzi}, A. and {Salvignol}, J. -C. and {Secroun}, A. and {Seidel}, G. and {Seiffert}, M. and {Sirignano}, C. and {Sirri}, G. and {Strada}, P. and {Smadja}, G. and {Stanco}, L. and {Wachter}, S. and {Anselmi}, S. and {Borsato}, E. and {Caillat}, L. and {Cogato}, F. and {Colodro-Conde}, C. and {Crouzet}, P. -E. and {Conforti}, V. and {D'Alessandro}, M. and {Copin}, Y. and {Cuillandre}, J. -C. and {Davies}, J.~E. and {Davini}, S. and {Derosa}, A. and {Diaz}, J.~J. and {Di Domizio}, S. and {Di Ferdinando}, D. and {Farinelli}, R. and {Ferrari}, A.~G. and {Fornari}, F. and {Gabarra}, L. and {Garcia}, R. and {Gutierrez}, C.~M. and {Giacomini}, F. and {Lagier}, P. and {Gianotti}, F. and {Krause}, O. and {Madrid}, F. and {Laudisio}, F. and {Macias-Perez}, J. and {Naletto}, G. and {Niclas}, M. and {Marpaud}, J. and {Mauri}, N. and {da Silva}, R. and {Passalacqua}, F. and {Paterson}, K. and {Patrizii}, L. and {Risso}, I. and {Solheim}, B.~G.~B. and {Scodeggio}, M. and {Stassi}, P. and {Steinwagner}, J. and {Tenti}, M. and {Testera}, G. and {Travaglini}, R. and {Tosi}, S. and {Troja}, A. and {Tubio}, O. and {Valieri}, C. and {Vescovi}, C. and {Ventura}, S. and {Aghanim}, N. and {Altieri}, B. and {Amara}, A. and {Amiaux}, J. and {Andreon}, S. and {Appleton}, P.~N. and {Aussel}, H. and {Baccigalupi}, C. and {Baldi}, M. and {Bardelli}, S. and {Basset}, A. and {Bonchi}, A. and {Bonino}, D. and {Branchini}, E. and {Brescia}, M. and {Brinchmann}, J. and {Camera}, S. and {Carbone}, C. and {Cardone}, V.~F. and {Carretero}, J. and {Casas}, S. and {Castellano}, M. and {Castignani}, G. and {Cavuoti}, S. and {Chabaud}, P. -Y. and {Cimatti}, A. and {Congedo}, G. and {Conselice}, C.~J. and {Conversi}, L. and {Courbin}, F. and {Courtois}, H.~M. and {Crocce}, M. and {Cropper}, M. and {Cuby}, J. -G. and {Da Silva}, A. and {Degaudenzi}, H. and {De Lucia}, G. and {Di Giorgio}, A.~M. and {Dinis}, J. and {Douspis}, M. and {Dubath}, F. and {Duncan}, C.~A.~J. and {Dupac}, X. and {Fabricius}, M. and {Farina}, M. and {Farrens}, S. and {Faustini}, F. and {Fosalba}, P. and {Fotopoulou}, S. and {Fourmanoit}, N. and {Frailis}, M. and {Franzetti}, P. and {Galeotta}, S. and {George}, K. and {Gillis}, B. and {Giocoli}, C. and {G{\'o}mez-Alvarez}, P. and {Granett}, B.~R. and {Grazian}, A. and {Guzzo}, L. and {Hailey}, M. and {Haugan}, S.~V.~H. and {Hoar}, J. and {Hoekstra}, H. and {Hook}, I. and {Hudelot}, P. and {Ili{\'c}}, S. and {Joachimi}, B. and {Keih{\"a}nen}, E. and {Kermiche}, S. and {Kiessling}, A. and {Kilbinger}, M. and {Kitching}, T. and {K{\"u}mmel}, M. and {Kunz}, M. and {Kurki-Suonio}, H. and {Lahav}, O. and {Liebing}, P. and {Lindholm}, V. and {Lorenzo Alvarez}, J. and {Mainetti}, G.},
        title = "{Euclid: III. The NISP Instrument}",
      journal = {\aap},
         year = 2025,
        month = may,
       volume = {697},
          eid = {A3},
        pages = {A3},
          doi = {10.1051/0004-6361/202450786},
archivePrefix = {arXiv},
       eprint = {2405.13493},
 primaryClass = {astro-ph.IM},
       adsurl = {https://ui.adsabs.harvard.edu/abs/2025A&A...697A...3E}
}

@article{Blanton2003,
   title={The Galaxy Luminosity Function and Luminosity Density at Redshiftz= 0.1},
   volume={592},
   ISSN={1538-4357},
   url={http://dx.doi.org/10.1086/375776},
   DOI={10.1086/375776},
   number={2},
   journal={ApJ},
   publisher={American Astronomical Society},
   author={Blanton, Michael R. and Hogg, David W. and Bahcall, Neta A. and Brinkmann, J. and Britton, Malcolm and Connolly, Andrew J. and Csabai, Istvan and Fukugita, Masataka and Loveday, Jon and Meiksin, Avery and Munn, Jeffrey A. and Nichol, R. C. and Okamura, Sadanori and Quinn, Thomas and Schneider, Donald P. and Shimasaku, Kazuhiro and Strauss, Michael A. and Tegmark, Max and Vogeley, Michael S. and Weinberg, David H.},
   year={2003},
   month=aug, 
   pages={819-–838} }

@ARTICLE{Ilbert2006,
       author = {{Ilbert}, O. and {Arnouts}, S. and {McCracken}, H.~J. and {Bolzonella}, M. and {Bertin}, E. and {Le F{\`e}vre}, O. and {Mellier}, Y. and {Zamorani}, G. and {Pell{\`o}}, R. and {Iovino}, A. and {Tresse}, L. and {Le Brun}, V. and {Bottini}, D. and {Garilli}, B. and {Maccagni}, D. and {Picat}, J.~P. and {Scaramella}, R. and {Scodeggio}, M. and {Vettolani}, G. and {Zanichelli}, A. and {Adami}, C. and {Bardelli}, S. and {Cappi}, A. and {Charlot}, S. and {Ciliegi}, P. and {Contini}, T. and {Cucciati}, O. and {Foucaud}, S. and {Franzetti}, P. and {Gavignaud}, I. and {Guzzo}, L. and {Marano}, B. and {Marinoni}, C. and {Mazure}, A. and {Meneux}, B. and {Merighi}, R. and {Paltani}, S. and {Pollo}, A. and {Pozzetti}, L. and {Radovich}, M. and {Zucca}, E. and {Bondi}, M. and {Bongiorno}, A. and {Busarello}, G. and {de La Torre}, S. and {Gregorini}, L. and {Lamareille}, F. and {Mathez}, G. and {Merluzzi}, P. and {Ripepi}, V. and {Rizzo}, D. and {Vergani}, D.},
        title = "{Accurate photometric redshifts for the CFHT legacy survey calibrated using the VIMOS VLT deep survey}",
      journal = {\aap},
         year = 2006,
        month = oct,
       volume = {457},
       number = {3},
        pages = {841--856},
          doi = {10.1051/0004-6361:20065138},
archivePrefix = {arXiv},
       eprint = {astro-ph/0603217},
 primaryClass = {astro-ph},
       adsurl = {https://ui.adsabs.harvard.edu/abs/2006A&A...457..841I}
}

@article{Polletta2007,
   title={Spectral Energy Distributions of Hard X‐Ray Selected Active Galactic Nuclei in theXMM‐NewtonMedium Deep Survey},
   volume={663},
   ISSN={1538-4357},
   url={http://dx.doi.org/10.1086/518113},
   DOI={10.1086/518113},
   number={1},
   journal={ApJ},
   publisher={American Astronomical Society},
   author={Polletta, M. and Tajer, M. and Maraschi, L. and Trinchieri, G. and Lonsdale, C. J. and Chiappetti, L. and Andreon, S. and Pierre, M. and Le Fevre, O. and Zamorani, G. and Maccagni, D. and Garcet, O. and Surdej, J. and Franceschini, A. and Alloin, D. and Shupe, D. L. and Surace, J. A. and Fang, F. and Rowan‐Robinson, M. and Smith, H. E. and Tresse, L.},
   year={2007},
   month=jul, pages={81--102} }

@ARTICLE{Hutterer_2006,
       author = {{Huterer}, Dragan and {Takada}, Masahiro and {Bernstein}, Gary and {Jain}, Bhuvnesh},
        title = "{Systematic errors in future weak-lensing surveys: requirements and prospects for self-calibration}",
      journal = {\mnras},
         year = 2006,
        month = feb,
       volume = {366},
       number = {1},
        pages = {101--114},
          doi = {10.1111/j.1365-2966.2005.09782.x},
archivePrefix = {arXiv},
       eprint = {astro-ph/0506030},
 primaryClass = {astro-ph},
       adsurl = {https://ui.adsabs.harvard.edu/abs/2006MNRAS.366..101H}
}

@ARTICLE{Ilbert_2021,
       author = {{Euclid Collaboration: Ilbert}, O. and {de la Torre}, S. and {Martinet}, N. and {Wright}, A.~H. and {Paltani}, S. and {Laigle}, C. and {Davidzon}, I. and {Jullo}, E. and {Hildebrandt}, H. and {Masters}, D.~C. and {Amara}, A. and {Conselice}, C.~J. and {Andreon}, S. and {Auricchio}, N. and {Azzollini}, R. and {Baccigalupi}, C. and {Balaguera-Antol{\'\i}nez}, A. and {Baldi}, M. and {Balestra}, A. and {Bardelli}, S. and {Bender}, R. and {Biviano}, A. and {Bodendorf}, C. and {Bonino}, D. and {Borgani}, S. and {Boucaud}, A. and {Bozzo}, E. and {Branchini}, E. and {Brescia}, M. and {Burigana}, C. and {Cabanac}, R. and {Camera}, S. and {Capobianco}, V. and {Cappi}, A. and {Carbone}, C. and {Carretero}, J. and {Carvalho}, C.~S. and {Casas}, S. and {Castander}, F.~J. and {Castellano}, M. and {Castignani}, G. and {Cavuoti}, S. and {Cimatti}, A. and {Cledassou}, R. and {Colodro-Conde}, C. and {Congedo}, G. and {Conversi}, L. and {Copin}, Y. and {Corcione}, L. and {Costille}, A. and {Coupon}, J. and {Courtois}, H.~M. and {Cropper}, M. and {Cuby}, J. and {Da Silva}, A. and {Degaudenzi}, H. and {Di Ferdinando}, D. and {Dubath}, F. and {Duncan}, C. and {Dupac}, X. and {Dusini}, S. and {Ealet}, A. and {Fabricius}, M. and {Farrens}, S. and {Ferreira}, P.~G. and {Finelli}, F. and {Fosalba}, P. and {Fotopoulou}, S. and {Franceschi}, E. and {Franzetti}, P. and {Galeotta}, S. and {Garilli}, B. and {Gillard}, W. and {Gillis}, B. and {Giocoli}, C. and {Gozaliasl}, G. and {Graci{\'a}-Carpio}, J. and {Grupp}, F. and {Guzzo}, L. and {Haugan}, S.~V.~H. and {Holmes}, W. and {Hormuth}, F. and {Jahnke}, K. and {Keihanen}, E. and {Kermiche}, S. and {Kiessling}, A. and {Kirkpatrick}, C.~C. and {Kunz}, M. and {Kurki-Suonio}, H. and {Ligori}, S. and {Lilje}, P.~B. and {Lloro}, I. and {Maino}, D. and {Maiorano}, E. and {Marggraf}, O. and {Markovic}, K. and {Marulli}, F. and {Massey}, R. and {Maturi}, M. and {Mauri}, N. and {Maurogordato}, S. and {McCracken}, H.~J. and {Medinaceli}, E. and {Mei}, S. and {Metcalf}, R. Benton and {Moresco}, M. and {Morin}, B. and {Moscardini}, L. and {Munari}, E. and {Nakajima}, R. and {Neissner}, C. and {Niemi}, S. and {Nightingale}, J. and {Padilla}, C. and {Pasian}, F. and {Patrizii}, L. and {Pedersen}, K. and {Pello}, R. and {Pettorino}, V. and {Pires}, S. and {Polenta}, G. and {Poncet}, M. and {Popa}, L. and {Potter}, D. and {Pozzetti}, L. and {Raison}, F. and {Renzi}, A. and {Rhodes}, J. and {Riccio}, G. and {Romelli}, E. and {Roncarelli}, M. and {Rossetti}, E. and {Saglia}, R. and {S{\'a}nchez}, A.~G. and {Sapone}, D. and {Schneider}, P. and {Schrabback}, T. and {Scottez}, V. and {Secroun}, A. and {Seidel}, G. and {Serrano}, S. and {Sirignano}, C. and {Sirri}, G. and {Stanco}, L. and {Sureau}, F. and {Tallada Cresp{\'a}}, P. and {Tenti}, M. and {Teplitz}, H.~I. and {Tereno}, I. and {Toledo-Moreo}, R. and {Torradeflot}, F. and {Tramacere}, A. and {Valentijn}, E.~A. and {Valenziano}, L. and {Valiviita}, J. and {Vassallo}, T. and {Wang}, Y. and {Welikala}, N. and {Weller}, J. and {Whittaker}, L. and {Zacchei}, A. and {Zamorani}, G. and {Zoubian}, J. and {Zucca}, E.},
        title = "{Euclid preparation. XI. Mean redshift determination from galaxy redshift probabilities for cosmic shear tomography}",
      journal = {\aap},
         year = 2021,
        month = mar,
       volume = {647},
          eid = {A117},
        pages = {A117},
          doi = {10.1051/0004-6361/202040237},
archivePrefix = {arXiv},
       eprint = {2101.02228},
 primaryClass = {astro-ph.CO},
       adsurl = {https://ui.adsabs.harvard.edu/abs/2021A&A...647A.117E}
}

@ARTICLE{Merz_2024,
       author = {{Merz}, Grant and {Liu}, Xin and {Schmidt}, Samuel and {Malz}, Alex I. and {Zhang}, Tianqing and {Branton}, Doug and {Burke}, Colin J. and {Delucchi}, Melissa and {Sai Ejjagiri}, Yaswant and {Kubica}, Jeremy and {Liu}, Yichen and {Lynn}, Olivia and {Oldag}, Drew and {The LSST Dark Energy Science Collaboration}},
        title = "{DeepDISC-photoz: Deep Learning-Based Photometric Redshift Estimation for Rubin LSST}",
      journal = {arXiv e-prints},
         year = 2024,
        month = nov,
          eid = {arXiv:2411.18769},
        pages = {arXiv:2411.18769},
          doi = {10.48550/arXiv.2411.18769},
archivePrefix = {arXiv},
       eprint = {2411.18769},
 primaryClass = {astro-ph.IM},
       adsurl = {https://ui.adsabs.harvard.edu/abs/2024arXiv241118769M}
}

@ARTICLE{Li_2023,
       author = {{Li}, Shun-Sheng and {Hoekstra}, Henk and {Kuijken}, Konrad and {Asgari}, Marika and {Bilicki}, Maciej and {Giblin}, Benjamin and {Heymans}, Catherine and {Hildebrandt}, Hendrik and {Joachimi}, Benjamin and {Miller}, Lance and {van den Busch}, Jan Luca and {Wright}, Angus H. and {Kannawadi}, Arun and {Reischke}, Robert and {Shan}, HuanYuan},
        title = "{KiDS-1000: Cosmology with improved cosmic shear measurements}",
      journal = {\aap},
         year = 2023,
        month = nov,
       volume = {679},
          eid = {A133},
        pages = {A133},
          doi = {10.1051/0004-6361/202347236},
archivePrefix = {arXiv},
       eprint = {2306.11124},
 primaryClass = {astro-ph.CO},
       adsurl = {https://ui.adsabs.harvard.edu/abs/2023A\&A...679A.133L}
}

@article{Campos_2024,
  author       = {Campos, A. and Yin, B. and Dodelson, S. and Amon, A. and Alarcon, A. and Sánchez, C. and Bernstein, G. M. and Giannini, G. and Myles, J. and Samuroff, S. and others},
  title        = {Enhancing weak lensing redshift distribution characterization by optimizing the Dark Energy Survey Self-Organizing Map Photo-z method},
  url          = {https://www.osti.gov/biblio/2432453},
  journal      = {No journal information},
  place        = {United States},
  year         = {2024},
  month        = {08}}

@ARTICLE{Giannini_2024,
       author = {{Giannini}, G. and {Alarcon}, A. and {Gatti}, M. and {Porredon}, A. and {Crocce}, M. and {Bernstein}, G.~M. and {Cawthon}, R. and {S{\'a}nchez}, C. and {Doux}, C. and {Elvin-Poole}, J. and {Raveri}, M. and {Myles}, J. and {Lin}, H. and {Amon}, A. and {Allam}, S. and {Alves}, O. and {Andrade-Oliveira}, F. and {Baxter}, E. and {Bechtol}, K. and {Becker}, M.~R. and {Blazek}, J. and {Camacho}, H. and {Campos}, A. and {Carnero Rosell}, A. and {Carrasco Kind}, M. and {Choi}, A. and {Cordero}, J. and {De Vicente}, J. and {DeRose}, J. and {Diehl}, H.~T. and {Dodelson}, S. and {Drlica-Wagner}, A. and {Eckert}, K. and {Fang}, X. and {Farahi}, A. and {Fosalba}, P. and {Friedrich}, O. and {Gruen}, D. and {Gruendl}, R.~A. and {Gschwend}, J. and {Harrison}, I. and {Hartley}, W.~G. and {Huff}, E.~M. and {Jarvis}, M. and {Krause}, E. and {Kuropatkin}, N. and {Lemos}, P. and {MacCrann}, N. and {McCullough}, J. and {Muir}, J. and {Pandey}, S. and {Prat}, J. and {Rodriguez-Monroy}, M. and {Ross}, A.~J. and {Rykoff}, E.~S. and {Samuroff}, S. and {Secco}, L.~F. and {Sevilla-Noarbe}, I. and {Sheldon}, E. and {Troxel}, M.~A. and {Tucker}, D.~L. and {Weaverdyck}, N. and {Yanny}, B. and {Yin}, B. and {Zhang}, Y. and {Abbott}, T.~M.~C. and {Aguena}, M. and {Bacon}, D. and {Bertin}, E. and {Bocquet}, S. and {Brooks}, D. and {Burke}, D.~L. and {Carretero}, J. and {Castander}, F.~J. and {Costanzi}, M. and {da Costa}, L.~N. and {Pereira}, M.~E.~S. and {Desai}, S. and {Doel}, P. and {Ferrero}, I. and {Flaugher}, B. and {Friedel}, D. and {Frieman}, J. and {Garc{\'\i}a-Bellido}, J. and {Gerdes}, D.~W. and {Gutierrez}, G. and {Hinton}, S.~R. and {Hollowood}, D.~L. and {Honscheid}, K. and {James}, D.~J. and {Kent}, S. and {Kuehn}, K. and {Lahav}, O. and {Lidman}, C. and {Lima}, M. and {Melchior}, P. and {Mena-Fern{\'a}ndez}, J. and {Menanteau}, F. and {Miquel}, R. and {Ogando}, R.~L.~C. and {Paterno}, M. and {Paz-Chinch{\'o}n}, F. and {Pieres}, A. and {Plazas Malag{\'o}n}, A.~A. and {Roodman}, A. and {Sanchez}, E. and {Scarpine}, V. and {Smith}, M. and {Suchyta}, E. and {Swanson}, M.~E.~C. and {Tarle}, G. and {Thomas}, D. and {To}, C. and {Vincenzi}, M. and {DES Collaboration}},
        title = "{Dark Energy Survey Year 3 results: redshift calibration of the MAGLIM lens sample from the combination of SOMPZ and clustering and its impact on cosmology}",
      journal = {\mnras},
         year = 2024,
        month = jan,
       volume = {527},
       number = {2},
        pages = {2010--2036},
          doi = {10.1093/mnras/stad2945},
archivePrefix = {arXiv},
       eprint = {2209.05853},
 primaryClass = {astro-ph.CO},
       adsurl = {https://ui.adsabs.harvard.edu/abs/2024MNRAS.527.2010G}
}

@ARTICLE{Myles_2021,
       author = {{Myles}, J. and {Alarcon}, A. and {Amon}, A. and {S{\'a}nchez}, C. and {Everett}, S. and {DeRose}, J. and {McCullough}, J. and {Gruen}, D. and {Bernstein}, G.~M. and {Troxel}, M.~A. and {Dodelson}, S. and {Campos}, A. and {MacCrann}, N. and {Yin}, B. and {Raveri}, M. and {Amara}, A. and {Becker}, M.~R. and {Choi}, A. and {Cordero}, J. and {Eckert}, K. and {Gatti}, M. and {Giannini}, G. and {Gschwend}, J. and {Gruendl}, R.~A. and {Harrison}, I. and {Hartley}, W.~G. and {Huff}, E.~M. and {Kuropatkin}, N. and {Lin}, H. and {Masters}, D. and {Miquel}, R. and {Prat}, J. and {Roodman}, A. and {Rykoff}, E.~S. and {Sevilla-Noarbe}, I. and {Sheldon}, E. and {Wechsler}, R.~H. and {Yanny}, B. and {Abbott}, T.~M.~C. and {Aguena}, M. and {Allam}, S. and {Annis}, J. and {Bacon}, D. and {Bertin}, E. and {Bhargava}, S. and {Bridle}, S.~L. and {Brooks}, D. and {Burke}, D.~L. and {Carnero Rosell}, A. and {Carrasco Kind}, M. and {Carretero}, J. and {Castander}, F.~J. and {Conselice}, C. and {Costanzi}, M. and {Crocce}, M. and {da Costa}, L.~N. and {Pereira}, M.~E.~S. and {Desai}, S. and {Diehl}, H.~T. and {Eifler}, T.~F. and {Elvin-Poole}, J. and {Evrard}, A.~E. and {Ferrero}, I. and {Fert{\'e}}, A. and {Flaugher}, B. and {Fosalba}, P. and {Frieman}, J. and {Garc{\'\i}a-Bellido}, J. and {Gaztanaga}, E. and {Giannantonio}, T. and {Hinton}, S.~R. and {Hollowood}, D.~L. and {Honscheid}, K. and {Hoyle}, B. and {Huterer}, D. and {James}, D.~J. and {Krause}, E. and {Kuehn}, K. and {Lahav}, O. and {Lima}, M. and {Maia}, M.~A.~G. and {Marshall}, J.~L. and {Martini}, P. and {Melchior}, P. and {Menanteau}, F. and {Mohr}, J.~J. and {Morgan}, R. and {Muir}, J. and {Ogando}, R.~L.~C. and {Palmese}, A. and {Paz-Chinch{\'o}n}, F. and {Plazas}, A.~A. and {Rodriguez-Monroy}, M. and {Samuroff}, S. and {Sanchez}, E. and {Scarpine}, V. and {Secco}, L.~F. and {Serrano}, S. and {Smith}, M. and {Soares-Santos}, M. and {Suchyta}, E. and {Swanson}, M.~E.~C. and {Tarle}, G. and {Thomas}, D. and {To}, C. and {Varga}, T.~N. and {Weller}, J. and {Wester}, W.},
        title = "{Dark Energy Survey Year 3 results: redshift calibration of the weak lensing source galaxies}",
      journal = {\mnras},
         year = 2021,
        month = aug,
       volume = {505},
       number = {3},
        pages = {4249--4277},
          doi = {10.1093/mnras/stab1515},
archivePrefix = {arXiv},
       eprint = {2012.08566},
 primaryClass = {astro-ph.CO},
       adsurl = {https://ui.adsabs.harvard.edu/abs/2021MNRAS.505.4249M}
}

@article{Eriksen_2020,
   title={The PAU Survey: Photometric redshifts using transfer learning from simulations},
   volume={497},
   ISSN={1365-2966},
   url={http://dx.doi.org/10.1093/mnras/staa2265},
   DOI={10.1093/mnras/staa2265},
   number={4},
   journal={MNRAS},
   publisher={Oxford University Press (OUP)},
   author={Eriksen, M and Alarcon, A and Cabayol, L and Carretero, J and Casas, R and Castander, F J and De Vicente, J and Fernandez, E and Garcia-Bellido, J and Gaztanaga, E and Hildebrandt, H and Hoekstra, H and Joachimi, B and Miquel, R and Padilla, C and Sanchez, E and Sevilla-Noarbe, I and Tallada, P},
   year={2020},
   month=aug, pages={4565--4579} }

@ARTICLE{Tanaka2018,
       author = {{Tanaka}, Masayuki and {Coupon}, Jean and {Hsieh}, Bau-Ching and {Mineo}, Sogo and {Nishizawa}, Atsushi J. and {Speagle}, Joshua and {Furusawa}, Hisanori and {Miyazaki}, Satoshi and {Murayama}, Hitoshi},
        title = "{Photometric redshifts for Hyper Suprime-Cam Subaru Strategic Program Data Release 1}",
      journal = {\pasj},
         year = 2018,
        month = jan,
       volume = {70},
          eid = {S9},
        pages = {S9},
          doi = {10.1093/pasj/psx077},
archivePrefix = {arXiv},
       eprint = {1704.05988},
 primaryClass = {astro-ph.GA},
       adsurl = {https://ui.adsabs.harvard.edu/abs/2018PASJ...70S...9T}
}

@ARTICLE{Wong_2025,
       author = {{Wong}, J.~H.~W. and {Brown}, M.~L. and {Duncan}, C.~A.~J. and {Amara}, A. and {Andreon}, S. and {Baccigalupi}, C. and {Baldi}, M. and {Bardelli}, S. and {Bonino}, D. and {Branchini}, E. and {Brescia}, M. and {Brinchmann}, J. and {Caillat}, A. and {Camera}, S. and {Capobianco}, V. and {Carbone}, C. and {Carretero}, J. and {Casas}, S. and {Castellano}, M. and {Castignani}, G. and {Cavuoti}, S. and {Cimatti}, A. and {Colodro-Conde}, C. and {Congedo}, G. and {Conselice}, C.~J. and {Conversi}, L. and {Copin}, Y. and {Courbin}, F. and {Courtois}, H.~M. and {Da Silva}, A. and {Degaudenzi}, H. and {De Lucia}, G. and {Di Giorgio}, A.~M. and {Dinis}, J. and {Dubath}, F. and {Dupac}, X. and {Dusini}, S. and {Farina}, M. and {Farrens}, S. and {Faustini}, F. and {Ferriol}, S. and {Frailis}, M. and {Franceschi}, E. and {Galeotta}, S. and {George}, K. and {Gillard}, W. and {Gillis}, B. and {Giocoli}, C. and {Grazian}, A. and {Grupp}, F. and {Guzzo}, L. and {Haugan}, S.~V.~H. and {Holmes}, W. and {Hook}, I. and {Hormuth}, F. and {Hornstrup}, A. and {Ili{\'c}}, S. and {Jahnke}, K. and {Jhabvala}, M. and {Keih{\"a}nen}, E. and {Kermiche}, S. and {Kiessling}, A. and {Kubik}, B. and {Kunz}, M. and {Kurki-Suonio}, H. and {Ligori}, S. and {Lilje}, P.~B. and {Lindholm}, V. and {Lloro}, I. and {Mainetti}, G. and {Maiorano}, E. and {Mansutti}, O. and {Marggraf}, O. and {Markovic}, K. and {Martinelli}, M. and {Martinet}, N. and {Marulli}, F. and {Massey}, R. and {Medinaceli}, E. and {Mei}, S. and {Melchior}, M. and {Mellier}, Y. and {Meneghetti}, M. and {Merlin}, E. and {Meylan}, G. and {Moresco}, M. and {Moscardini}, L. and {Neissner}, C. and {Niemi}, S. -M. and {Padilla}, C. and {Paltani}, S. and {Pasian}, F. and {Pedersen}, K. and {Pettorino}, V. and {Pires}, S. and {Polenta}, G. and {Poncet}, M. and {Popa}, L.~A. and {Raison}, F. and {Renzi}, A. and {Rhodes}, J. and {Riccio}, G. and {Romelli}, E. and {Roncarelli}, M. and {Rossetti}, E. and {Saglia}, R. and {Sakr}, Z. and {S{\'a}nchez}, A.~G. and {Sapone}, D. and {Sartoris}, B. and {Schneider}, P. and {Schrabback}, T. and {Secroun}, A. and {Seidel}, G. and {Serrano}, S. and {Sirignano}, C. and {Sirri}, G. and {Stanco}, L. and {Steinwagner}, J. and {Tallada-Cresp{\'\i}}, P. and {Taylor}, A.~N. and {Tereno}, I. and {Toledo-Moreo}, R. and {Torradeflot}, F. and {Tutusaus}, I. and {Valenziano}, L. and {Vassallo}, T. and {Verdoes Kleijn}, G. and {Veropalumbo}, A. and {Wang}, Y. and {Weller}, J. and {Zamorani}, G. and {Zucca}, E. and {Burigana}, C. and {Calabrese}, M. and {Pezzotta}, A. and {Scottez}, V. and {Spurio Mancini}, A. and {Viel}, M.},
        title = "{Euclid: Optimising tomographic redshift binning for 3$\times$2pt power spectrum constraints on dark energy}",
      journal = {arXiv e-prints},
         year = 2025,
        month = jan,
          eid = {arXiv:2501.07559},
        pages = {arXiv:2501.07559},
          doi = {10.48550/arXiv.2501.07559},
archivePrefix = {arXiv},
       eprint = {2501.07559},
 primaryClass = {astro-ph.CO},
       adsurl = {https://ui.adsabs.harvard.edu/abs/2025arXiv250107559W}
}

@article{Gatti_2021,
   title={Dark Energy Survey Year 3 Results: clustering redshifts – calibration of the weak lensing source redshift distributions with redMaGiC and BOSS/eBOSS},
   volume={510},
   ISSN={1365-2966},
   url={http://dx.doi.org/10.1093/mnras/stab3311},
   DOI={10.1093/mnras/stab3311},
   number={1},
   journal={MNRAS},
   publisher={Oxford University Press (OUP)},
   author={Gatti, M and Giannini, G and Bernstein, G M and Alarcon, A and Myles, J and Amon, A and Cawthon, R and Troxel, M and DeRose, J and Everett, S and Ross, A J and Rykoff, E S and Elvin-Poole, J and Cordero, J and Harrison, I and Sanchez, C and Prat, J and Gruen, D and Lin, H and Crocce, M and Rozo, E and Abbott, T M C and Aguena, M and Allam, S and Annis, J and Avila, S and Bacon, D and Bertin, E and Brooks, D and Burke, D L and Rosell, A Carnero and Kind, M Carrasco and Carretero, J and Castander, F J and Choi, A and Conselice, C and Costanzi, M and Crocce, M and da Costa, L N and Pereira, M E S and Dawson, K and Desai, S and Diehl, H T and Eckert, K and Eifler, T F and Evrard, A E and Ferrero, I and Flaugher, B and Fosalba, P and Frieman, J and García-Bellido, J and Gaztanaga, E and Giannantonio, T and Gruendl, R A and Gschwend, J and Hinton, S R and Hollowood, D L and Honscheid, K and Hoyle, B and Huterer, D and James, D J and Kuehn, K and Kuropatkin, N and Lahav, O and Lima, M and MacCrann, N and Maia, M A G and March, M and Marshall, J L and Melchior, P and Menanteau, F and Miquel, R and Mohr, J J and Morgan, R and Ogando, R L C and Palmese, A and Paz-Chinchón, F and Percival, W J and Plazas, A A and Rodriguez-Monroy, M and Roodman, A and Rossi, G and Samuroff, S and Sanchez, E and Scarpine, V and Secco, L F and Serrano, S and Sevilla-Noarbe, I and Smith, M and Soares-Santos, M and Suchyta, E and Swanson, M E C and Tarle, G and Thomas, D and To, C and Varga, T N and Weller, J and Wilkinson, R D},
   year={2021},
   month=nov, pages={1223--1247} }

@INPROCEEDINGS{Carretero_2017,
       author = {{Carretero}, J. and {Tallada}, P. and {Casals}, J. and {Caubet}, M. and {Castander}, F. and {Blot}, L. and {Alarc{\'o}n}, A. and {Serrano}, S. and {Fosalba}, P. and {Acosta-Silva}, C. and {Tonello}, N. and {Torradeflot}, F. n. and {Eriksen}, M. and {Neissner}, C. and {Delfino}, M.},
        title = "{CosmoHub and SciPIC: Massive cosmological data analysis, distribution and generation using a Big Data platform}",
    booktitle = {Proceedings of the European Physical Society Conference on High Energy Physics. 5-12 July},
         year = 2017,
        month = jul,
          eid = {488},
        pages = {488},
          doi = {10.22323/1.314.0488},
       adsurl = {https://ui.adsabs.harvard.edu/abs/2017ehep.confE.488C}
}

@ARTICLE{wright2025,
       author = {{Wright}, Angus H. and {St{\"o}lzner}, Benjamin and {Asgari}, Marika and {Bilicki}, Maciej and {Giblin}, Benjamin and {Heymans}, Catherine and {Hildebrandt}, Hendrik and {Hoekstra}, Henk and {Joachimi}, Benjamin and {Kuijken}, Konrad and {Li}, Shun-Sheng and {Reischke}, Robert and {von Wietersheim-Kramsta}, Maximilian and {Yoon}, Mijin and {Burger}, Pierre and {Chisari}, Nora Elisa and {de Jong}, Jelte and {Dvornik}, Andrej and {Georgiou}, Christos and {Harnois-D{\'e}raps}, Joachim and {Jalan}, Priyanka and {William}, Anjitha John and {Joudaki}, Shahab and {Lesci}, Giorgio Francesco and {Linke}, Laila and {Loureiro}, Arthur and {Mahony}, Constance and {Maturi}, Matteo and {Miller}, Lance and {Moscardini}, Lauro and {Napolitano}, Nicola R. and {Porth}, Lucas and {Radovich}, Mario and {Schneider}, Peter and {Tr{\"o}ster}, Tilman and {Valentijn}, Edwin and {Wittje}, Anna and {Yan}, Ziang and {Zhang}, Yun-Hao},
        title = "{KiDS-Legacy: Cosmological constraints from cosmic shear with the complete Kilo-Degree Survey}",
      journal = {\aap},
         year = 2025,
        month = nov,
       volume = {703},
          eid = {A158},
        pages = {A158},
          doi = {10.1051/0004-6361/202554908},
archivePrefix = {arXiv},
       eprint = {2503.19441},
 primaryClass = {astro-ph.CO},
       adsurl = {https://ui.adsabs.harvard.edu/abs/2025A&A...703A.158W}
}

@ARTICLE{wright2025_calibration,
       author = {{Wright}, Angus H. and {Hildebrandt}, Hendrik and {van den Busch}, Jan Luca and {Bilicki}, Maciej and {Heymans}, Catherine and {Joachimi}, Benjamin and {Mahony}, Constance and {Reischke}, Robert and {St{\"o}lzner}, Benjamin and {Wittje}, Anna and {Asgari}, Marika and {Chisari}, Nora Elisa and {Dvornik}, Andrej and {Georgiou}, Christos and {Giblin}, Benjamin and {Hoekstra}, Henk and {Jalan}, Priyanka and {William}, Anjitha John and {Joudaki}, Shahab and {Kuijken}, Konrad and {Lesci}, Giorgio Francesco and {Li}, Shun-Sheng and {Linke}, Laila and {Loureiro}, Arthur and {Maturi}, Matteo and {Moscardini}, Lauro and {Porth}, Lucas and {Radovich}, Mario and {Tr{\"o}ster}, Tilman and {von Wietersheim-Kramsta}, Maximilian and {Yan}, Ziang and {Yoon}, Mijin and {Zhang}, Yun-Hao},
        title = "{KiDS-Legacy: Redshift distributions and their calibration}",
      journal = {\aap},
         year = 2025,
        month = nov,
       volume = {703},
          eid = {A144},
        pages = {A144},
          doi = {10.1051/0004-6361/202554909},
archivePrefix = {arXiv},
       eprint = {2503.19440},
 primaryClass = {astro-ph.CO},
       adsurl = {https://ui.adsabs.harvard.edu/abs/2025A&A...703A.144W}
}

@ARTICLE{Stolzner_2021,
       author = {{St{\"o}lzner}, B. and {Joachimi}, B. and {Korn}, A. and {Hildebrandt}, H. and {Wright}, A.~H.},
        title = "{Self-calibration and robust propagation of photometric redshift distribution uncertainties in weak gravitational lensing}",
      journal = {\aap},
         year = 2021,
        month = jun,
       volume = {650},
          eid = {A148},
        pages = {A148},
          doi = {10.1051/0004-6361/202040130},
archivePrefix = {arXiv},
       eprint = {2012.07707},
 primaryClass = {astro-ph.CO},
       adsurl = {https://ui.adsabs.harvard.edu/abs/2021A&A...650A.148S}
}

@ARTICLE{dAssignies25,
       author = {{d'Assignies}, W. and {Manera}, M. and {Padilla}, C. and {Ilbert}, O. and {Hildebrandt}, H. and {Reynolds}, L. and {Chaves-Montero}, J. and {Wright}, A.~H. and {Tallada-Cresp{\'\i}}, P. and {Eriksen}, M. and {Carretero}, J. and {Roster}, W. and {Kang}, Y. and {Naidoo}, K. and {Miquel}, R. and {Altieri}, B. and {Amara}, A. and {Andreon}, S. and {Auricchio}, N. and {Baccigalupi}, C. and {Bagot}, D. and {Baldi}, M. and {Balestra}, A. and {Bardelli}, S. and {Battaglia}, P. and {Biviano}, A. and {Branchini}, E. and {Brescia}, M. and {Camera}, S. and {Capobianco}, V. and {Carbone}, C. and {Cardone}, V.~F. and {Casas}, S. and {Castander}, F.~J. and {Castellano}, M. and {Castignani}, G. and {Cavuoti}, S. and {Chambers}, K.~C. and {Cimatti}, A. and {Colodro-Conde}, C. and {Congedo}, G. and {Conselice}, C.~J. and {Conversi}, L. and {Copin}, Y. and {Courbin}, F. and {Courtois}, H.~M. and {Crocce}, M. and {Da Silva}, A. and {Degaudenzi}, H. and {de la Torre}, S. and {De Lucia}, G. and {Douspis}, M. and {Dupac}, X. and {Ealet}, A. and {Escoffier}, S. and {Farina}, M. and {Faustini}, F. and {Ferriol}, S. and {Finelli}, F. and {Fosalba}, P. and {Fotopoulou}, S. and {Frailis}, M. and {Franceschi}, E. and {Fumana}, M. and {Galeotta}, S. and {George}, K. and {Gillis}, B. and {Giocoli}, C. and {G{\'o}mez-Alvarez}, P. and {Gracia-Carpio}, J. and {Grazian}, A. and {Grupp}, F. and {Holmes}, W. and {Hook}, I.~M. and {Hornstrup}, A. and {Jahnke}, K. and {Jhabvala}, M. and {Joachimi}, B. and {Keih{\"a}nen}, E. and {Kermiche}, S. and {Kiessling}, A. and {Kubik}, B. and {K{\"u}mmel}, M. and {Kunz}, M. and {Kurki-Suonio}, H. and {Lahav}, O. and {Le Brun}, A.~M.~C. and {Ligori}, S. and {Lilje}, P.~B. and {Lindholm}, V. and {Lloro}, I. and {Mainetti}, G. and {Maino}, D. and {Maiorano}, E. and {Mansutti}, O. and {Marcin}, S. and {Marggraf}, O. and {Markovic}, K. and {Martinelli}, M. and {Martinet}, N. and {Marulli}, F. and {Massey}, R. and {Masters}, D.~C. and {Medinaceli}, E. and {Mei}, S. and {Melchior}, M. and {Mellier}, Y. and {Meneghetti}, M. and {Merlin}, E. and {Meylan}, G. and {Mora}, A. and {Moresco}, M. and {Moscardini}, L. and {Neissner}, C. and {Niemi}, S.-M. and {Paltani}, S. and {Pasian}, F. and {Pedersen}, K. and {Pettorino}, V. and {Pires}, S. and {Polenta}, G. and {Poncet}, M. and {Popa}, L.~A. and {Pozzetti}, L. and {Raison}, F. and {Rebolo}, R. and {Renzi}, A. and {Rhodes}, J. and {Riccio}, G. and {Romelli}, E. and {Roncarelli}, M. and {Rossetti}, E. and {Saglia}, R. and {Sakr}, Z. and {Sapone}, D. and {Sartoris}, B. and {Schewtschenko}, J.~A. and {Schneider}, P. and {Schrabback}, T. and {Secroun}, A. and {Sefusatti}, E. and {Seidel}, G. and {Seiffert}, M. and {Serrano}, S. and {Simon}, P. and {Sirignano}, C. and {Sirri}, G. and {Spurio Mancini}, A. and {Stanco}, L. and {Steinwagner}, J. and {Tavagnacco}, D. and {Taylor}, A.~N. and {Teplitz}, H.~I. and {Tereno}, I. and {Tessore}, N. and {Toft}, S. and {Toledo-Moreo}, R. and {Torradeflot}, F. and {Tsyganov}, A. and {Tutusaus}, I. and {Valenziano}, L. and {Valiviita}, J. and {Vassallo}, T. and {Verdoes Kleijn}, G. and {Wang}, Y. and {Weller}, J. and {Zamorani}, G. and {Zucca}, E. and {Bolzonella}, M. and {Burigana}, C. and {Gabarra}, L. and {Mart{\'\i}n-Fleitas}, J. and {Risso}, I. and {Scottez}, V. and {Viel}, M.},
        title = "{Euclid: Photometric redshift calibration performance with the clustering-redshifts technique in the Flagship 2 simulation}",
      journal = {\aap},
         year = 2025,
        month = oct,
       volume = {702},
          eid = {A155},
        pages = {A155},
          doi = {10.1051/0004-6361/202555551},
archivePrefix = {arXiv},
       eprint = {2505.10416},
 primaryClass = {astro-ph.CO},
       adsurl = {https://ui.adsabs.harvard.edu/abs/2025A&A...702A.155D}
}

@ARTICLE{Cunha_2009,
       author = {{Cunha}, Carlos E. and {Lima}, Marcos and {Oyaizu}, Hiroaki and {Frieman}, Joshua and {Lin}, Huan},
        title = "{Estimating the redshift distribution of photometric galaxy samples - II. Applications and tests of a new method}",
      journal = {\mnras},
         year = 2009,
        month = jul,
       volume = {396},
       number = {4},
        pages = {2379-2398},
          doi = {10.1111/j.1365-2966.2009.14908.x},
archivePrefix = {arXiv},
       eprint = {0810.2991},
 primaryClass = {astro-ph},
       adsurl = {https://ui.adsabs.harvard.edu/abs/2009MNRAS.396.2379C}
}

@ARTICLE{Zuntz_2021,
       author = {{Zuntz}, Joe and {Lanusse}, Fran{\c{c}}ois and {Malz}, Alex I. and {Wright}, Angus H. and {Slosar}, An{\v{z}}e and {Abolfathi}, Bela and {Alonso}, David and {Bault}, Abby and {Bom}, Cl{\'e}cio R. and {Brescia}, Massimo and {Broussard}, Adam and {Campagne}, Jean-Eric and {Cavuoti}, Stefano and {Cypriano}, Eduardo S. and {Fraga}, Bernardo M.~O. and {Gawiser}, Eric and {Gonzalez}, Elizabeth J. and {Green}, Dylan and {Hatfield}, Peter and {Iyer}, Kartheik and {Kirkby}, David and {Nicola}, Andrina and {Nourbakhsh}, Erfan and {Park}, Andy and {Teixeira}, Gabriel and {Heitmann}, Katrin and {Kovacs}, Eve and {Mao}, Yao-Yuan and {LSST Dark Energy Science Collaboration}},
        title = "{The LSST-DESC 3x2pt Tomography Optimization Challenge}",
      journal = {The Open Journal of Astrophysics},
         year = 2021,
        month = oct,
       volume = {4},
       number = {1},
          eid = {13},
        pages = {13},
          doi = {10.21105/astro.2108.13418},
archivePrefix = {arXiv},
       eprint = {2108.13418},
 primaryClass = {astro-ph.IM},
       adsurl = {https://ui.adsabs.harvard.edu/abs/2021OJAp....4E..13Z}
}

@article{Duncan_2022,
    author = {Duncan, Kenneth J},
    title = {All-purpose, all-sky photometric redshifts for the Legacy Imaging Surveys Data Release 8},
    journal = {MNRAS},
    volume = {512},
    number = {3},
    pages = {3662-3683},
    year = {2022},
    month = {03},
    issn = {0035-8711},
    doi = {10.1093/mnras/stac608},
    url = {https://doi.org/10.1093/mnras/stac608},
    eprint = {https://academic.oup.com/mnras/article-pdf/512/3/3662/43286902/stac608.pdf},
}

@article{Buchs_2019,
    author = {Buchs, R and Davis, C and Gruen, D and DeRose, J and Alarcon, A and Bernstein, G M and Sánchez, C and Myles, J and Roodman, A and Allen, S and Amon, A and Choi, A and Masters, D C and Miquel, R and Troxel, M A and Wechsler, R H and Abbott, T M C and Annis, J and Avila, S and Bechtol, K and Bridle, S L and Brooks, D and Buckley-Geer, E and Burke, D L and Carnero Rosell, A and Carrasco Kind, M and Carretero, J and Castander, F J and Cawthon, R and D’Andrea, C B and da Costa, L N and De Vicente, J and Desai, S and Diehl, H T and Doel, P and Drlica-Wagner, A and Eifler, T F and Evrard, A E and Flaugher, B and Fosalba, P and Frieman, J and García-Bellido, J and Gaztanaga, E and Gruendl, R A and Gschwend, J and Gutierrez, G and Hartley, W G and Hollowood, D L and Honscheid, K and James, D J and Kuehn, K and Kuropatkin, N and Lima, M and Lin, H and Maia, M A G and March, M and Marshall, J L and Melchior, P and Menanteau, F and Ogando, R L C and Plazas, A A and Rykoff, E S and Sanchez, E and Scarpine, V and Serrano, S and Sevilla-Noarbe, I and Smith, M and Soares-Santos, M and Sobreira, F and Suchyta, E and Swanson, M E C and Tarle, G and Thomas, D and Vikram, V and (DES Collaboration)},
    title = {Phenotypic redshifts with self-organizing maps: A novel method to characterize redshift distributions of source galaxies for weak lensing},
    journal = {MNRAS},
    volume = {489},
    number = {1},
    pages = {820-841},
    year = {2019},
    month = {08},
    issn = {0035-8711},
    doi = {10.1093/mnras/stz2162},
    url = {https://doi.org/10.1093/mnras/stz2162},
    eprint = {https://academic.oup.com/mnras/article-pdf/489/1/820/29218653/stz2162.pdf},
}

@ARTICLE{Coupon_2009,
       author = {{Coupon}, J. and {Ilbert}, O. and {Kilbinger}, M. and {McCracken}, H.~J. and {Mellier}, Y. and {Arnouts}, S. and {Bertin}, E. and {Hudelot}, P. and {Schultheis}, M. and {Le F{\`e}vre}, O. and {Le Brun}, V. and {Guzzo}, L. and {Bardelli}, S. and {Zucca}, E. and {Bolzonella}, M. and {Garilli}, B. and {Zamorani}, G. and {Zanichelli}, A. and {Tresse}, L. and {Aussel}, H.},
        title = "{Photometric redshifts for the CFHTLS T0004 deep and wide fields}",
      journal = {\aap},
         year = 2009,
        month = jun,
       volume = {500},
       number = {3},
        pages = {981-998},
          doi = {10.1051/0004-6361/200811413},
archivePrefix = {arXiv},
       eprint = {0811.3326},
 primaryClass = {astro-ph},
       adsurl = {https://ui.adsabs.harvard.edu/abs/2009A&A...500..981C}
}

@ARTICLE{Tucci_2025,
       author = {{Euclid Collaboration: Tucci}, M. and {Paltani}, S. and {Hartley}, W.~G. and others},
        title = "{Euclid Quick Data Release (Q1). Photometric redshifts and physical properties of galaxies through the PHZ processing function}",
      journal = {A\&A, in press (Euclid Q1 SI), \url{https://doi.org/10.1051/0004-6361/202554588}},
         year = 2025,
        month = mar,
          eid = {arXiv:2503.15306},
        pages = {arXiv:2503.15306},
archivePrefix = {arXiv},
       eprint = {2503.15306},
 primaryClass = {astro-ph.GA},
       adsurl = {https://ui.adsabs.harvard.edu/abs/2025arXiv250315306E}
}

@ARTICLE{Scaramella-EP1,
       author = {{Euclid Collaboration: Scaramella}, R. and {Amiaux}, J. and {Mellier}, Y. and others},
        title = "{Euclid preparation. I. The Euclid Wide Survey}",
      journal = {\aap},
         year = 2022,
        month = jun,
       volume = {662},
          eid = {A112},
        pages = {A112},
          doi = {10.1051/0004-6361/202141938},
archivePrefix = {arXiv},
       eprint = {2108.01201},
 primaryClass = {astro-ph.CO},
       adsurl = {https://ui.adsabs.harvard.edu/abs/2022A&A...662A.112E}
}

%
%

\begin{appendix}

\section{Self-organising maps}
\label{TA}

The self-organising map can be used as a highly effective tool for calibrating redshift distributions. Its mechanism, initially proposed by \cite{Lima_2008}, is based on the assumption that galaxy samples with similar colour-space distributions resemble alike underlying redshift distributions. In this section, we introduce the working principle of the SOM.

A SOM consists of a fixed number of cells arranged on a grid of arbitrary, but commonly two, dimensions. At initialisation, the nodes, neurons, or cells of the (2D) map, which typically are arranged as a grid of hexagonal units, are appointed a (random) weight vector $w$ of the same dimension as an input training vector $x$ \citep{H_johnston2021}. In the case of a galaxy survey for example, $x$ could comprise five measurements of $ugriz$ photometry \citep{Geach_2011}. This vector denotes the relative position of each node in the multi-dimensional parameter space occupied by the data. Training commences by presenting the map with a set of data points which the cells compete for. Distance metrics determine the winner neuron for any given training object \citep{Masters_2015}. The distance metric $d_{j}(t)$ of the cell weight vector $w$ to an input vector $x$, often referred to as Euclidean distance \citep{Geach_2011}, is defined as
\begin{equation}
    \label{ed}
    d_{j}(t) \coloneqq \sqrt{\sum_{k=1}^{n} \left[x_{k}(t) - w_{j,k}(t)\right]^{2}}
\end{equation}
at a given time step $t$, with $j$ referring to the neuron in question, $k$ being the data attribute, and $n$ the total number of attributes for any input vector $x \in \mathbb{R}^{n}$. The neuron whose vector best resembles any given data point is henceforth referred to as best matching unit (BMU). A visual display in form of a sketch is given by \cref{fig:2.01}.

\begin{figure}[ht!]
   \centering
   \includegraphics[width=9.6 cm]{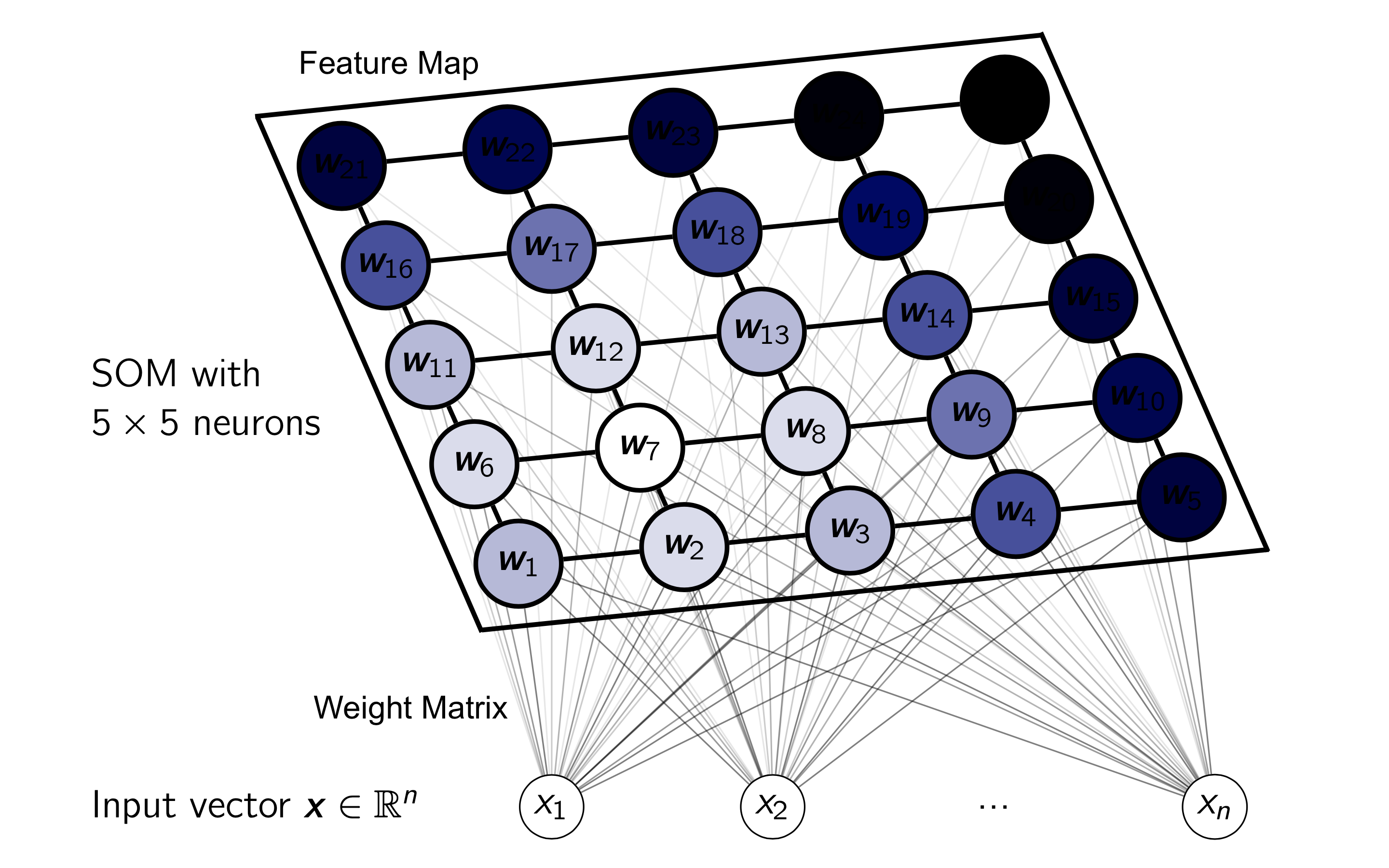}
   \caption[]{Architecture of a 5$\times$5 neuron SOM with a Kohonen layer, depicted here as `Feature Map'. The input vector $x$ is mapped onto the neurons, represented by their weight vectors $w_{i}$. The best matching unit (here $w_{7}$) and the surrounding weights are adjusted according to the current learning rate and neighbourhood function. The neurons in this example are coloured based on the magnitude of the adjustment in the current training step, where white are high magnitudes of adjustment and black are low magnitudes. Adapted from \cite{vdb_2021}.}
    \label{fig:2.01}
    \end{figure}

Because this metric treats all input dimensions equally, features with inherently larger uncertainties end up disproportionately influencing the results. To address this, the algorithm reduces the high-dimensional input space, regardless of its original dimensionality, into a single scalar value. This dimensionality reduction ensures that relative distances between neurons meaningfully reflect the structure of the data, enabling an effective 2D mapping of the populated SOM \citep{H_johnston2021}. Even so, the fundamental property which distinctly separates SOMs from other unsupervised learning techniques, is the fact that at each iteration, the weights of the BMU are adjusted, along with smaller updates applied to the neighboring neurons’ weights \citep{Carrasco_Kind_2013, Davidzon_2019}. By reweighing the importance of every neuron per data point, neurons of better representation receive greater updates than negligible neurons. This evaluation of the relative influence between neurons is expressed via the topological neighbourhood function
\begin{equation}
     T_{\text{$a,B(x)$}}(e) = \text{exp} \ \left(-\frac{S^{2}_{\text{$a,B(x)$}}}{2\sigma(t)^{2}}\right)\,
\end{equation}
which, depending on the epoch $e$, can be encoded as a normalised Gaussian kernel centred on the BMU \citep{Masters_2015,Campos_2024}. The parameter $S$ denotes the lateral distance between cell $a$ and the winning neuron $B(x)$, 
\begin{equation}
    S_{\textrm{$a,B(x)$}} = |w_{a} - w_{B(x)}|\;.
\end{equation}
The width of the Gaussian neighbourhood $\sigma(t)$ is fixed to 
\begin{equation}
    \sigma(t) = \sigma_{\mathrm{init}}\left(\frac{1}{\sigma_{\mathrm{init}}}\right)^{\frac{t}{N}}\;,
\end{equation}
where $N$ refers to the number of training iterations. Initially, $\sigma_{\mathrm{init}}$ is chosen large enough to cover most of the map. The width itself shrinks during training so that by the end of this process, only the BMU and directly adjacent cells are (significantly) affected by new data \citep{Masters_2015}. After identifying the BMU per iteration, the weight vector is updated by equations
\begin{equation}
    \Delta w_{a} (t) = \eta (e) \, T_{a,B(x)}(e) \, \left[x (t)-w_{a}(t)\right]\;
\end{equation}
and 
\begin{equation}
    w_{a} (t+1) = w_{a} (t) + \Delta w_{a}(t)\;,
\end{equation}
with $a$ resembling a neuron, $\eta$ the monotonically time decreasing learning rate, and $B(x)$ the winning neuron. As such, the SOM becomes progressively less responsive to new training data per iteration. The topological neighbourhood $T$ is then a normalised measure for every neurons weight adaptation. This becomes clearer if one assumes two neurons, $a$ and $b$, to both be the winner neuron. In this case, the lateral distance is $S=0$ and therefore $T(e)=1$. After training, the map can be used to categorise previously unseen data, by appointing the closest weight vector to a given data point, based on the learned representation \citep{Davidzon_2019}.

\section{Covariate shift}
\label{cv}
One of the most frequent sources of error encountered in statistical analysis with ML is caused by covariate shift \citep{luo2022bayesian}. This shift refers to the difference in feature space between the calibration and validation samples. In redshift calibration, this difference arises from the fact that the methods with which the samples are selected are fundamentally different \citep{Beck_2017}. It is important to note that the definition of the redshift tomography and subsequent selection of the samples introduce additional bias. Using Flagship2 mock data (see \cref{flagship}), the upper panel in \cref{fig:2.02} features the true $n(z)$ of a wide-like validation sample represented by the orange histogram. In contrast, the grey histogram depicts the $n(z)$ of a spectroscopic calibration sample, taking into account a non-trivial selection function, specifically cuts made in colour-colour space. The purple and green histograms in the upper panel respectively portray the $n(z)$ of galaxies of a random cell that belong to the validation sample and to the calibration sample, respectively. Clearly, the green distribution in \cref{fig:2.02} is inadequate for calibrating the purple distribution attributed to the selected cell. This highlights the need for caution when considering the mean spec-$z$s of SOM cells as metric to assign galaxies to their respective tomographic redshift bins, given that redshift bias is driven by the discrepancy between the green and purple histograms. However, if an identical photo-$z$ binning (here 0.5 $\leq z_{\rm p} \leq$ 0.7) is applied to both samples, as done for the brown and red distributions in the lower panel of \cref{fig:2.02}, only representative sources in redshift are used for calibration purposes. In doing so, the discrepancy in distributions is drastically reduced. Applying a one-sided (spectroscopic-only) cut would change the feature space of the training sample \citep{autenrieth2021stratified}. Because the spectroscopic $n(z)$ aligns with the colour-based structure of the SOM grid, cells with low mean photo-$z$s may inadvertently be used to calibrate high-redshift galaxies if photo-$z$ binning is not enforced. Instead, when both cell distributions are divided into photo-$z$ bins first, every cell can, in principle, be used for calibration.  

\begin{figure}[t!]
\centering
\includegraphics[width=9cm]{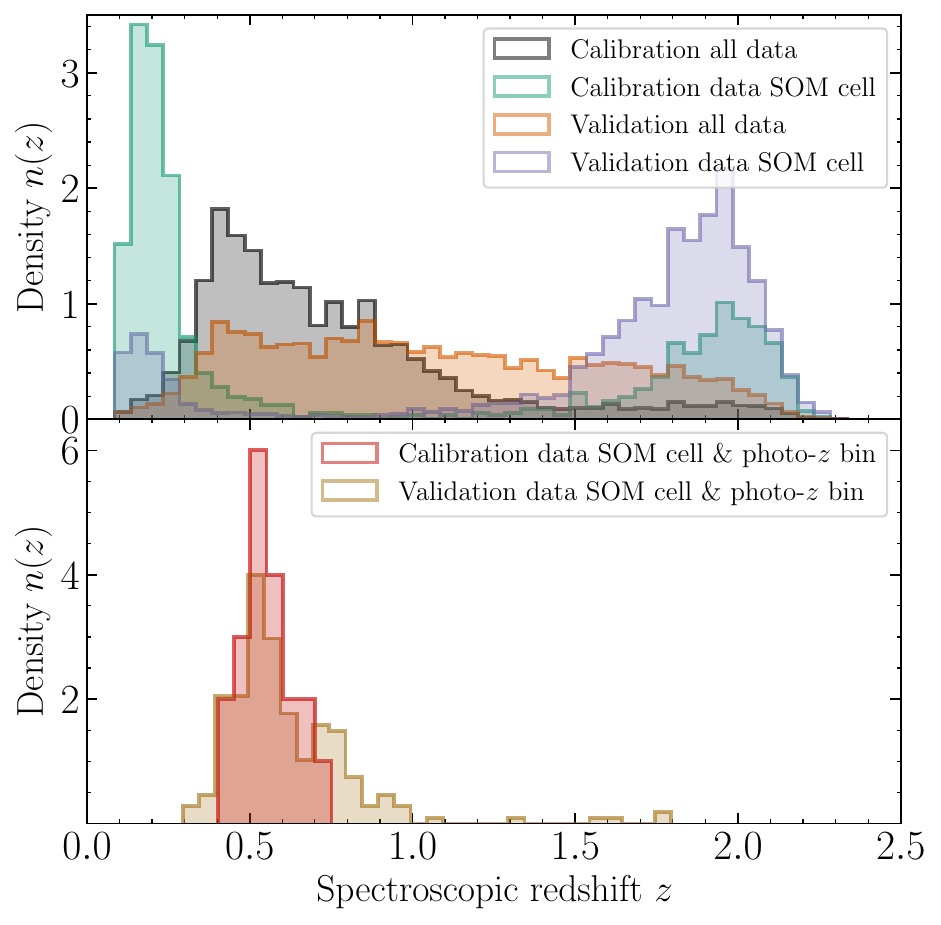}
\caption[]{\emph{Top}: Mean spec-$z$ distributions for the validation and calibration sample subject to a non-trivial selection function. The respective distributions are also depicted for a single SOM cell chosen at random. \emph{Bottom}: Display of the distributions found for the same randomly selected cell after applying photo-$z$ binning.}
\label{fig:2.02}
\end{figure}

\section{Improved realism}
\label{IR}

While the majority of bins listed in \cref{tab:6.9} meet the requirement set by \Euclid, their biases are somewhat optimistic. The calibration sample used to train the SOM undergoes a less complex selection compared to the selection boundaries faced by real observations, as addressed by \cite{Masters_2015, Masters_2017, Masters_2019}. This pertains to the calibration sample generation process, where sources are selected from a SOM trained on noiseless and therefore unrealistic photometry, resulting in a more uniform distribution in colour-colour space. Consequently, the current calibration sample used for training includes distribution counts for certain areas in the colour-colour space with limited or no reliably observed sources. Hence, it would be imprudent to assume that \Euclid will achieve the same level of accuracy as indicated by these predictions. It is therefore important to enhance the realism of the calibration data to a point where the predictions made within this work become reliably applicable to \Euclid. Given that C3R2 \citep{Masters_2017} currently stands as the most extensive spectroscopic sample used for calibration efforts, we test our pipeline on various varied calibration samples by changing the sampling process itself in order to compare subsequent distributions to those found with C3R2. To achieve this, we utilise SOM statistics extracted from \cite{Masters_2019}, including cell ID, median spec-$z$, associated colours and the count of reliable spectra where the quality flag ($Q$) is set to $Q=4$ in each cell. It is worth noting that this initial approach is somewhat conservative, as it does not yet incorporate spectra from other surveys beyond C3R2, which could contribute to the calibration, nor does it include less reliable spectra currently classified as $Q=3$. By leveraging the available information, we can simulate a set of more realistic spectroscopic calibration samples. In the pursuit of robust calibration, we propose four distinct approaches, based in part on the distributions found in C3R2, to alternate samples for calibration purposes.

\begin{figure}[t!]
\centering
\includegraphics[width=9cm]{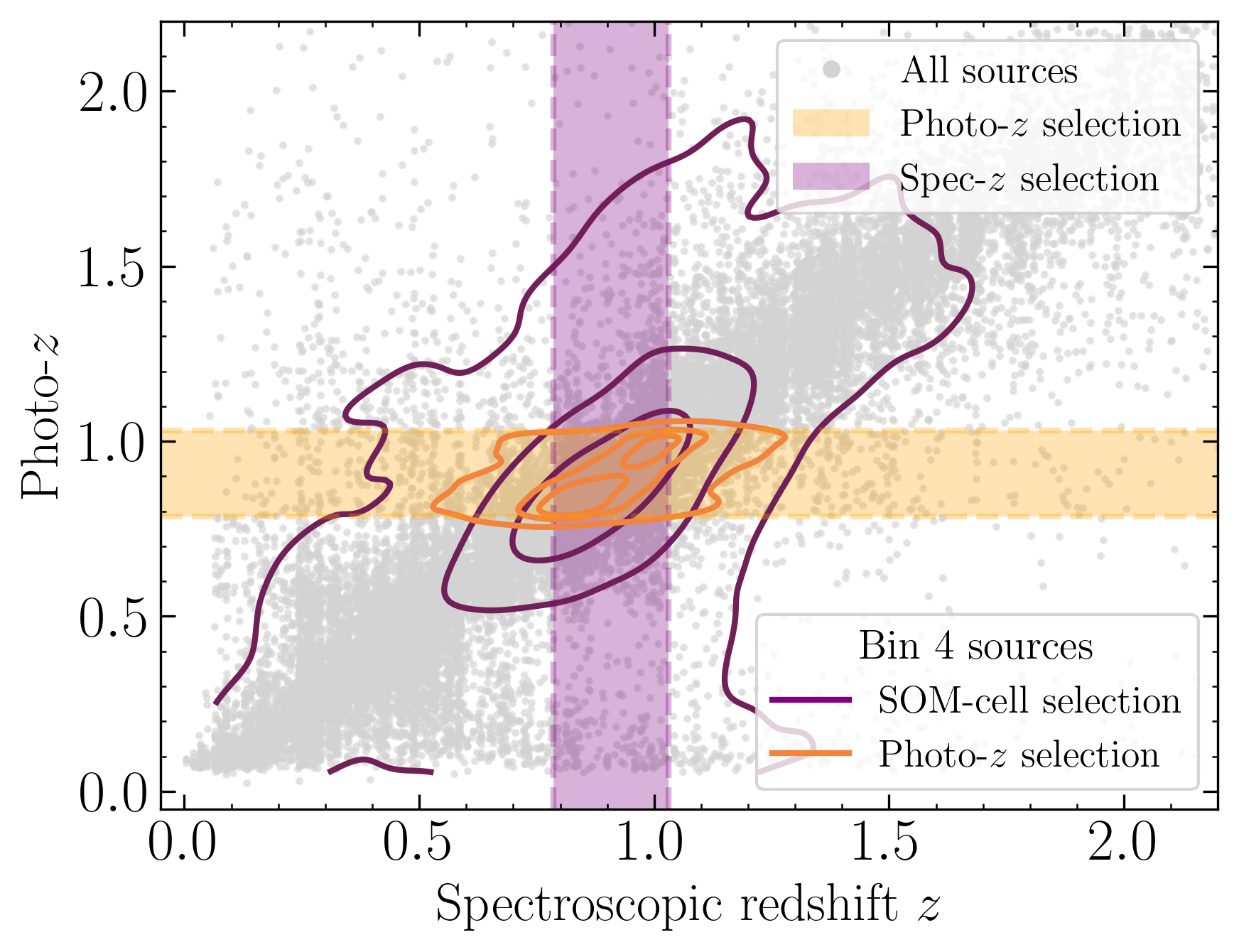}
\caption[]{Distribution of photo-$z$ over spec-$z$, where the full sample is shown in grey, and the shaded areas represent the fourth tomographic bin listed in \cref{tab:6.4} for photometric and spectroscopic redshift (orange and purple bands, respectively). Overlaid are 1, 2, and 3$\sigma$ completeness contours for the two selection strategies: photo-$z$ binning within the SOM (orange) and mean spec-$z$ binning per SOM cell (purple).}
\label{fig:6.7}
\end{figure}

\subsection{Pathological spec-\textit{z} calibration sample \& mean spec-\textit{z} SOM cell tomography}
\label{m1}

The first among these samples is what we colloquially refer to as the pathological calibration sample. This sample intentionally comprises catastrophic instances that deviate markedly from the characteristics observed in both the established Masters SOM calibration sample and our prior calibration sample (see \cref{SC}). Notably, it introduces challenges by incorporating a substantial redshift desert where no data points exist. By examining the calibration performance on this extreme sample, we aim to assess the SOMs robustness and its ability to handle unconventional data distributions. The redshift distributions are depicted in \cref{fig:6.13}. This new calibration sample is then used to train a noisy SOM with bias calculations based on tomography defined on the mean spec-$z$ of SOM cells as introduced in \cref{SC}. As with \cref{fig:6.10}, we observe that none of the bins satisfy the \Euclid requirement (see \cref{fig:A.1} and \cref{tab:A.1}). Notably, we observe larger mean biases accompanied by even larger standard deviations with distributions at times characterised by extreme outliers.

\subsection{Masters-like photo-\textit{z} calibration sample \& mean spec-\textit{z} SOM cell tomography}
\label{m2}

Compared to \cref{m1}, we adjust the $n(z)$ of our calibration sample to mimic the mean photo-$z$ distribution of the Masters calibration sample. This requires binning the redshift range [$0.07 \leq z_{\rm obs} \leq 1.41$] spanned by the sample into 245 equidistant slices, referred to as redshift blocks. This ensures that each block contains $\geq$ 1 cell, aiming for the highest level of fidelity. Reshaping the calibration sample $n(z)$ by updating the way in which we sample the trained noiseless SOM, involves examining the ratio of sources found for each redshift block, 
\begin{equation}
    f_{\rm rb} = \frac{\text{Number of galaxies noiseless SOM}}{\text{Number of galaxies Masters SOM}}\, , 
\end{equation}
for both, the Masters and noiseless SOM. Based on $f_{\rm rb}$, we proceed to resample the noiseless SOM per redshift block. In the cases where $f_{\rm rb}<1$, up to four adjacent blocks are included, increasing the number of spectra considered for sampling, provided their combined respective fraction is $f_{\rm rb}\geq 1$. Finally, any duplicates are removed from the modified distribution of the calibration sample. The new training sample (see \cref{fig:6.13}) is used to train a noisy SOM with bias calculation based on tomography following \cref{SC}. The bias results are shown in \cref{fig:A.2} and \cref{tab:A.2}. We drastically reduce the biases per tomographic redshift bin compared to \cref{m1}, showcasing the importance of both a representative training sample in ML applications and spectroscopic coverage in the commonly less accessible colour-colour space. 

\begin{figure}[t!]
\centering
\includegraphics[width=9cm]{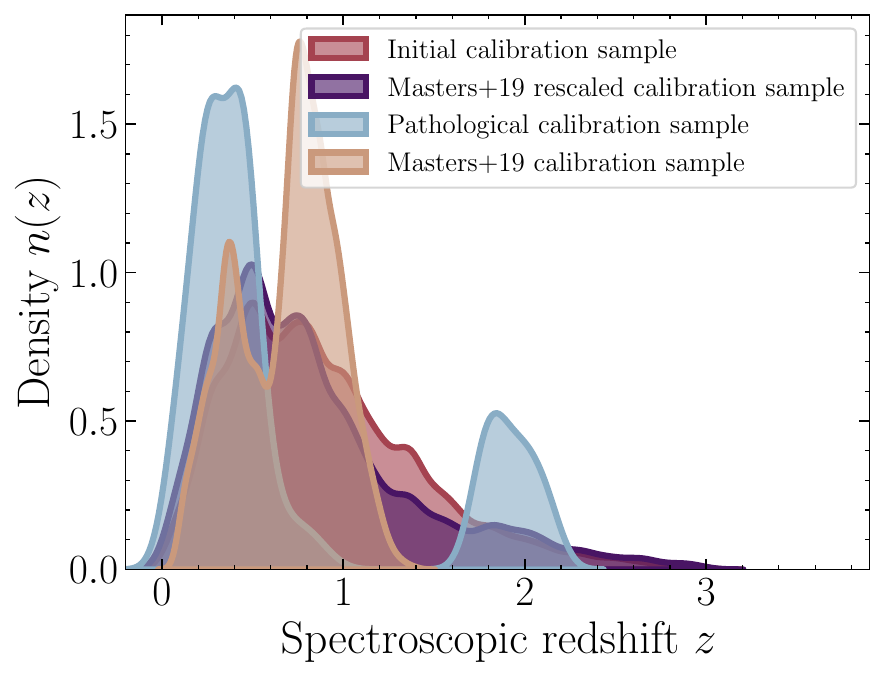}
\caption[]{Display of the $n(z)$ distributions of the calibration sample introduced in \cref{SC} (red), the Masters SOM cell mean specz-$z$ (brown) as well as \cref{m2} (blue) and \cref{m3} (purple) calibration samples.}
\label{fig:6.13}
\end{figure}

\subsection{Masters-like photo-\textit{z} calibration sample \& photo-\textit{z} tomography}
\label{m3}

The calibration sample remains the same as defined in \cref{m2}. However, in this instance, the calibration sample (consistent with that of \cref{m2}) is used to train a noisy SOM with bias calculation determined by tomography defined on photo-$z$ rather than mean spec-$z$ of SOM cells, as elucidated in \cref{PC}. The obtained biases are presented in \cref{fig:A.3} and \cref{tab:A.3}. Considering QC, we achieve sufficient biases in six bins, with results comparable to those obtained in \cref{tab:6.9}. Contrary to before, however, we observe smaller fluctuations, potentially due to the calibration sample distribution. We attribute the notable increase in performance compared to \cref{m2} to the implementation of photo-$z$ tomography, which in all scenarios presented delivers much reduced biases compared to the tomography most commonly defined on the SOM cell mean spec-$z$.

\subsection{Colour-colour sampling \& photo-\textit{z} tomography}

While redshift biases achieved in \cref{m3} outperform respective biases found in \cref{m1} and \cref{m2}, the selection and approach of simulating a ``very realistic” spec-$z$ calibration sample is based exclusively on the Masters SOM statistics. Since SOM cells can provide the mean or median of the underlying data distributions they represent, deviations and rounding are introduced, which prevent the new calibration sample from being a perfect representation of the Masters SOM distribution. In other words, the available Masters data provides feature-averaged statistics only. An improved method would therefore require the entire calibration sample catalogue, containing all galaxy attributes, used to train the Masters SOM in order to use multi-dimensional colour-colour space information. However, this poses a challenging task, as it entails aligning two multi-dimensional colour-colour spaces between SOMs and subsequently translating the colour-redshift relation.

\subsection{Future optimisation}
\label{FI}

A strategy to mitigate data loss due to QC involves the clustering of cells. This method avoids the exclusion of EWS-like data located in cells with no spectroscopic calibration data available, by merging them with cells carrying contributions from both samples in nearby colour-colour space \citep{van_den_Busch_2022}. As demonstrated in the Appendix of \cite{Wright_2020b}, biases per bin can be computed as a function of clusters ($N$\textsubscript{clust}), selecting $N$\textsubscript{clust} to maximise data completeness while minimising the bias. One promising approach involves reconstructing the true $n(z)$ distribution for each SOM cell. This two-step process involves initially distributing sources based on the maps topology, followed by refining each cell by uncovering the true $n(z)$ of its sources. Lastly, it seems that the relationship between biases may contain additional information that could be valuable in constraining and mitigating spectroscopic bias. Finally, it may be useful to examine redshifted SEDs of individual low-$Q$ galaxies, as they impose an extra selection on the true $n(z)$. Removing or correcting them via SED-based training selection could potentially improve bias estimates.

\begin{table*}[ht!]
\centering
\caption[]{Biases in the mean redshift estimation per bin according to \cref{m1}.}
\vspace{10pt}
\begin{tabular}{cccccc} 
  \hline\hline\noalign{\vskip 1.5pt}
  Bias ($\Delta\langle z \rangle$) & bin 1  & bin 2  & bin 3  &bin 4  &bin 5  \\ 
 \hline
 QC no & 0.219 $\pm$ 0.093 & 0.182 $\pm$ 0.069 & 0.173 $\pm$ 0.043 & 0.211 $\pm$ 0.205 &0.171 $\pm$ 0.177 \\
QC yes & 0.083 $\pm$ 0.014 & 0.103 $\pm$ 0.033 & 0.152 $\pm$ 0.028 & 0.137 $\pm$ 0.076 & 0.113 $\pm$ 0.142
\end{tabular}

\vspace{10pt}
\begin{tabular}{ccccc} 
\hline\hline\noalign{\vskip 1.5pt}
 Bias ($\Delta\langle z \rangle$)& bin 6  & bin 7  & bin 8  &bin 9   \\
 \hline
 QC no & 0.006 $\pm$ 0.223 & $-$0.139 $\pm$ 0.163 & $-$0.308 $\pm$ 0.199 & $-$0.275 $\pm$ 0.199 \\
 QC yes &  $-$0.025 $\pm$ 0.176 &$-$0.097 $\pm$ 0.160 & $-$0.205 $\pm$ 0.153 & $-$0.096 $\pm$ 0.141 \\
\end{tabular}
\tablefoot{The values shown are the mean biases over 60 different lines of sight as well as standard deviations for the lowest nine bins. Under the recognition of QC, none of the bins lie within the \Euclid bias requirement.}
\label{tab:A.1}
\end{table*}


\begin{table*}[ht!]
\centering
\caption[]{Biases in the mean redshift estimation per bin according to \cref{m2}. }
\vspace{10pt}
\begin{tabular}{cccccc} 
  \hline\hline\noalign{\vskip 1.5pt}
  Bias ($\Delta\langle z \rangle$) & bin 1  & bin 2  & bin 3  &bin 4  &bin 5  \\ 
 \hline
 QC no & 0.077 $\pm$ 0.023 & 0.035 $\pm$ 0.028 & 0.033 $\pm$ 0.028 & 0.006 $\pm$ 0.017 &0.003 $\pm$ 0.014 \\
QC yes & 0.059 $\pm$ 0.012 & 0.030 $\pm$ 0.025 & 0.031 $\pm$ 0.028 & 0.006 $\pm$ 0.016 & 0.007 $\pm$ 0.018
\end{tabular}

\vspace{10pt}
\begin{tabular}{ccccc} 
\hline\hline\noalign{\vskip 1.5pt}
 Bias ($\Delta\langle z \rangle$)& bin 6  & bin 7  & bin 8  &bin 9   \\
 \hline
 QC no & 0.009 $\pm$ 0.018 & $-$0.034 $\pm$ 0.060 & $-$0.044 $\pm$ 0.048 & $-$0.063 $\pm$ 0.024 \\
 QC yes &  0.020 $\pm$ 0.014 &$-$0.009 $\pm$ 0.035 & $-$0.018 $\pm$ 0.032 & $-$0.033 $\pm$ 0.021 \\
\end{tabular}
\tablefoot{The values shown are the mean biases over 100 different lines of sight as well as standard deviations for the lowest nine bins. Under the recognition of QC, bins 2, 3, and 5 lie within the \Euclid bias requirement.}
\label{tab:A.2}
\end{table*}

\begin{table*}[ht!]
\centering
\caption[]{Biases in the mean redshift estimation per bin according to \cref{m3}. }
\vspace{10pt}
\begin{tabular}{cccccc} 
  \hline\hline\noalign{\vskip 1.5pt}
  Bias ($\Delta\langle z \rangle$) & bin 1  & bin 2  & bin 3  &bin 4  &bin 5  \\ 
 \hline
 QC no & 0.003 $\pm$ 0.012 & 0.003 $\pm$ 0.011 & $\emptyset$ $\pm$ 0.008 & 0.001 $\pm$ 0.014 &0.001 $\pm$ 0.011 \\
QC yes & 0.017 $\pm$ 0.008 & 0.007 $\pm$ 0.008 & $\emptyset$ $\pm$ 0.007 & 0.002 $\pm$ 0.014 & $\emptyset$ $\pm$ 0.010
\end{tabular}

\vspace{10pt}
\begin{tabular}{cccccc} 
\hline\hline\noalign{\vskip 1.5pt}
 Bias ($\Delta\langle z \rangle$)& bin 6  & bin 7  & bin 8  &bin 9   \\
 \hline
 QC no & $\emptyset$ $\pm$ 0.013 & 0.009 $\pm$ 0.038 & $-$0.007 $\pm$ 0.054 & $-$0.007 $\pm$ 0.067 & 0.008 $\pm$ 0.015 \\
 QC yes &  0.001 $\pm$ 0.013 &0.009 $\pm$ 0.036 & 0.012 $\pm$ 0.053 & $-$0.006 $\pm$ 0.007 & 0.009 $\pm$ 0.015 \\
\end{tabular}
\tablefoot{The values shown are the mean biases over 50 different lines of sight as well as standard deviations for all ten bins. Under the recognition of QC, six bins lie within the \Euclid bias requirement, while bins 1, 8, and 10 fail to satisfy the requirement by errors on the order of $\sim$ 0.001 only, showcasing the improvement achievable with photo-$z$ based tomography.}
\label{tab:A.3}
\end{table*}

\begin{figure*}[ht!]
\centering
\includegraphics[width=18cm]{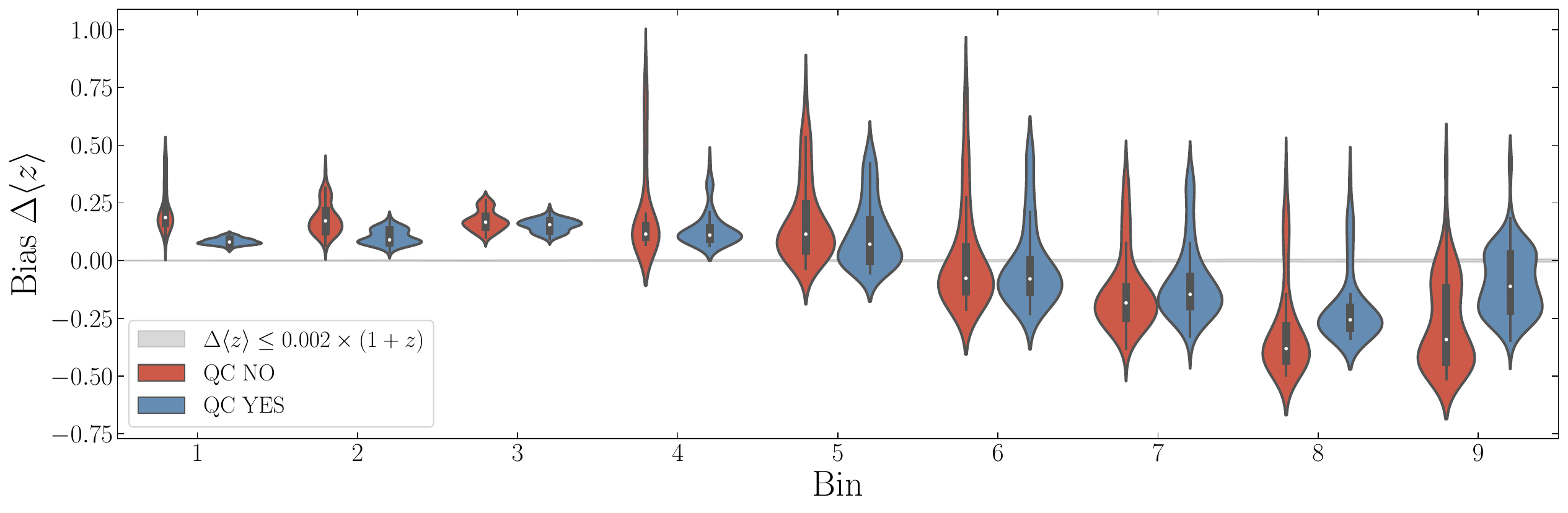}
\caption[]{Violin plot of biases per bin according to \cref{m1}.}
\label{fig:A.1}
\end{figure*}

\begin{figure*}[ht!]
\centering
\includegraphics[width=18cm]{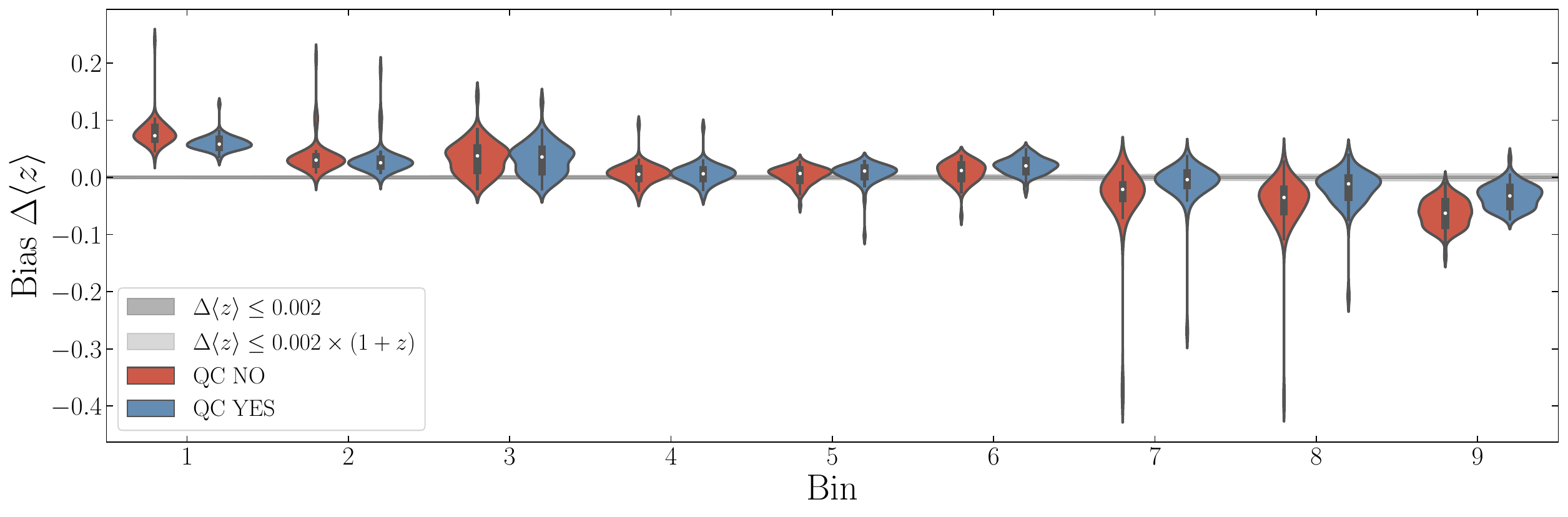}
\caption[]{Violin plot of biases per bin according to \cref{m2}.}
\label{fig:A.2}
\end{figure*}


\begin{figure*}[ht!]
\centering
\includegraphics[width=18cm]{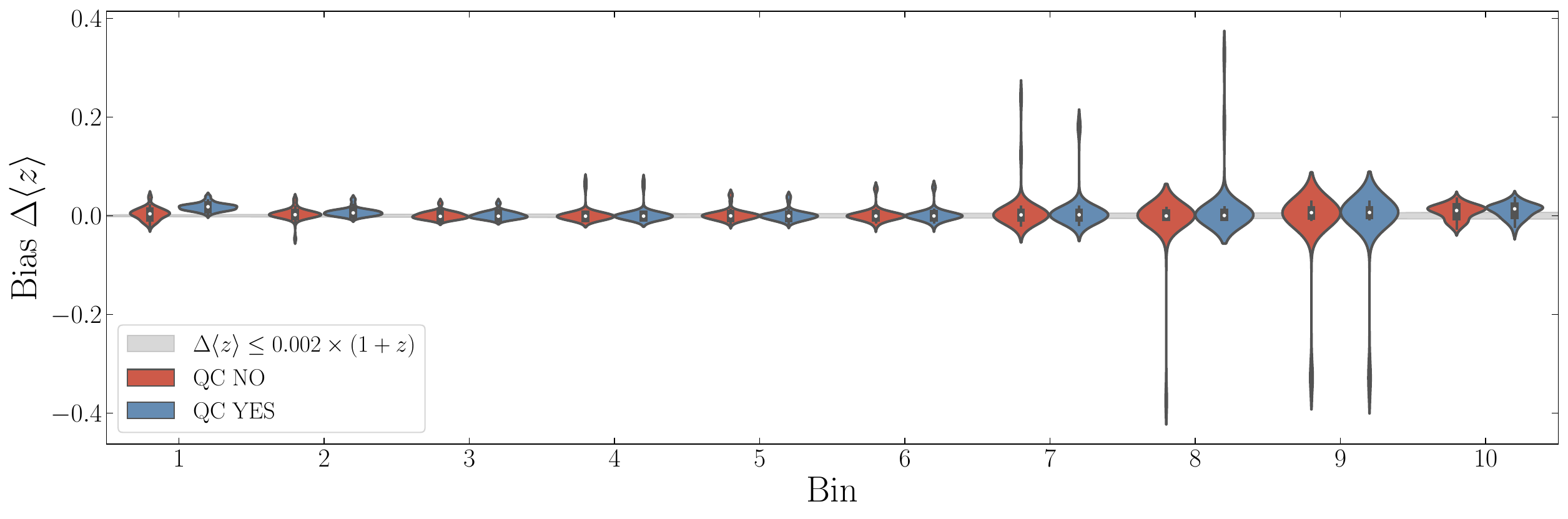} 
\includegraphics[width=18cm]{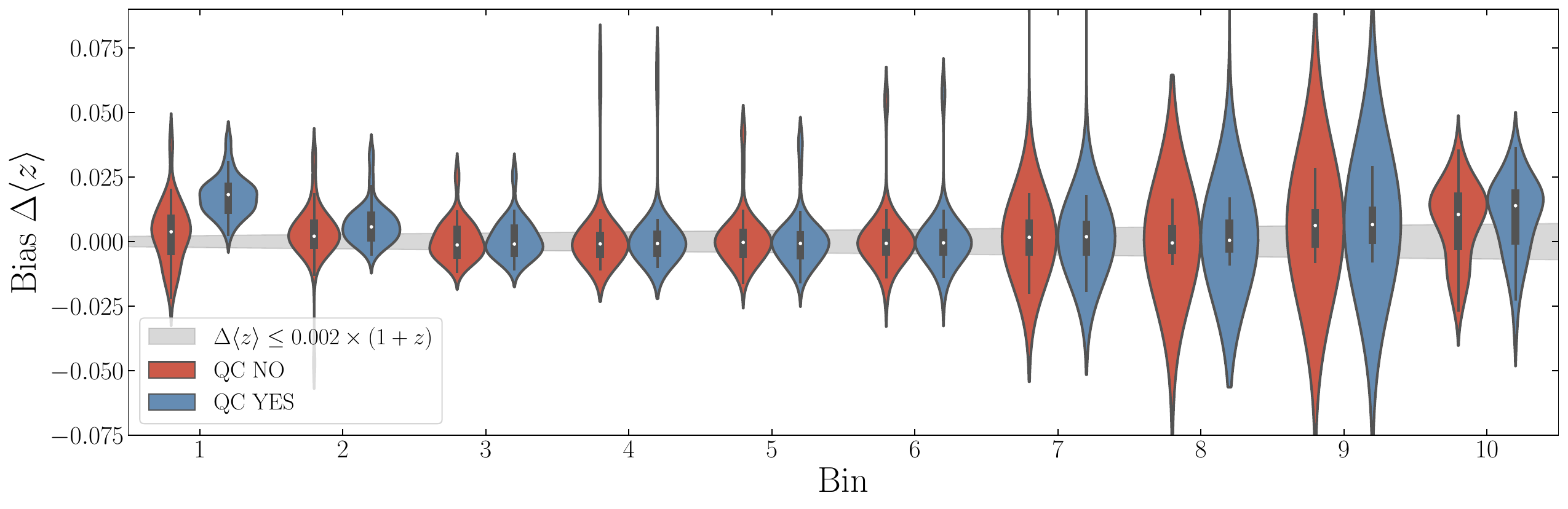}
\caption[]{Violin plot of biases per bin according to \cref{m3}. \emph{Top}: Includes the entire distributions. \emph{Bottom}: $y$-axis limited version, effectively removing 2\% of extreme outliers from the sample to uncover the mean biases with respect to the requirement.}
\label{fig:A.3}
\end{figure*}

\newpage
\section{Calibration coverage per tomography}

Depending on the chosen tomography, the fractional use of the spectroscopic sample for calibration varies significantly. As displayed in \cref{fig:43}, photo-$z$ based tomography consistently displays fractional coverage above $\sim 98\%$ per bin, as its selection is no longer linked to the mean spec-$z$ of SOM cells. This cannot be said for SOM-based tomography, where underrepresented bins at low and high redshift face up to $\sim 25\%$ loss of calibration galaxies. 

\begin{figure}[ht!]
\centering
\includegraphics[width=9cm]{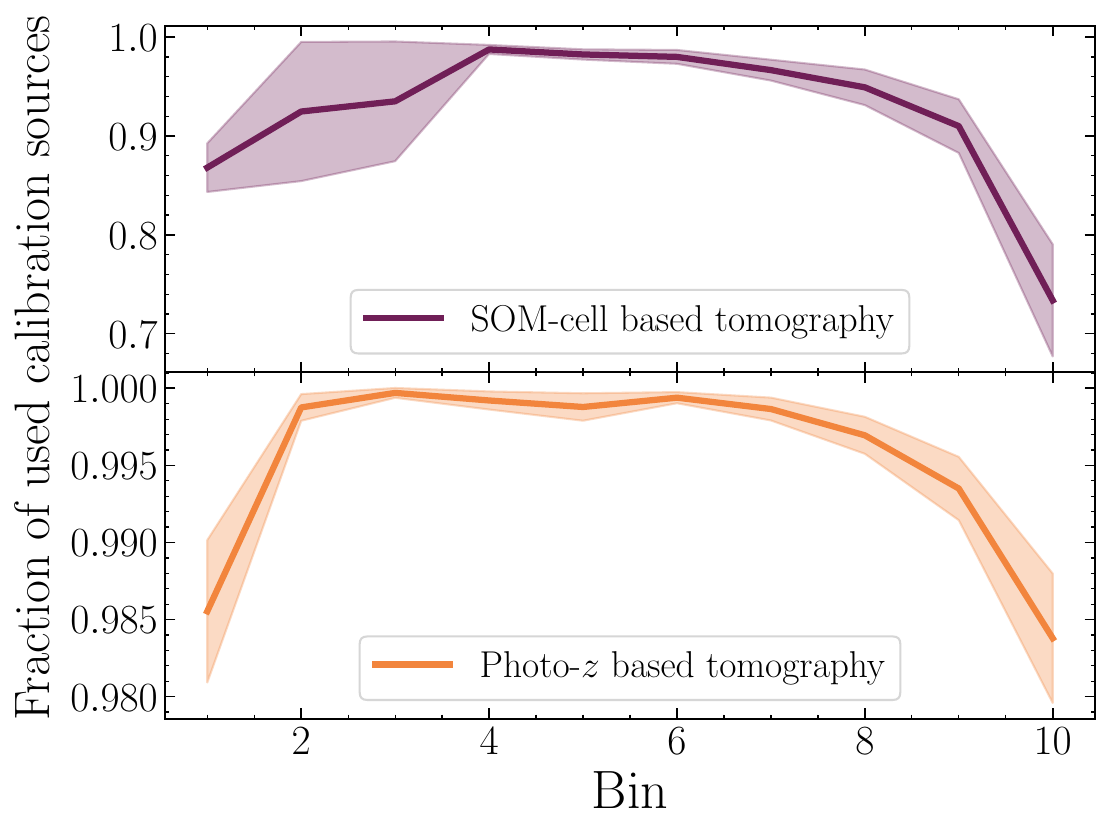}
\caption[]{Visualisation of the effective fraction of calibration data considered after applying QC and gold sample selection for the tomography defined on the mean spec-$z$ of SOM cells (purple) and photo-$z$ (orange).}
\label{fig:43}
\label{LastPage}
\end{figure}

\end{appendix}

\end{document}